\documentclass[times,twocolumn,final]{elsarticle}

\usepackage{medima}
\usepackage{framed,multirow}

\usepackage[normalem]{ulem}
\usepackage{url}
\usepackage{xcolor}
\usepackage{hyperref}
\usepackage{longtable, array, booktabs}
\usepackage{pdflscape}
\usepackage{comment}
\usepackage{todonotes}
\usepackage[export]{adjustbox}
\usepackage{graphicx}
\usepackage{bm}
\usepackage{subfigure}
\usepackage{caption} 
\usepackage{longtable}
\usepackage{multirow}
\usepackage[utf8]{inputenc}
\usepackage[T1]{fontenc}
\usepackage{gensymb}
\usepackage{amssymb}
\usepackage{makecell}
\usepackage{pgfplots}
\usepackage{siunitx}
\usepackage{graphicx}
\usepackage{tikz}

\definecolor{newcolor}{rgb}{.8,.349,.1}
\definecolor{bronze}{HTML}{cd7f32}
\definecolor{Carmine}{HTML}{960018}
\definecolor{LightPink}{HTML}{E1CDDB}
\definecolor{LightCyan}{HTML}{A2E2DC}
\definecolor{Savanna}{HTML}{D1C592}
\definecolor{Burgundy}{HTML}{C6A3A7}
\definecolor{Grass}{HTML}{A1BA8E}
\definecolor{Lavender}{HTML}{B0A1BF}
\definecolor{Lime}{HTML}{CEE2C6}
\definecolor{Blue}{HTML}{A7BED8}
\definecolor{Green}{HTML}{A8CA86}
\definecolor{Rose}{HTML}{E1ACB5}
\definecolor{Tangerine}{HTML}{FCB891}
\definecolor{Gold}{HTML}{D0D090}
\definecolor{Gray}{HTML}{B5B5A7}
\definecolor{Purple}{HTML}{B8B5EF}
\definecolor{Yellow}{HTML}{E1D195}
\definecolor{CiteColor}{HTML}{45A3BA}
\definecolor{LightGreen}{HTML}{7aeb7a}
\definecolor{ForestGreen}{HTML}{2ec205}
\definecolor{RoyalPurple}{HTML}{7851A9}
\definecolor{ChromeYellow}{HTML}{FFA700}
\definecolor{lightbrown}{HTML}{b5651d}
\definecolor{Royalazure}{HTML}{0038A8}
\definecolor{r1col}{rgb}{0, 0, 0}
\definecolor{r2col}{rgb}{0, 0, 0}
\definecolor{r3col}{rgb}{0, 0, 0}
\definecolor{r4col}{rgb}{0, 0, 0}
\definecolor{r5col}{rgb}{0, 0, 0}
\usepackage{color, colortbl}
\usepackage{array}
\usepackage{here}
\newcolumntype{P}[1]{>{\scriptsize\raggedright\arraybackslash}p{#1}}
\newcolumntype{D}[1]{>{\scriptsize\raggedright\arraybackslash}p{#1}}
\newcolumntype{C}[1]{>{\hangindent=1em\scriptsize\raggedright\arraybackslash}p{#1}
}
\newcolumntype{E}[1]{>{\tiny\raggedright\arraybackslash}p{#1}}

\newcommand{\CPath}{CPath}

\newcommand\ro[1]{\textcolor{cyan}{\textbf{Babak}}}

\begin{document}
\begin{frontmatter}
\title{Computational Pathology: A Survey Review and The Way Forward}%
\author[1]{Mahdi {S. Hosseini}\corref{cor1}}
\cortext[cor1]{Corresponding author.\newline
\textit{E-mail address:} \href{mahdi.hosseini@concordia.ca}{mahdi.hosseini@concordia.ca}}
\author[2]{Babak {Ehteshami Bejnordi}\fnref{fn1}}
\fntext[fn1]{Qualcomm AI Research is an initiative of Qualcomm Technologies, Inc.}
\author[3]{Vincent Quoc-Huy {Trinh }}
\author[4]{Lyndon {Chan}}
\author[4]{Danial {Hasan}}
\author[4]{Xingwen {Li}}
\author[4]{Stephen {Yang}}
\author[4]{Taehyo {Kim}}
\author[4]{Haochen {Zhang}}
\author[4]{Theodore {Wu}}
\author[4]{Kajanan {Chinniah}}
\author[1]{Sina {Maghsoudlou}}
\author[4]{Ryan {Zhang}}
\author[4]{\newline Jiadai {Zhu}}
\author[4]{Samir  {Khaki}}
\author[5]{Andrei {Buin}}
\author[1]{Fatemeh {Chaji}}
\author[6]{Ala {Salehi}}
\author[7]{Bich Ngoc {Nguyen}}
\author[8]{\newline Dimitris  {Samaras}}
\author[4]{Konstantinos {N. Plataniotis}}
\address[1]{Department of Computer Science and Software Engineering (CSSE), Concordia Univeristy, Montreal, QC H3H 2R9, Canada}
\address[2]{Qualcomm AI Research, Qualcomm Technologies Netherlands B.V., Amsterdam, The Netherlands}
\address[3]{Institute for Research in Immunology and Cancer of the University of Montreal, Montreal, QC H3T 1J4, Canada}
\address[4]{The Edward S. Rogers Sr. Department of Electrical \& Computer Engineering (ECE), University of Toronto, Toronto, ON M5S 3G4, Canada}
\address[5]{Huron Digitial Pathology, St. Jacobs, ON N0B 2N0, Canada}
\address[6]{Department of Electrical and Computer Engineering, University of New Brunswick, Fredericton, NB E3B 5A3, Canada}
\address[7]{University of Montreal Hospital Center, Montreal, QC H2X 0C2, Canada}
\address[8]{Department of Computer Science, Stony Brook University, Stony Brook, NY 11794, United States}

\begin{abstract}
\textbf{Abstract.} Computational Pathology ({\CPath}) is an interdisciplinary science that augments developments of computational approaches to analyze and model medical histopathology images. The main objective for {\CPath} is to develop infrastructure and workflows of digital diagnostics as an assistive CAD system for clinical pathology, facilitating transformational changes in the diagnosis and treatment of cancer \textcolor{r2col}{that are mainly address by {\CPath} tools}. With evergrowing developments in deep learning and computer vision algorithms, and the ease of the data flow from digital pathology, currently {\CPath} is witnessing a paradigm shift. Despite the sheer volume of engineering and scientific works being introduced for cancer image analysis, there is still a considerable gap of adopting and integrating these algorithms in clinical practice. This raises a significant question regarding the direction and trends that are undertaken in {\CPath}. In this article we provide a comprehensive review of more than 700 papers to address the challenges faced in problem design all-the-way to the application and implementation viewpoints. We have catalogued each paper into a model-card by examining the key works and challenges faced to layout the current landscape in {\CPath}. We hope this helps the community to locate relevant works and facilitate understanding of the field's future directions. In a nutshell, we oversee the {\CPath} developments in cycle of stages which are required to be cohesively linked together to address the challenges associated with such multidisciplinary science. We overview this cycle from different perspectives of data-centric, model-centric, and application-centric problems. We finally sketch remaining challenges and provide directions for future technical developments and clinical integration of {\CPath}. For updated information on this survey review paper and accessing to the original model cards repository, please refer to \href{https://github.com/AtlasAnalyticsLab/CPath_Survey}{GitHub}. \newline

\end{abstract}

\end{frontmatter}

{\small\tableofcontents}

\section{Introduction}
\label{sec:introduction}
April 2017 marked a turning point for digital pathology when the Philips IntelliSite digital scanner received \textcolor{r5col}{the US Food \& Drugs Administration} (FDA) approval (with limited use case) for diagnostic applications in clinical pathology \cite{FDA_Philips_Intellisite,evans2018us}. A subsequent validation guideline was created to help ensure the produced Whole Slide Image (WSI) scans could be used in clinical settings without compromising patient care, while maintaining similar results to the current gold standard of optical microscopy \cite{araujo2019performance,williams2018digital,531,kuo2019optical}. The use of WSIs offers significant advantages to the pathologist's workflow: digitally captured images, compared to tissue slides, are immune from accidental physical damage and maintain their quality over time \cite{pell2019use, al2012digital}. Clinics and practices can share and store these high-resolution images digitally enabling asynchronous viewing/collaboration worldwide \cite{griffin2017digital,saco2016current}. The development of \textit{digital pathology} shows great promise as a framework to improve work efficiency in the practice of pathology \cite{kaushal2021validation,saco2016current}. Adopting a digital workflow also opens immense opportunities for using computational methods to augment and expedite their workflow--the field of \textit{Computational Pathology (CPath)} is dedicated to researching and developing these methods \cite{van2021deep,cui2021artificial,echle2021deep,acs2020artificial,SALVI2021104129,SRINIDHI2021101813}. 

However, despite the aforementioned advantages, the adoption of digital pathology, and hence computational pathology, has been slow. Some pathologists consider the analysis of WSIs as opposed to glass slides as an unnecessary change in their workflow \cite{lujan2022challenges,liu2019digital,griffin2017digital,jara2010digital} and recent surveys indicate that the switch to digital pathology does not provide enough financial incentive \cite{smith2022road,sundar2020awareness,buabbas2021evaluating,baidoshvili2018evaluating, Dennis2005TheUO, al2012digital}. This is where advances from CPath can address or overpower many of the concerns in adopting a digital workflow. For example, CPath models to identify morphological features that correlate with breast cancer \cite{209} provide substantial benefits to clinical accuracy. Further, CPath models that identify lymph node metastases with better sensitivity while reducing diagnostic time \cite{138} can streamline workflows to increase pathologist throughput and generate more revenue \cite{cooper2012digital, kim2022application}. 

Similar to digital pathology, the adoption of CPath methods has also lagged despite the many benefits it offers to improve efficiency and accuracy in pathology \cite{CPsurvey1, kumar2020whole, evans2018us, 370}. This lack of adoption and integration into clinical practice raises a significant question regarding the direction and trends of current work in CPath. This survey looks to review the field of CPath in a systematic fashion by breaking down the various steps involved in a CPath workflow and categorizing CPath works to both determine trends in the field and provide a resource for the community to reference when creating new works.

Existing survey papers in the field of CPath can be clustered into a few groups. The first focuses on the design and applications of smart diagnosis tools \cite{139, niazi2019digital, sobhani2021artificial, klein2021artificial, go2022digital, rony2019deep, SRINIDHI2021101813, SALVI2021104129, acs2020artificial, abinaya2022systematic, bilal2022role, schneider2022integration, jiang2020emerging, lancellotti2021artificial}. These works focus on designing novel architectures for \textcolor{r5col}{artificial intelligence} (AI) models with regards to specific clinical tasks, although they may briefly discuss clinical challenges and limitations. A second group of works focus on clinical barriers for AI integration discussing specific certifications and regulations required for the development of medical devices under clinical settings \cite{colling2019artificial,baxi2022digital, sakamoto2020narrative, cheng2021challenges, brixtel2022whole, shmatko2022artificial}. Lastly, the final group focuses on both the design and the integration of AI tools with clinical applications \cite{makhlouf2022general, huo2021ai, van2021deep, cui2021artificial, serag2019translational, kim2022application, wong2022current, alamir2022role, echle2021deep, cifci2022artificial, haggenmuller2021skin}. These works speak to both the computer vision and pathology communities in developing \textcolor{r1col}{machine learning (ML)} models that can satisfy clinical use cases. 

\begin{figure}[t]
\centering
\includegraphics[width=0.48\textwidth]{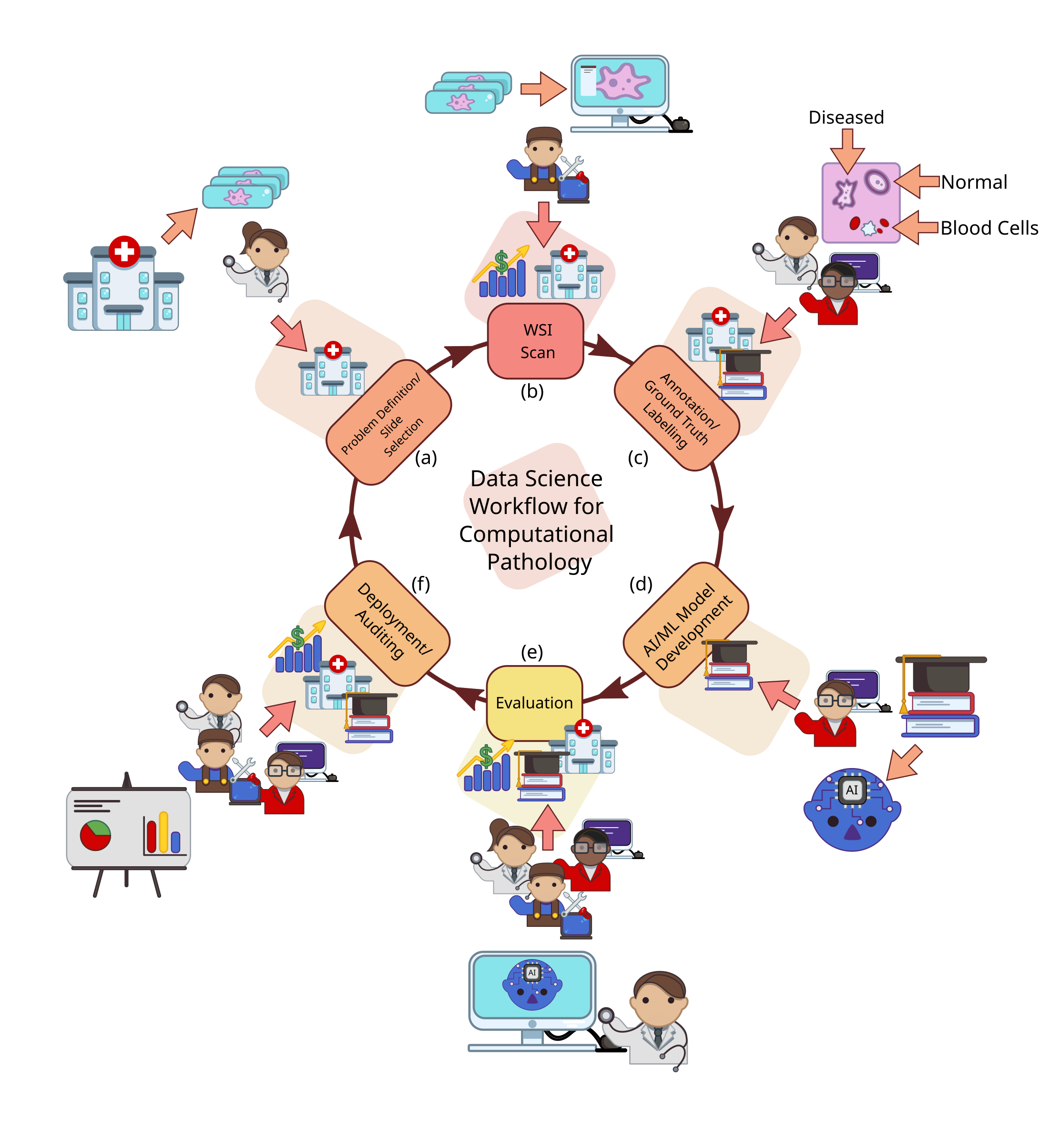} 
\caption{We divide the data science workflow for pathology into multiple stages, wherein each brings a different level of experience. For example, the annotation/ground truth labelling stage (c) is where domain expert knowledge is consulted as to augment images with associated metadata. Meanwhile, in the evaluation phase (e), we have computer vision scientists, software developers, and pathologists working in concert to extract meaningful results and implications from the representation learning.}
\label{fig:data_science_workflow}
\end{figure}

Our work is situated in this final group as we breakdown the end-to-end CPath workflow into stages and systematically review works related to and addressing those stages. We oversee this as a workflow for CPath research that breaks down the process of problem definition, data collection, model creation, and clinical validation into a cycle of stages. A visual representation of this cycle is provided in Figure \ref{fig:data_science_workflow}. We review over 700 papers from all areas of the CPath field to examine key works and challenges faced. By reviewing the field so comprehensively, our goal is to layout the current landscape of key developments to allow computer scientists and pathologists alike to situate their work in the overall CPath workflow, locate relevant works, and facilitate an understanding of the field's future directions. We also adopt the idea of generating model cards from \cite{mitchell2019model} and designed a card format specifically tailored for CPath. Each paper we reviewed was catalogued as a model card that concisely describes (1) the organ of application, (2) the compiled dataset, (3) the machine learning model, and (4) the target task. The complete model card categorization of the reviewed publications is provided in Appendix \ref{subsec:supp_model_card} for the reader's use. 

In our review of the CPath field, we find that two main approaches emerge in works: 1) a data-centric approach and 2) a model-centric approach. Considering a given application area, such as specific cancers, e.g. breast ductal carcinoma in-situ (DCIS), or a specific task, e.g. segmentation of benign and malignant regions of tissue, researchers in the CPath field focus generally on either improving the data or innovating on the model used. 

Works with data-centric approaches focus on collecting pathology data and compiling datasets to train models on certain tasks based on the premise that the transfer of domain expert knowledge to models is captured by the process of collecting and labeling high-quality data \cite{gu2021lessons, huo2021ai, tomaszewski2021overview}. The motivation behind this approach in CPath is driven by the need to 1) address the lack of labeled WSI data representing both histology and histopathology cases due to the laborious annotation process \cite{baidoshvili2018evaluating} and 2) capture a predefined pathology ontology provided by domain expert pathologists for the class definitions and relations in tissue samples. Regarding the lack of labeled WSI data our analysis reveals that there are a larger number of datasets with granular labels, but there is a larger total amount of data available for a given organ and disease application that have weakly supervised labels at the Slide or Patient-level. Although some tasks, such as segmentation and detection, require WSI data to have more granular labels at the region-of-interest (ROI) or image mosaic/tiles (known as patch) levels, to capture more precise information for training models, there is a potential gap to leverage the large amount of weakly-supervised data to train models that can be later \textcolor{r1col}{used} downstream on smaller strongly-supervised datasets for those tasks. \textcolor{r2col}{When considering the ontology of pathology as compared to the field of computer vision, we note that pathology datasets have far fewer classes than computer vision (e.g. ImageNet-20K contains 20,000 class categories for natural images \cite{ImageNet_Dataset} whereas CAMELYON17 has four annotated classes for breast cancer metastases \cite{Camelyon17}), but has much more variation within each of these classes in terms of representations and fuzzy boundaries around the \textit{grade} of cancers which subdivides each class into many more in reality.} There are also very rare classes in the form of rare diseases and cancers, as presented in Figure \ref{fig:cancer-statistics} and discussed in Section \ref{sec:application}, which present a class imbalance challenge when compiling data or training models. If one considers the complexities involved in representational learning of related tissues and diseases, it raises the question of whether there is a clear understanding and consensus in the field of how an efficient dataset should be compiled for model development. Our survey analyzes the availability of CPath datasets along with what area of application they address and their annotation level in detail in Section \ref{data-avail}, and the complete table of datasets we have covered is available in Appendix \ref{subsec:supp_dat}. Section \ref{sec:AnnotationLabellingWorkflow} goes into more depth about the various levels of annotation, the annotation process, and selecting the appropriate annotation level for a task. 

The  model-centric approach, by contrast, is favoured by computer scientists and engineers, who design algorithmic approaches based on the available pathology data. Selection of a modelling approach, such as self-supervised, weakly-supervised, or strongly-supervised learning, is dictated directly by the amount of data available for a given annotation level and task. Currently, many models are developed on datasets with strongly-supervised labels at the ROI, Patch, or Pixel-levels to address tasks such as tissue type classification or disease detection. However, a recent trend is developing to apply self-supervised and weakly-supervised learning methods to leverage the large amount of data with Slide and Patient-level annotations \cite{35}. Models are trained in a self or weakly supervised manner to learn representations on a wider range of pathology data across organs and diseases, which can be leveraged for other tasks requiring more supervision but without the need for massive labeled datasets \cite{60, 279, chen2022scaling}. This trend points to the future direction of CPath models following a similar trend to that in computer vision, where large-scale models are being pre-trained using self-supervised techniques to achieve state-of-the-art performance in downstream tasks \cite{simclr, ViT1}. 

Although data and model centric approaches are both important in advancing the performance of models and tools in CPath, we note a need for much more \textit{application} centric work in CPath. We define a study to be application centric if the primary focus is on addressing a particularly impactful task or need in the clinical workflow, ideally including clinical validation of the method or tool. To this end, Section \ref{sec:application} details the clinical pathology workflow from specimen collection to report generation, major task categories in CPath, and specific applications per organ. Particularly, we find that very few works focus on the pre or post-analytical phases of the pathology workflow where many errors can occur, instead focusing on the analytical phase where interpretation tasks take place. Additionally, certain types of cancer with deadly survival rates are underrepresented in CPath datasets and works. Very few CPath models and tools have been validated in a clinical setting by pathologists, suggesting that there may still be massive barriers to actually using CPath tools in practice. All of this points to a severe oversight by the CPath community towards considering the actual application and implementation of tools in a clinical setting. We suspect this to be a major reason as to why there is a slow uptake in adopting CPath tools by pathology labs.

The contributions of this survey include the provision of an end-to-end workflow for developing CPath work which outlines the various stages involved and is reflected within the survey sections. Further, we propose and provide a comprehensive conceptual model card framework for CPath that clearly categorizes works by their application of interest, dataset usage, and model, enabling consistent and easy comparison and retrieval of papers in relevant areas. Based on our analysis of the field, we highlight several challenges and trends, including the availability of datasets, focus on models leveraging existing data, disregard of impactful application areas, and lack of clinical validation. Finally, we give suggestions for addressing these aforementioned challenges and provide directions for future work in the hopes of aiding the adoption and implementation of CPath tools in clinical settings.

The structure of this survey closely follows the CPath data workflow illustrated in Figure \ref{fig:data_science_workflow}. Section \ref{sec:application} begins by outlining the clinical pathology workflow and covers the various task domains in CPath, along with organ specific tasks and diseases. The next step of the workflow involves the processes and methods of histopathology data collection, which is outlined in Section \ref{sec:HistoDataCollection}. Following data collection, Section \ref{sec:AnnotationLabellingWorkflow} details the corresponding annotation and labeling methodology and considerations. Section \ref{sec:models-list} covers deep learning designs and methodologies for CPath applications. Section \ref{sec:EvalReg} focuses on regulatory measures and clinical validation of CPath tools. Section \ref{sec:emerging-trends} explores emerging trends in recent CPath research. Finally, we provide our perceived challenges and future outlook of CPath in Section \ref{sec:Outlook}.
\section{Clinical Applications for CPath} \label{sec:application}
The field of CPath is dedicated to the creation of tools that  address and aid steps in the clinical pathology workflow. Thus, a grounded understanding of the clinical workflow is of paramount importance before development of any CPath tool. The outcomes of clinical pathology are diagnostics, prognostics, and predictions of therapy response. Computational pathology systems that focus on diagnostic tasks aim to assist the pathologists in tasks such as tumour detection, tumour grading, quantification of cell numbers, etc. Prognostic systems aim to predict survival for individual patients while therapy response predictive models aid personalized treatment decisions based on histopathology images. Figure \ref{fig:task_ex} visualizes the goals pertaining to these tasks. In this section, we provide detail on the clinical pathology workflow, the major application areas in diagnostics, prognostics, and therapy response, and finally detail the cancers and CPath applications in specific organs. The goal is to outline the tasks and areas of application in pathology where CPath tools and systems can be developed and implemented.

\subsection{Clinical Pathology Workflow} \label{clinical_workflow_section}
\begin{figure}[t]
\centering
\includegraphics[width=0.48\textwidth]{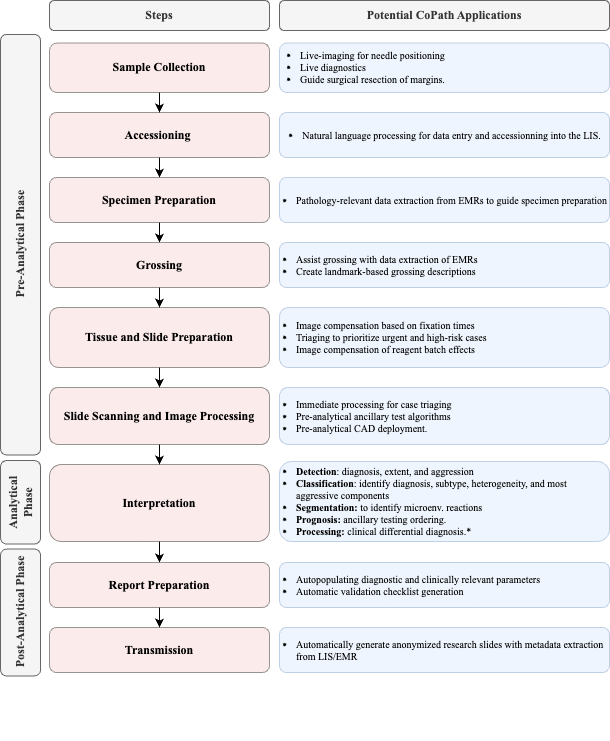}
\caption{Quality assurance and control phases developed by pathologists to oversee the clinical pathology workflow into three main phases of pre-analytical, analytical, and post-analytica phases. We further show how each of these processes can be augmented under the potential CPath applications in an end-to-end pipeline.}
\label{fig:clinical_workflow}
\end{figure}

\textcolor{r1col}{
This subsection provides a general overview of the clinical workflow in pathology covering the collection of a tissue sample, its subsequent processing into a slide, inspection by a pathologist, and compilation of the analysis and diagnosis into a pathology. Figure \ref{fig:clinical_workflow} summarizes these steps at a high level and provides suggestions for corresponding CPath applications. The steps are organized under the conventional pathology phases for samples: pre-analytical, analytical, and post-analytical. These phases were developed to categorize quality control measures, as each phase has its own set of potential sources of errors \cite{njoroge2014risk}, and thus potential sources of corrections during which CPath and healthcare artificial intelligence tools could prove useful. For details about each step of the workflow, please refer to the Appendix \ref{sec:appendix_clinical_workflow}.
}

\textbf{\textit{Pre Analytical Phase}}
\textcolor{r1col}{
The first step of the \textit{pre-analytical} phase is a biopsy performed to collect a tissue sample, where the biopsy method is dependent on the type of sample required and the tissue characteristics. Sample collection is followed by accessioning of the sample which involves entering of the patient and specimen information into a Laboratory Information System (LIS) and linking to the Electronic Medical Records (EMR) and potentially a Slide Tracking System (STS). After accessioning, smaller specimens that have not already been preserved by fixation in formalin are fixated. Once the basic specimen preparation has occurred, the tissue is analyzed by the pathology team without the use of a microscope; a step called grossing. Grossing involves cross-referencing the clinical findings and the EMR reports, with the operator localizing the disease, locating the pathological landmarks, describing these landmarks, and measuring disease extent. Specific sampling of these landmarks is performed, and these samples are then put into cassettes for the final fixation. Subsequently, the samples are then sliced using a microtome, stained using the relevant stains for diagnosis, and covered with a glass slide.
}

\textbf{\textit{Analytical Phase}}
\textcolor{r1col}{
After a slide is processed and prepared, a pathologist views the slide to analyze and interpret the sample. The approach to interpretation varies depending on the specimen type. Interpretation of smaller specimens is focused on diagnosis of any disease. Analysis is performed in a decision-tree style approach to add diagnosis-specific parameters, e.g. esophagus biopsy $\rightarrow$ type of sampled mucosa $\rightarrow$ presence of folveolar-type mucosa $\rightarrow$ identify Barrett’s metaplasia $\rightarrow$ identify degree of dysplasia. Once the main diagnosis has been identified and characterized, the pathologist sweeps the remaining tissue for secondary diagnoses which can also be characterized depending on their nature. Larger specimens are more complex and usually focus on characterizing the tissue and identifying unexpected diagnoses beyond the prior diagnosis from a small specimen biopsy. Microscopic interpretation of large specimens is highly dependent on the quality of the grossing and the appropriate detection and sampling of landmarks. Each landmark (e.g., tumor surface, tumor at deepest point, surgical margins, lymph node in mesenteric fat) is characterized either according to guidelines, if available, or according to the pathologist’s judgment. After the initial microscopic interpretation additional deeper cuts (``levels''), special stains, immunohistochemistry (IHC), and/or molecular testing may be performed to hone the diagnosis by generating new material or slides from the original tissue block.
}

\textbf{\textit{Post-Analytical Phase}}
\textcolor{r1col}{
The pathologist synthesizes a diagnosis by aggregating their findings from grossing and microscopic examination in combination with the patient’s clinical information, all of which are included in a final pathology report. The classic sections of a pathology report are patient information, a list of specimens included, clinical findings, grossing report, microscopic description, final diagnosis, and comment. The length and degree of complexity of the report again depends on the specimen type. Small specimen reports are often succinct, clearly and unambiguously listing relevant findings which guide treatment and follow-up. Large specimen reports depend on the disease, for example, in cancer resection specimens the grossing landmarks are specifically targeted at elements that will guide subsequent treatment.
}

\textcolor{r1col}{In the past, pathology reports had no standardized format, usually taking a narrative-free text form. Free text reports can omit necessary data, include irrelevant information, and contain inconsistent descriptions \cite{renshaw_synoptic}. To combat this, synoptic reporting was introduced to provide a structured and standardized reporting format specific to each organ and cancer of interest \cite{renshaw_synoptic, hewer2020oncologist}. Over the last 15 years, synoptic reporting has enabled pathologists to communicate information to surgeons, oncologists, patients, and researchers in a consistent manner across institutions and even countries. The College of American Pathologists (CAP) and the International Collaboration on Cancer Reporting (ICCR) are the two major institutions publishing synoptic reporting protocols. The parameters included in these protocols are determined and updated by CAP and ICCR respectively to remain up-to-date and relevant for diagnosis of each cancer type. For the field of computational pathology, synoptic reporting provides a significant advantage in dataset and model creation, as a pre-normalized set of labels exist across a variety of cases and slides in the form of the synoptic parameters filled out in each report. Additionally, suggestion or prediction of synoptic report values are a possible CPath application area.
}

\subsection{Diagnostic Tasks}
\begin{figure}[t]
\centering
\includegraphics[width=0.45\textwidth]{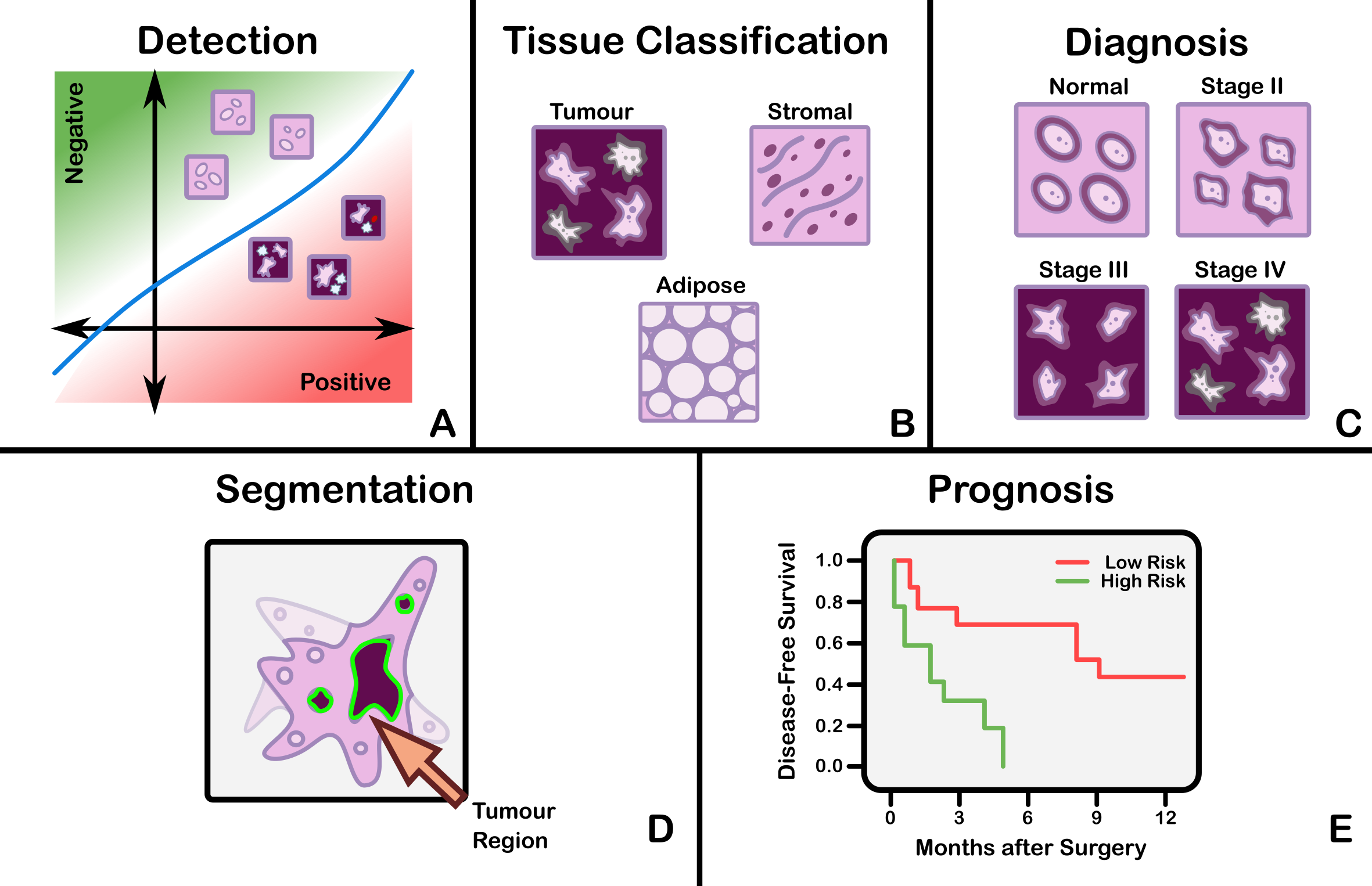}
\caption{The categorization of diagnostic tasks in computational pathology along with examples A) \textbf{Detection:} common detection task such as differentiating positive from negative classes like malignant from benign, B) \textbf{Tissue Subtype Classification:} classification task for tumorous tissue, Stroma, and adipose tissue, C) \textbf{Disease Diagnosis:} common disease diagnosis task like cancer staging, D) \textbf{Segmentation:} tumor segmentation in WSIs, and E) \textbf{Prognosis tasks:} shows a graph comparing survival rate and months after surgery.}
\label{fig:task_ex}
\end{figure}

Computational pathology systems that focus on diagnostic tasks can broadly be categorized as: (1) disease detection, (2) tissue subtype classification, (3) disease diagnosis, and (4) segmentation\textcolor{r4col}{.} These tasks are visually depicted in Figure \ref{fig:task_ex}. Note how the detection tasks all involve visual analysis of the tissue in WSI format. Thus computer vision approach is primarily adopted towards tackling diagnostic tasks in \textcolor{r5col}{computer aided diagnosis} (CAD). For additional detail on some previous works on these diagnostic tasks, we refer the reader to Appendix \ref{sec:appendix_diagnostic}

\textit{\textbf{Detection}}
We define the detection task as a binary classification problem where inputs are labeled as positive or negative, indicating the presence or absence of a certain feature. There may be variations in the level of annotation required, e.g. slide-level, patch-level, pixel-level detection depending on the feature in question. Although detection tasks may not provide an immediate disease diagnosis, it is a highly relevant task in many pathology workflows as pathologists incorporate the presence or absence of various histological features into synoptic reports that lead to diagnosis. Broadly, detection tasks fall into two main categories: (1) screening the presence of cancers and (2) detecting histopathological features specific to certain diagnoses. 

\textcolor{r1col}{Cancer detection algorithms can assist the pathologists by filtering obviously normal WSIs and directing pathologist’s focus to metastatic regions \cite{15}. Although pathologists have to review all the slides to check for multiple conditions regardless of the clinical diagnosis, an accurate cancer detection CAD would expedite the workflow by pinpointing the ROIs and summarizing results into synoptic reports, ultimately leading to a reduces time per slide. Due to this potential impact, cancer detection tasks have been explored in a broad set of organs. Additionally, the simple labeling in binary detection tasks allows for deep learning methods to generalize across different organs where similar cancers form \cite{141,147,171}. }

\textit{\textbf{Tissue Subtype Classification}}
\textcolor{r1col}{Treatment and patient prognosis can vary widely depending on the stage of cancer, and finely classifying specific tissue structures associated with a specific disease type provides essential diagnostic and prognostic information \cite{10}. Accordingly, accurately classifying tissue subtypes is a crucial component of the disease diagnosis process. As an example, discriminating between two forms of glioma (a type of brain cancer), glioblastoma multiforme and lower grade glioma, is critical as they differ by over $45\%$ in patient survival rates \cite{34}. Additionally, accurate classification is key in colorectal cancer (CRC) diagnosis, as high morphological variation in tumor cells \cite{55} makes certain forms of CRC difficult to diagnose by pathologists \cite{52}. We define this classification of histological features as tissue subtype classification. }


\textit{\textbf{Disease Diagnosis}}
\textcolor{r1col}{The most frequently explored design of deep learning in digital pathology involves emulating pathologist diagnosis. We define this multi-class diagnosis problem as a disease diagnosis task. Note the similarity with detection--disease diagnosis can be considered a fine-grained classification problem which subdivides the general positive disease class into finer disease-specific labels based on the organ and patient context.}

\textit{\textbf{Segmentation}}
The segmentation task moves one step beyond classification by adding an element of spatial localization to the predicted label(s). \textcolor{r4col}{In semantic segmentation, objects of interest are delineated in an image by assigning class labels to every pixel. These class labels can be discrete or non-discrete, the latter being a more difficult task \cite{chan2021comprehensive}}. Another variant of the segmentation task is instance segmentation, which aims to achieve both pixel-level segmentation accuracy as well as clearly defined object (instance) boundaries. Segmentation approaches can accurately capture many morphological statistics \cite{29} and textural features \cite{107}, both of which are relevant for cancer diagnosis and prognosis. Most frequently, segmentation is used to capture characteristics of individual glands, nuclei, and tumor regions in WSIs. For instance, glandular structure is a critical indicator of the severity of colorectal carcinoma \cite{241}, thus accurate segmentation could highlight particularly abnormal glands to the pathologist as demonstrated in \cite{241, 96, 105}. Overall, segmentation provides localization and classification of cancer-specific tumors and of specific histological features that can be meaningful for the pathologist's clinical interpretation.

\subsection{Prognosis}
Prognosis involves predicting the likely development of a disease based on given patient features. For accurate survival prediction, models must learn to both identify and infer the effects of histological features on patient risk. Prognosis represents a merging of the diagnosis classification task and the disease-survivability regression task. 

Training a model for prognosis requires a comprehensive set of both histopathology slides and patient survival data (i.e. a variant of multi-modal representation learning). Despite the complexity of the input data, ML models are still capable of extracting novel histological patterns for disease-specific survivability \cite{37,26,111}. Furthermore, strong models can discover novel prognostically-relevant histological features from WSI analysis \cite{42,240}. As the quality and comprehensiveness of data improves, additional clinical factors could be incorporated into deep learning analysis to improve prognosis.

\subsection{Prediction of treatment response}
With the recent advances in targeted therapy for cancer treatment, clinicians are able to use treatment options that precisely identify and attack certain types of cancer cells. While the number of options for targeted therapy are constantly increasing, it becomes increasingly important to identify patients who are potential therapy responders to a specific therapy option and avoid treating non-responding patients who may experience severe side effects. Deep learning can be used to detect structures and transformations in tumour tissue that could be used as predictive markers of a positive treatment response. Training such deep learning models usually requires large cohorts of patient data for whom the specific type of treatment option and the corresponding response is known. 

\subsection{Organs and Diseases}\label{sec:organs_diseases}
\begin{figure}[t]
\centering
\includegraphics[width=0.35\textwidth]{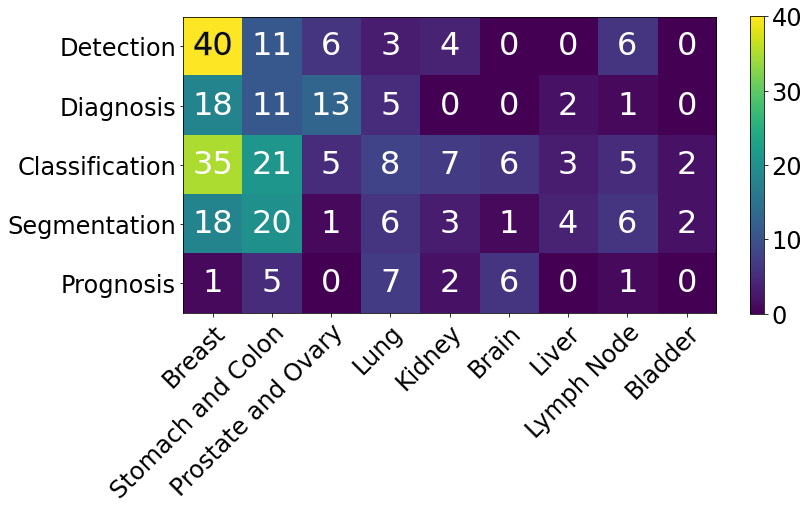}
\caption{
Distribution of diagnostic tasks in CPath for different organs from Table \ref{OV}. This distribution includes more than 400 cited works from 2018 to 2022 inclusive. The x-axis covers different organs, the y-axis displays different diagnostic tasks, and the height of the bars along the vertical axis measures the number of works that have examined the specific task and organ. Please refer to Table \ref{OV} in the supplementary section for more information.}
\label{fig:organ_overview}
\end{figure}

This section presents an overview of the various anatomical application areas for computational pathology grouped by the targeted organ. Each organ section gives a brief overview of the types of cancers typically found and the content of the pathology report as noted from the corresponding CAP synoptic reporting outline (discussed at \ref{clinical_workflow_section}). Figure \ref{fig:organ_overview} highlights the intersection between the major diagnostic tasks and the anatomical focuses in state-of-the-art research. The majority of papers are dedicated to the four most common cancer sites: breast, colon, prostate, and lung \cite{Cancer2020}. Additionally, a significant amount of research is also done on cancer types with highest mortality, brain and liver. \cite{Cancer2020}. Note that details of some additional works that may be of interest for each organ type can be found in Appendix \ref{sec:organs_and_diseases}

\textit{\textbf{Breast}}
Breast cancers can start from different parts of the breast and majorly consist of 1) Lobular cancers that start from lobular glands, 2) Ductal cancers, 3) Paget cancer which involves the nipple, 4) Phyllodes tumor that stems from fat and connective tissue surrounding the ducts and lobules, and 5) Angiosarcoma which starts in the lining of the blood and lymph vessels. In addition, based on whether the cancer has spread or not, breast cancers can be categorized into \textit{in situ} or \textit{invasive}/\textit{infiltrating} forms. DCIS is a precancerous state and is still confined to the ducts. Once the cancerous cells grow out of the ducts, the carcinoma is now considered \textit{invasive} or \textit{infiltrative} and can metastasize \cite{what-breast}.

Synoptic reports for breast cancer diagnosis are divided based on the type of cancers mentioned above. For DCIS and invasive breast cancers, synoptic reports focus on the histologic type and grade, along with the nuclear grade, evidence of necrosis, margin, involvement of regional lymph nodes, and biomarker status.  Notably, architectural patterns are no longer a valuable predictive tool compared to nuclear grade and necrosis to determine a relative ordering of diagnostic importance for DCIS \cite{architect-DCIS}. In contrast to DCIS and invasive cancers, Phyllodes tumours vary due to their differing origin in the fat and connective tissue, focusing on analyzing the stroma characteristics, existence of heterologous elements, mitotic rate, along with the involvement of lymph nodes. Finally, to determine therapy response and treatments, biomarker tests for Estrogen, Progesterone \cite{allison2020estrogen} and HER-2 \cite{wolff2018her2} receptors are recommended, along with occasional tests for Ki67 antigens \cite{dowsett2011assessment, healey2017assessment}. 

Most breast cancer-focused works in CPath propose various solutions for carcinoma detection and metastasis detection, an important step for assessing cancer stage and morbidity. Metastasis detection using deep learning methods was shown to outperform pathologists' exhaustive diagnosis by $9\%$ \textcolor{r5col}{free-response receiver operating characteristic} (FROC) in \cite{43}.

\textit{\textbf{Prostate}}
Prostate cancer is the second most prevalent cancer among the total population and the most common cancer among men (both excluding non-melanoma skin cancers). However, most prostate cancers are not lethal. Prostate cancer can occur in any of the three prostate zones: Central (CZ), Peripheral (PZ), and Transition (TZ), in increasing order of aggressiveness. Prostate cancers are almost always adenocarcinomas, which develop from the gland cells that make prostate fluid. The other types of prostate cancers are small cell carcinomas, neuroendocrine tumors, transitional cell carcinomas, isolated intraductal carcinoma, and sarcomas (which are very rare). Other than cancers, there are multiple conditions that are important to identify or diagnose as precursors to cancer or not. Prostatic intraepithelial neoplasia (PIN) is diagnosed as either low-grade PIN or high-grade PIN. Men with high-grade PIN need closely monitored follow-up sessions to screen for prostate cancer. Similarly, atypical small acinar proliferation (ASAP) is another precancerous condition requiring follow-up biopsies. \cite{what-prostate}


To grade and score tumours, pathologists use a Tumor, Nodes, Metastasis (TNM) framework. In the synoptic report, pathologists identify and report the histologic type and grades, and involvement of regional lymph nodes to help grade and provide prognosis for any tumours. Specifically for prostate analysis, tumour size and volume are both important factors in prognosis according to multiple studies \cite{ito2019organ, epstein2011prognostic, salomon2003prognostic, stamey1999biological}. Similarly, location is important to note for both prognosis and therapy response \cite{lee2015biologic}. Invasion to nearby (except perineural invasion) tissues is noted and can correlate to TMN classification \cite{prostate-invasion}. Additionally, margin analysis is especially important in prostate cancers as \textcolor{r4col}{the presence of} a positive margin \textcolor{r4col}{increases the risk of} cancer recurrence \textcolor{r4col}{and} metastasis \cite{wright2010positive}. Finally, intraductal carcinoma (IDC) must be identified and distinguished from PIN and PIA; as it is strongly associated with a high Gleason score, a high-volume tumor, and metastatic disease \cite{varma2021intraductal, montironi2018features, zhou2013intraductal, cohen2007proposal, guo2006intraductal}.

After a prostate cancer diagnosis is established, pathologists assign a Gleason Score to determine the cancer’s grade: a grade from 1 to 5 is assigned to the two most common areas and those two grades are summed to make a final Gleason Score \cite{prostate-diagnosis}. For Gleason scores of 7, where survival and clinical outcomes demonstrate large variance, the identification of Cribriform glands is key in helping to narrow possible outcomes \cite{kweldam2015cribriform, lee2018spectrum}.

\textit{\textbf{Ovary}}
Ovarian cancer is the deadliest gynecologic malignancy and accounts for more than $14,000$ deaths each year \cite{bachert2020serous}. Ovarian cancer manifests in three types: 1) epithelial cell tumors that start from the epithelial cells covering the outer surface of the ovary, 2) germ cell tumors which start from the cells that produce eggs, and 3) stromal tumors which start from cells that hold the ovary together and produce the hormones estrogen and progesterone. Each of these cancer types can be classified into benign, intermediate and malignant categories. Overall, epithelial cell tumors are the most common ovarian cancer and have the worst prognosis \cite{what-ovary}.
 
When compiling a synoptic report for ovarian cancer diagnosis, pathologists focus on histologic type and grade, extra-organ involvement, regional lymph nodes, T53 gene mutations, and serous tubal intraeptithelial carconma (STIC). Varying histologic tissue types are vital to determine the pathology characteristics and determining eventual prognosis. For example, generally endometrioid, mucinous, and clear cell carcinomas have better outcomes than serous carcinomas \cite{yang2021long}. Additionally, lymph node involvement and metastasis in both regional and distant nodes has a direct correlation to patient survival, grading, and treatment. Determining the presence of STICs correlates directly to the presence of ovarian cancer, as $60\%$ of ovarian cancer patients will also have an associated STIC \cite{bachert2020serous}. Finally, T53 gene mutations are the most common in epithelial ovarian cancer; which has the worst prognosis among ovarian cancers, so determining their presence is critical to patient cancer risk and therapy response \cite{vermij2022p53,zhang2016tp53}. There are not a large number of works dedicated to the ovary specifically, but most works on ovary focus on classification of its five most common cancer subtypes: high-grade serous (HGSC), low-grade serous (LGSC), endometriod (ENC), clear cell (CCC), and mucinous (MUC) \cite{253,280}.

\textit{\textbf{Lung}}
Lung cancer is the third most common cancer type, next to breast and prostate cancer \cite{national_cancer_institute}. Lung cancers mostly start in the bronchi, bronchioles, or alveoli and are divided into two major types, non-small cell lung carcinomas (NSCLC) ($80-85\%$) and small cell lung carcinomas (SCLC) ($10-15\%$). Although NSCLS cancers are different in terms of origin, they are grouped because they have similar outcomes and treatment plans. Common NSCLS cancers are 1) adenocarcinoma, 2) squamous cell carcinoma 3) large cell carcinoma, and some other uncommon subtypes \cite{what-lung}.


For reporting, histologic type helps determine NSCLC vs SCLC and the subtype of NSCLC. Although NSCLC generally has favourable survival rates and prognosis as compared to SCLC, certain subtypes of NSCLC can have lower survival rates due to co-factors \cite{yoshizawa2011impact}. Histologic patterns are applicable in adenocarcinomas, consisting of favourable types: lepidic, intermediate: acinar and papillary, and unfavourable: micropapillary and solid \cite{moreira2020grading}. Grading each histologic type aids in categorization but is differentiated based on each type, and thus is out of scope for this paper. Importantly for lung cancers, tumour size is an independent prognostic factor for early cancer stages, lymph node positivity, and locally invasive disease. Additionally, the size of the invasive portion is an important factor for prognosis of nonmucinous adenocarcinoma with lepidic pattern \cite{tsutani2012prognostic, maeyashiki2013size, amin2017eighth, wang2018tumor, zhang2015relationship, yoshizawa2011impact}. Other important lung specific features are visceral pleural invasion, which is associated with worse prognosis in early-stage lung cancer even with tumors < 3cm \cite{gao2022peripheral}, and lymphatic invasion which indicates an unfavourable prognostic finding \cite{tsutani2012prognostic, amin2017ajcc}.

\textit{\textbf{Colon and Rectum}}
Colorectal cancers are two of the five most common cancer types \cite{Cancer2020}. Cancer cells usually start to develop in the innermost layer of the colon and rectum walls, known as the mucosa, and continue their way up to other layers. In other layers, there are lymph and blood vessels that can be used by cancer cells to travel to nearby lymph nodes or other organs \cite{american2020colorectal}.  Colorectal cancers usually start with the creation of different types of polyps, each possessing a unique risk of developing into cancer. Most colorectal cancers are adenocarcinomas, which are split into three well-studied subtypes: classic adenocarcinoma (AC), signet ring cell carcinoma (SRCC), and mucinous adenocarcinoma (MAC). In most cases, AC has a better prognosis than MAC or SRCC. Other types, albeit uncommon, of colorectal cancers are: carcinoid tumors, gastrointestinal stromal tumors (GISTs), lymphomas, and sarcomas \cite{what-colon}.

As in other cancers, histologic grade is the most important factor in cancer prognosis along with regional lymph node status and metastasis. The tumor site is also important in determining survival rates and prognosis \cite{you2020american}. Vascular invasion of both small and large vessels are important factors in adverse outcomes and metastasis \cite{lim2010prognostic, santos2013clinicopathological, gomez2014critical}, and perineural invasion has been shown in multiple studies to be an indicator of poor prognosis \cite{liebig2009perineural, ueno2013study, gomez2014critical}. Additionally, microsatellite instability (MSI) is shown to be a good indicator of prognosis and is divided into three stages in decreasing adversity of Stable (MSI-S), Low (MSI-L), and High (MSI-H) \cite{phipps2013colon}. Finally, some studies have indicated the usefulness of biomarkers in colorectal cancer treatment, with biomarkers such as BRAF mutations, KRAS mutations, MSI, APC, Micro-RNA, and PIK3CA \cite{vacante2018biomarkers}. 

Works are relatively well-distributed among various tasks including disease diagnosis, segmentation, and detection. Expanding on colorectal cancer detection, work from \cite{144} used feature analysis for colorectal and mucinous adenocarcinomas using heatmap visualizations. They discovered that adenocarcinoma is often detected by ill-shaped epithelial cells and that misclassification can occur due to lumen regions that resemble the malformed epithelial cells. Similarly for mucinous carcinoma, the model again recognizes the dense epithelium, but this time ignores the primary characteristic of the carcinoma (abundance of extracellular mucin). These findings suggest that a thorough analysis of class activation maps can be helpful for improving the classifier's accuracy and intuitiveness.

\textit{\textbf{Bladder}}
There are several layers within the bladder wall with most cancers starting in the internal layer, called the urothelium or transitional epithelium. Cancers remaining in the inner layer are non-invasive or carcinoma in situ (CIS) or stage 0. If they grow into other layers such as the muscle or fatty layer, the cancer is now \textit{invasive}. Nearly all bladder cancers are urothelial carcinomas or transitional cell carcinomas (TCC). However, there are other types of cancer such as squamous cell carcinomas, adenocarcinomas, small cell carcinomas, and sarcomas which all are very rare. In the early stages, all types of bladder cancers are treated similarly but as their stage progresses, and chemotherapy is needed, different drugs might be used based on the type of the cancer \cite{what-bladder}. As with other organs, histologic type and grade also play a role in prognosis and treatment \cite{chalasani2009histologic}, and lymphovascular invasion is independently associated with poor prognosis and recurrence \cite{lotan2005lymphovascular}.



Works focusing on the bladder display promising results that could lead to rapid clinical application. For example, a prediction method for four molecular subtypes (basal, luminal, luminal p53, and double negative) of muscle-invasive bladder cancer was proposed in \cite{61},  outperforming pathologists by $30\%$ in classification accuracy when restricted to a tissue morphology-based assessment. Further improvements in accuracy could help expedite diagnosis by complementing traditional molecular testing methods.

\textit{\textbf{Kidney}}
Each kidney is made up of thousands of glomeruli which feed into the renal tubules. Kidney cancer can occur in the cells that line the tubules (renal cell carcinoma (RCC)), blood vessels and connective tissue (sarcomas), or urothelial cells (Urothelial carcinoma). RCC accounts for about $90\%$ of kidney cancers and comes in two types: 1) clear cell renal carcinoma, which are most common and 2) non-clear cell renal carcinoma consisting of papillary, chromophobe and some very rare subtypes \cite{kidney_acs_2020}. The \textcolor{r5col}{CAP's} cancer protocol template for the kidney is solely focused on RCCs \cite{srigley2020protocol}, likely due to their high probability. Tumour size is directly associated with malignancy rates, with $1$cm size increases resulting in $16\%$ increases in malignancy \cite{bonsib2006renal}. Additionally, the RCC histologic type \textcolor{r4col}{is} correlated with metastasis, with clear cell, capillary, collecting ducts (Bellini), and medullary being the most aggressive \cite{muglia2015renal}. 

Many works are focused on glomeruli segmentation, as the number of glomeruli and glomerulosclerosis constitute standard components of a renal pathology report~\cite{kannan_morgan}. In addition to glomeruli detection, some works have also detected other relevant features such as tubules, Bowman's capsules, and arteries \cite{225}. The results display strong performance on PAS-stained nephrectomy samples and tissue transplant biopsies, and there seems to be a strong correlation between the visual elements identified by the network and those identified by renal pathologists.

\textit{\textbf{Brain}}
There are two main types of brain tumors: malignant and non-malignant. Malignant tumors can be classified as primary tumors (originating from the brain) or secondary (metastatic) \cite{gehan1977prognostic, li2021systematic}. The most common type of brain cancers are gliomas, occurring $50.4\%$ of the time, and are classified into four grades \cite{gliomapercentage}. In the synoptic reporting, tumour location is noted as it has some impact on the prognosis, with parietal tumours showing better prognosis compared to other locations \cite{gehan1977prognostic}. Additionally, focality of glioblastomas (a subtype of gliomas) is important to determine as multifocal glioblastoma is far more aggressive and resistant to chemotherapy as compared to unifocal \cite{li2021systematic}. A recent summary of the World Health Organization's (WHO) classification of tumors of the central nervous system has indicated that biomarkers as both ancillary and diagnostic predictive tools \cite{louis20212021}. Additionally, in a recent WHO edition of classification of tumours of the central nervous system, molecular information is now integrated with histologic information into tumor diagnosis for cases such as diffuse gliomas and embryonal tumors \cite{cns_tumours_who}.

Accordingly, most works focus on gliomas and more specifically glioblastoma, the most aggressive and invasive form of glioma. Due to glioblastoma's extremely low survival rate of $5\%$ after 5 years, compared to a low grade glioma's survival rate of over $50\%$ after 5 years \cite{www.cancer.ca,34}, it is critical to distinguish the two forms for improved patient care and prognosis.

\textit{\textbf{Liver}}
Liver cancer is one of the most common causes of cancer death \cite{world_health_organization}. In particular, hepatocellular carcinoma (HCC) is the most common type of primary liver cancer and has various subtypes, but they generally have little impact on treatment \cite{kim2020histopathological}. Histogolical grade is divided into nuclear features and differentiation, which directly correlate to tumour size, presentation, and metastatic rate \cite{lauwers2002prognostic, spolverato2015hepatic}. Notably, high-grade dysplastic nodules are included in synoptic reports for HCC but are difficult to assess and have high inter-observer disagreement \cite{wanless1995terminology}, and thus is an area where CAD systems could be leveraged to normalize assessments. Current grading of this cancer suffers from an unsatisfactory level of standardization \cite{martins_filho_paiva_azevedo_alves_2017}, likely due to the diversity and complexity of the tissue. This could explain why relatively low number of works are dedicated to liver disease diagnosis and prognosis. Instead, most works focus on the segmentation of cancerous tissues.

\textit{\textbf{Lymph Nodes}}
There are hundreds of lymph nodes in the human body that contain immune cells capable of fighting infections. Cancer manifests in lymph nodes in two ways: 1) cancer that originates in the lymph node itself known as lymphoma and 2) cancer cells from different origins that invade lymph nodes \cite{what-lymph}. As mentioned in the prior organ sections, lymphocytic infiltration is correlated with cancer recurrence on multiple organs and lymph nodes are the most common site for metastasis. The generalizable impact to multiple organs and importance of detecting lymphocytic infiltration is why many works focused on lymph nodes address metastasis detection \cite{wang2022lymph}.

\textit{\textbf{Organ Agnostic}}
The remaining papers focus on segmentation, diagnosis, and prognosis tasks that attempt to generalize to multiple organs, or target organ agnostic applications. An interesting approach to increase the generalization capability of deep learning in histopathology is proposed in \cite{110}. Currently, publicly available datasets with thorough histological tissue type annotations are organ specific or disease specific and thus constrain the generalizability of CPath research. To fill this gap, a novel dataset called Atlas of Digital Pathology (ADP) is proposed \cite{110}. This dataset contains multi-label patch-level annotations of Histological Tissue Types (HTTs) arranged in a hierarchical taxonomy. Through supervised training on ADP, high performance on multiple tasks is achieved even on unseen tissue types.

\section{Data Collection for CPath}\label{sec:HistoDataCollection}
One of the first steps in the workflow for any CPath research is the collection of a representative dataset. This procedure often requires large volumes of data that should be annotated with ground-truth labels for further analysis \cite{35,281,279}. However, creating a meaningful dataset with corresponding annotations is a significant challenge faced in the CPath community \cite{35,281,363,279,li2021pathal}.

This section outlines the entire process of the data-centric design approach in CPath, including tissue slide preparation and WSI scanning--the first two stages in the proposed workflow shown in Figure \ref{fig:data_science_workflow}. Additionally, the trend in dataset compilation across the 700 papers surveyed is discussed regarding dataset sizes, public availability, and annotation types; see Table \ref{DC} in the Supplementary Material for information regarding the derivations and investigation of said trends.

\subsection{Tissue Slide Preparation} \label{creation}
For the application development stages in CPath, the creation of a new WSI dataset must begin with selection of relevant glass slides. High quality WSIs are required for effective analysis, however, considerations must be made for potential slide artifacts and variations inherently present. As described in Section \ref{clinical_workflow_section}, pathology samples are categorized as either biopsy or resection samples, with most samples being prepared as \textit{permanent} samples and some intra-operative resection samples being prepared as \textit{frozen} samples. 

\textbf{\textit{Variations and Irregularities}}
\textcolor{r1col}{Throughout the slide sectioning process, artifacts and irregularities can occur which reduce the slide quality, including: uncovered portions, air bubbles in between the glass seal, tissue chatter artifacts, tissue folding and tears, ink markings present on the slide, and dirt, debris, microorganisms, or cross-contamination of slides by unrelated tissue from other organs\cite{Ghaznavi_2012,373,Rolls_2016}. Frozen sections can present unique irregularities and variations, such as freezing artifacts, cracks in the tissue specimen block, or delay of fixation causing drying artifacts\cite{Suvarna_2019, Peters_2016}. Beyond these irregularities, glass slides may vary in stain colouring, occurring due to differences in slide thickness, \textcolor{r5col}{tissue thickness,} fixation, tissue processing schedule, patient variation, stain variation, and lab variation\cite{Yagi_2011,Suvarna_2019,7243333, bejnordi2014quantitative,28, Zarella_2018}. }

\textcolor{r1col}{All such defects and variations are important to keep in mind when selecting glass slides for the development and application process in CPath, as they can both reduce the quality of the WSI as well as impact the performance of developed CAD tools trained with these WSIs \cite{Ghaznavi_2012,373,7243333}. A more detailed discussion on the surveyed works in CPath which seek to identify and correct issues in slide artifacts and colour variation in WSIs is found in Section \ref{secII:variation}. However, prior to digitization, artifacts, and irregularities can be kept at a minimum by following good pathology practices. While an in-depth discussion of this topic is outside the scope of this paper, some research provides an extensive list of recommendations for reducing such errors in slide sectioning \cite{Rolls_2016}.}

\subsection{Whole Slide Imaging (WSI)} \label{scan}
\textbf{\textit{WSI Scan}}
Once a glass slide is prepared, it must be digitized into a WSI. The digitization and processing workflow for WSIs can be summarized as a four-step process \cite{Pantanowitz_2010}: (1) Image acquisition via scanning; (2) Image storage; (3) Image editing and annotation; (4) Image display \cite{pantanowitz2018twenty}. As the first two steps of the digitization workflow are the most relevant for WSI collection and with regards to the CPath workflow, they are discussed to a greater extent below.

Slide scanning is carried out through a dedicated slide scanner device. A plethora of such devices currently exist or are in development; see Appendix Table \ref{tab:1ca} for a collection of commercially available WSI scanners. Additionally, some research has investigated and compared the capabilities and performances of various WSI scanners \cite{hossain2018practical, tabata2019validation, cheng2019assessing, lemaillet2021colorimetrical}.

\textcolor{r1col}{In order to produce a WSI that is in focus, which is especially important for CPath works, appropriate focal points must be chosen across the slide either using a depth map or by selecting arbitrarily spaced tiles in a subset \cite{Indu_2016}. Once focal points are chosen, the image is scanned by capturing tiles or linear scans of the image, which are stitched together to form the full image \cite{Zarella_2018,Indu_2016}. Slides can be scanned at various magnification levels depending on the downstream task and analysis required, with the vast majority being scanned at $20\times$ ($\sim0.5\mu{m}$/pixel) or $40\times$ ($\sim0.25\mu{m}$/pixel) magnification \cite{Zarella_2018}. }

\textbf{\textit{WSI Storage and Standards}}
\textcolor{r1col}{WSIs are in giga-pixel dimension format \cite{CPsurvey1,Herrmann_2018}. For instance a tissue in $1cm\times1{cm}$ size scanned $@0.25\mu{m}$/pixel resolution can produce a $4.8$GB image (uncompressed) with $50,000\times50,000$ pixels. Due to this large size, hardware constraints may not support viewing entire WSIs at full resolution, thus WSIs are most often stored in a tiled format so only the viewed portion of the image (tile) is loaded into memory and rendered \cite{MarquesGodinho_2017}. When building CAD tools for CPath, this large WSI dimensionality must be taken into account in determining how much compute is required to analyze a WSI. Alongside the WSI, metadata regarding patient, tissue specimen, scanner, and WSI information is stored for reference \cite{CPsurvey1, Herrmann_2018,Clunie_2020}. Due to their clinical use, it is important to develop effective storage solutions for WSI images and metadata, allowing for robust data management, querying of WSIs, and efficient data retrieval \cite{Wang_2012, Rao_2018}. Further details on WSI image formats and storage methods are discussed in Appendix \ref{appendix_wsi}}

\textcolor{r1col}{To develop CPath CAD tools in a widespread and general manner, a standardized format for WSIs and their corresponding metadata is essential \cite{Herrmann_2018}. However, there is a general lack of standardization for WSI formats outputted by various scanners, as shown in Table \ref{tab:1ca}, especially regarding metadata storage. The Digital Imaging and Communications in Medicine (DICOM) standard provides a format for CPath image formatting and data management through Supplement 145\cite{Clunie_2020,Singh_2011}, and has been shown in research to allow for efficient access and interoperability of data between varying medical centers and devices \cite{Herrmann_2018}. However, few scanners are DICOM-compliant and thus there are challenges to using different models of scanners, thus different image formats and metadata structures, in the context of dataset aggregation and processing. }


Apart from storage format, a general framework for storing and distributing WSIs is also an important pillar for CPath. In other medical imaging fields such as radiology, images are often stored in a picture archiving and communications systems (PACS) in a standardized DICOM format, with DICOM storage and retrieval protocols to interface with other systems \cite{MarquesGodinho_2017}. The need for standardization persists in pathology for WSI storage solutions; few works have proposed solutions to incorporate DICOM-based WSIs in a PACS, although some research has successfully implemented a WSI PACS consistent using the DICOM standard using a web-based service for viewing and image querying \cite{MarquesGodinho_2017}.

\textbf{\textit{WSI Defects and Variations}}
Certain aspects of the slide scanning process can introduce unfavorable irregularities and variations \cite{kanwal2022devil}. A major source of defects is out-of-focus regions in a generated WSI; often caused by glass slide artifacts, such as air bubbles and tissue folds, which interfere with selection of focus points for a slide \cite{Ghaznavi_2012,228}. Out-of-focus regions degrade WSI quality and are detrimental to the performance of CAD tools developed with these WSIs, presenting concerns for performance with studies showing high false-positive errors \cite{334, 373}. Additionally, as WSIs are scanned in strips or tiles, any misalignment between sections can introduce striping/stitching errors in the final image \cite{Cross_2018}. Another source of error may appear during tissue-background segmentation where the scanner may misidentify some tissue regions as background, potentially missing crucial tissue areas on the glass slide from being digitized \cite{133}. 

Variations in staining refers to differences in colour and contrast of the tissue structures in the final WSI occurring due to differences in the staining process, staining chemicals, and tissue state. Variations in colour can lead to difficulty in generalizing CAD tools to WSIs from different labs, institutions, and settings \cite{48,190}.  \textcolor{r5col}{Even identical staining techniques can yield different WSIs due to scanner differences in sensor design, light source and calibration \cite{Zarella_2018, Duenweg_2023_wsi_scanning}, creating challenges for cross-laboratory dataset generation}. These additional sources of variation add layers of complexity to the WSI processing workflow, and must be kept in mind during slide selection and dataset curation for CAD tool development and deployment.

\textbf{\textit{Addressing Irregularities and Variations}}\label{secII:variation}
Much work has gone into identifying areas of irregularities within WSIs, most notably blur and tissue fold detection \cite{228,334}. Some research has explored automated deep learning tools to identify these irregularities at a more efficient pace than manual inspection \cite{228,334}. Developing techniques for addressing staining variation has also been a significant research area \cite{54,154,187,7243333,627, 539,481} as the use of techniques addressing stain variation is important for all future works. We list some computational approaches proposed to address these issues: An example method proposed in \cite{54} uses a stain normalization technique, attempting to map the original WSI onto a target color profile. In this technique, a color deconvolution matrix is estimated to convert each image to a target \textcolor{r5col}{hematoxylin and eosin} (H$\&$E) color space and each image is normalized to a target image colour profile through spline interpolation \cite{54}. A second approach applies color normalization using the H channel with a threshold on the H channel on a Lymphocyte Detection dataset\cite{627}. Recent studies have shown promise in having deep neural networks accomplish the stain normalization in contrast to the previous classical approaches \cite{21,50,154,347}, commonly applying generative models such as generative adversarial networks (GANs) to stain normalization. Furthermore, Histogram Equalization (HE) technique for contrast enhancement is used in \cite{mehmood2022malignancy}, where novel preprocessing technique is proposed to select and enhance a portion of the images instead of the whole dataset, resulting in improved performance and computational efficiency.

An alternative approach to address the impact of stain variation on training CAD tools is data augmentation. Such methods augment the data with duplicates of the original data, containing adjustments to the color channels of the image, creating images of varying stain coloration, and training train models that are accustomed to stain variations\cite{190}. This method has been frequently used as a pre-processing step in the development of training datasets for deep learning \cite{83,216,286}. A form of medically-irrelevant data augmentation based on random style transfer, called STRAP, was proposed by researchers and outperformed stain normalization \cite{539}. Similar to style transfer, \cite{lin2022unpaired} proposes stain transfer which allows one to virtually generate multiple types of staining from a single stained WSI.

\newcommand\picdims[4][]{%
  \setbox0=\hbox{\includegraphics[#1]{#4}}%
  \clipbox{.5\dimexpr\wd0-#2\relax{} %
           .5\dimexpr\ht0-#3\relax{} %
           .5\dimexpr\wd0-#2\relax{} %
           .5\dimexpr\ht0-#3\relax}{\includegraphics[#1]{#4}}}
\begin{figure}
\setlength{\tabcolsep}{2pt}
\centering
    \begin{tabular}{m{0.23\linewidth} m{0.23\linewidth} m{0.23\linewidth} m{0.23\linewidth}}
    \begin{tikzpicture}
    \draw (0, 0) node[inner sep=0] {\picdims[width=\linewidth]{\linewidth}{0.75\linewidth}{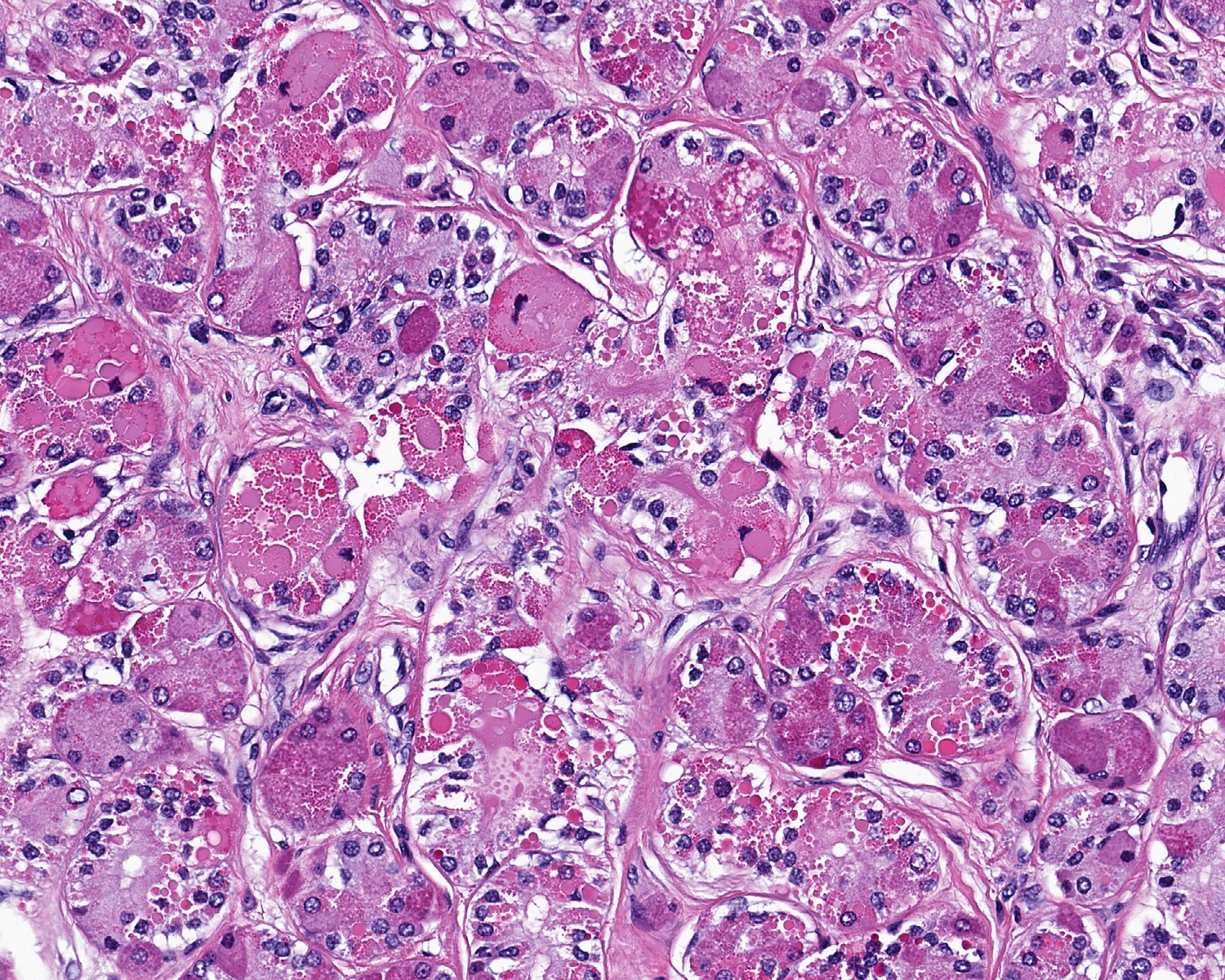}};
    \draw (0, 0.9) node {\tiny H\&E \cite{stain_h&e}};
    \end{tikzpicture} &
    \begin{tikzpicture}
    \draw (0, 0) node[inner sep=0] {\picdims[height=\linewidth]{\linewidth}{0.75\linewidth}{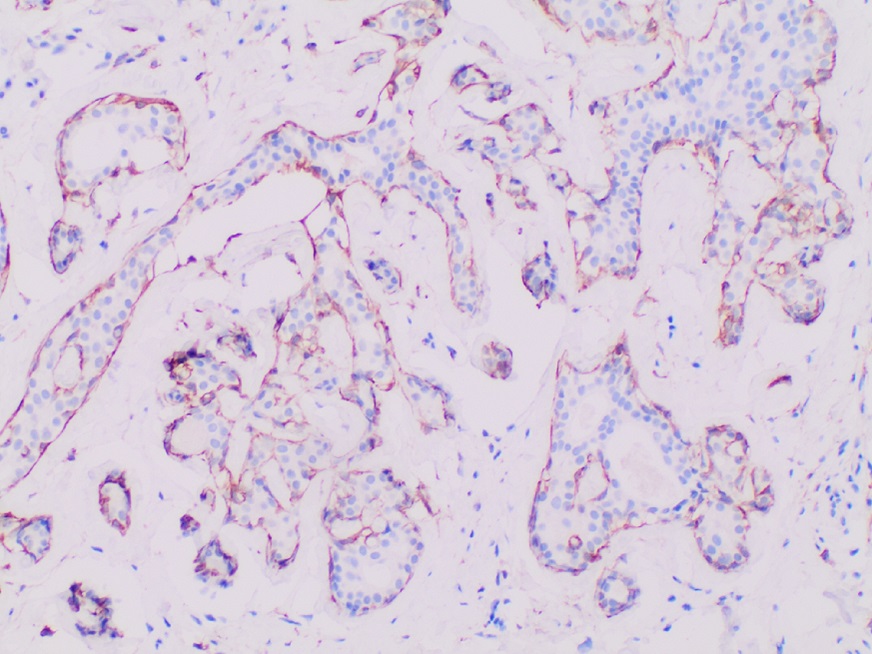}};
    \draw (0, 0.9) node {\tiny IHC \cite{stain_ihc}};
    \end{tikzpicture} &
    \begin{tikzpicture}
    \draw (0, 0) node[inner sep=0] {\picdims[height=\linewidth]{\linewidth}{0.75\linewidth}{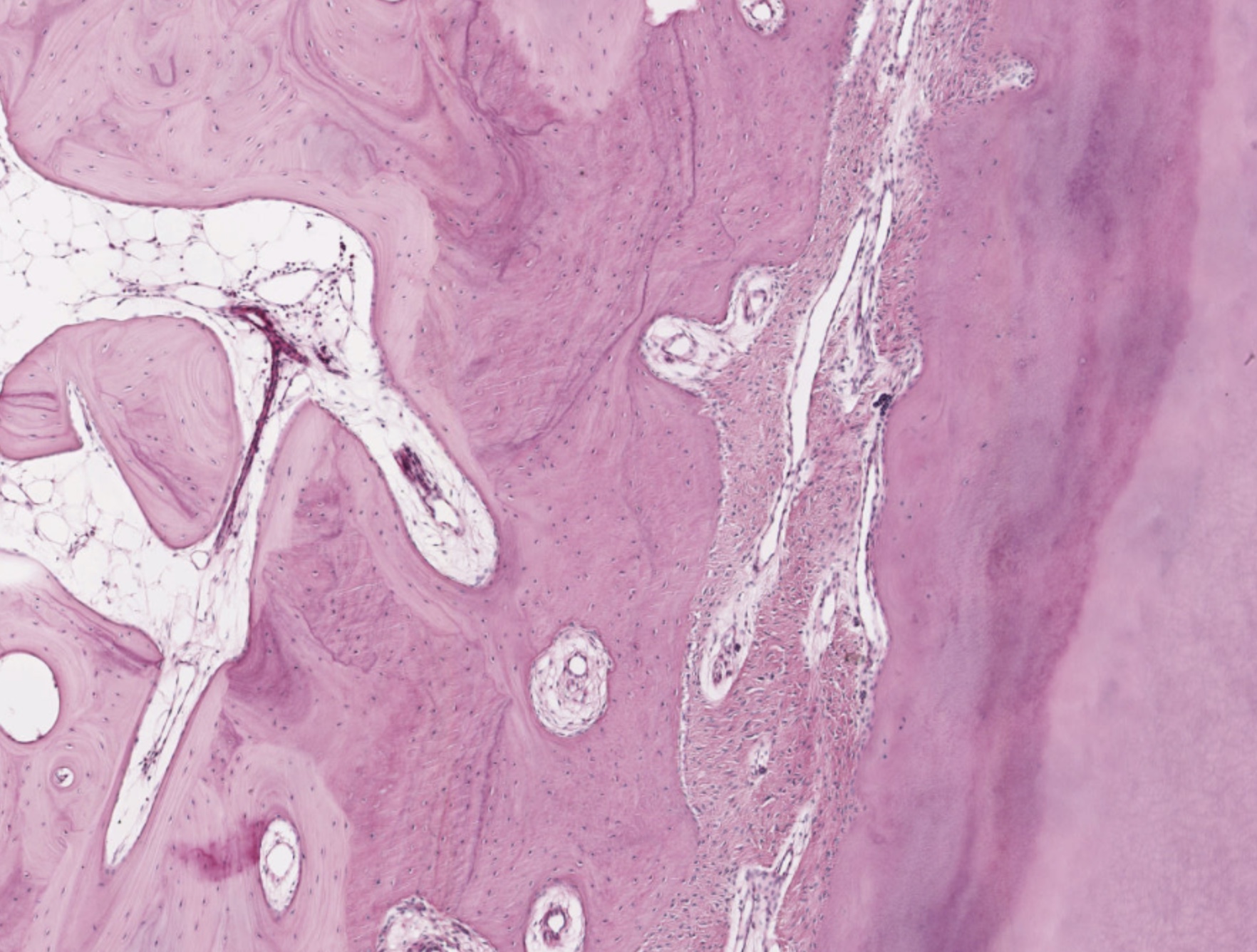}};
    \draw (0, 0.9) node {\tiny PAS \cite{umich_virtual_slide_list}};
    \end{tikzpicture} &
    \begin{tikzpicture}
    \draw (0, 0) node[inner sep=0] {\picdims[height=\linewidth]{\linewidth}{0.75\linewidth}{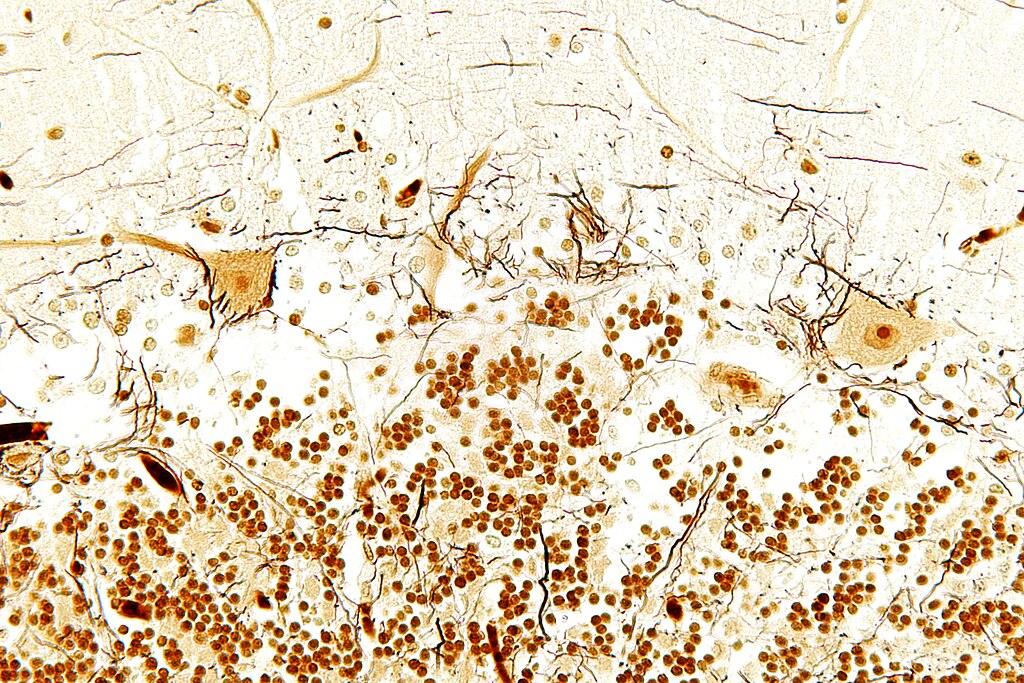}};
    \draw (0, 0.9) node {\tiny Bielschowsky  \cite{stain_bielschowsky}};
    \end{tikzpicture} \\
    \begin{tikzpicture}
    \draw (0, 0) node[inner sep=0] {\picdims[width=\linewidth]{\linewidth}{0.75\linewidth}{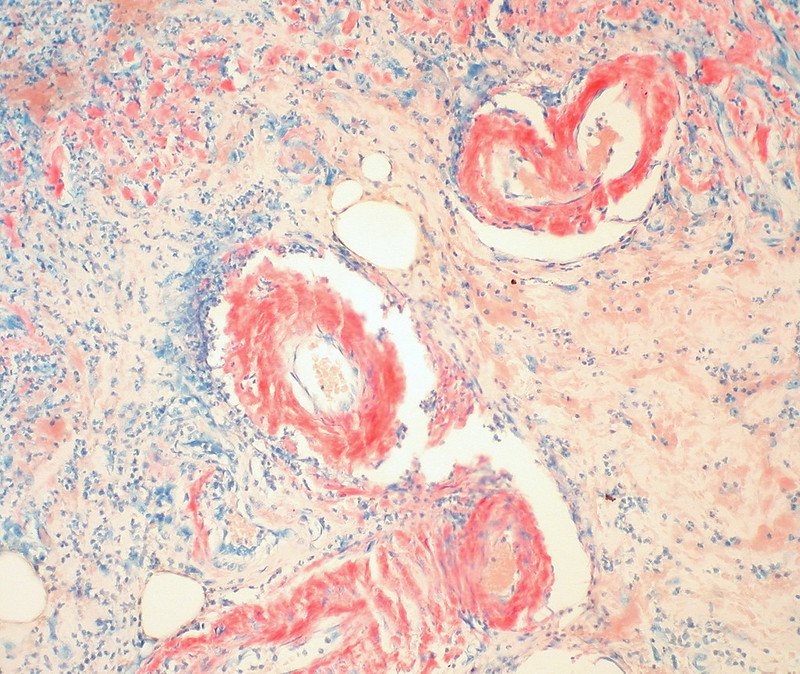}};
    \draw (0, 0.9) node {\tiny Congo Red \cite{stain_congored}};
    \end{tikzpicture} &
    \begin{tikzpicture}
    \draw (0, 0) node[inner sep=0] {\picdims[height=\linewidth]{\linewidth}{0.75\linewidth}{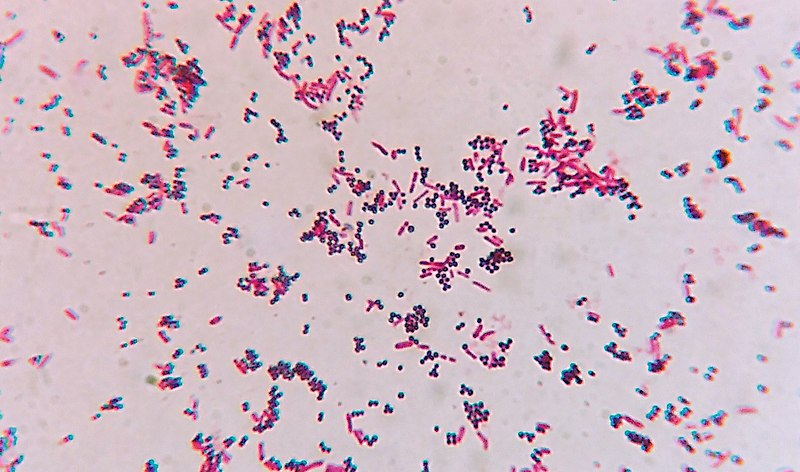}};
    \draw (0, 0.9) node {\tiny Gram \cite{stain_gram}};
    \end{tikzpicture} &
    \begin{tikzpicture}
    \draw (0, 0) node[inner sep=0] {\picdims[height=\linewidth]{\linewidth}{0.75\linewidth}{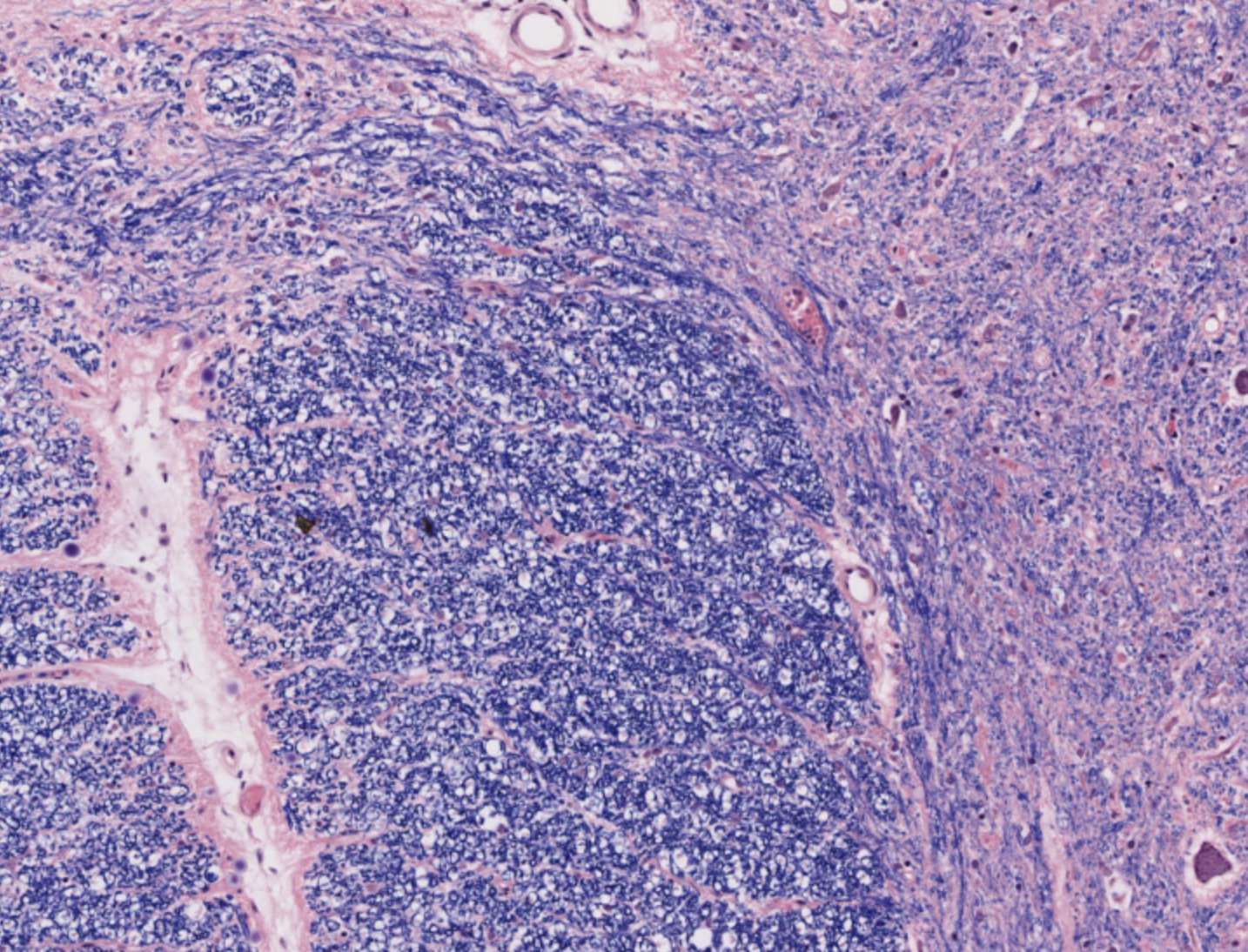}};
    \draw (0, 0.9) node {\tiny Luxol Blue \cite{umich_virtual_slide_list}};
    \end{tikzpicture} &
    \begin{tikzpicture}
    \draw (0, 0) node[inner sep=0] {\picdims[height=\linewidth]{\linewidth}{0.75\linewidth}{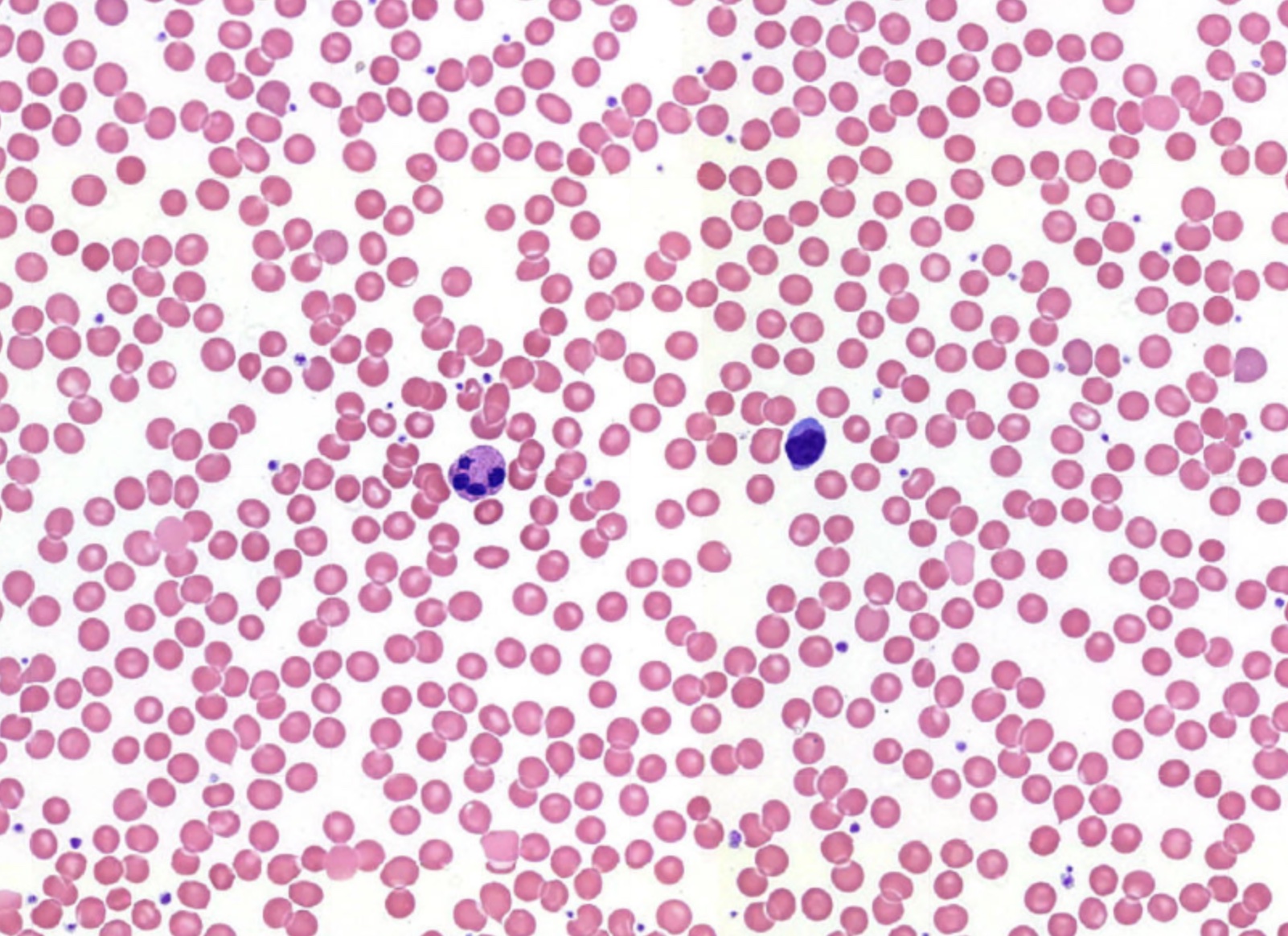}};
    \draw (0, 0.9) node {\tiny Giemsa \cite{umich_virtual_slide_list}};
    \end{tikzpicture} \\
    \begin{tikzpicture}
    \draw (0, 0) node[inner sep=0] {\picdims[height=\linewidth]{\linewidth}{0.75\linewidth}{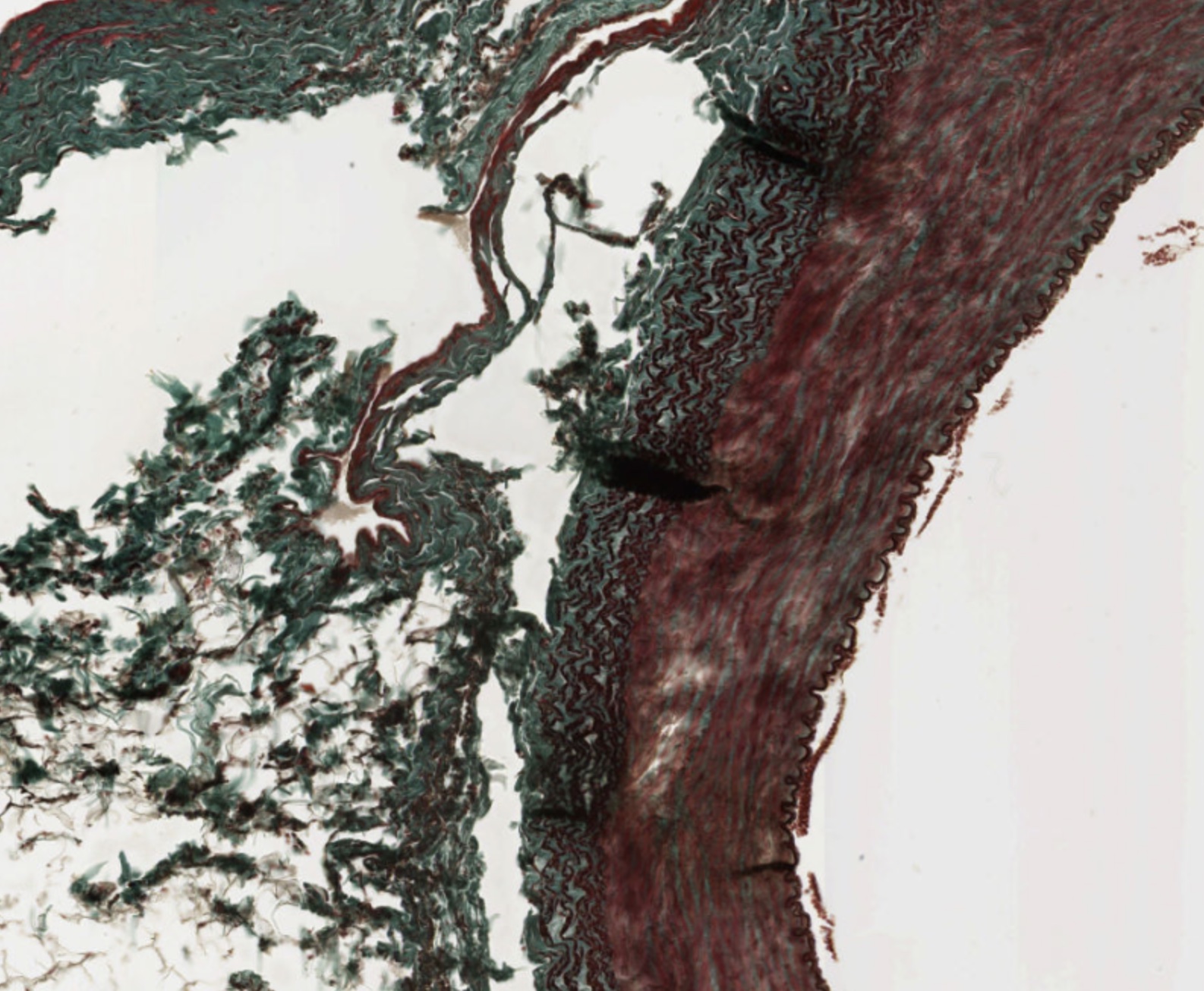}};
    \draw (0, 0.9) node {\tiny Masson \cite{umich_virtual_slide_list}};
    \end{tikzpicture} &
    \begin{tikzpicture}
    \draw (0, 0) node[inner sep=0] {\picdims[height=\linewidth]{\linewidth}{0.75\linewidth}{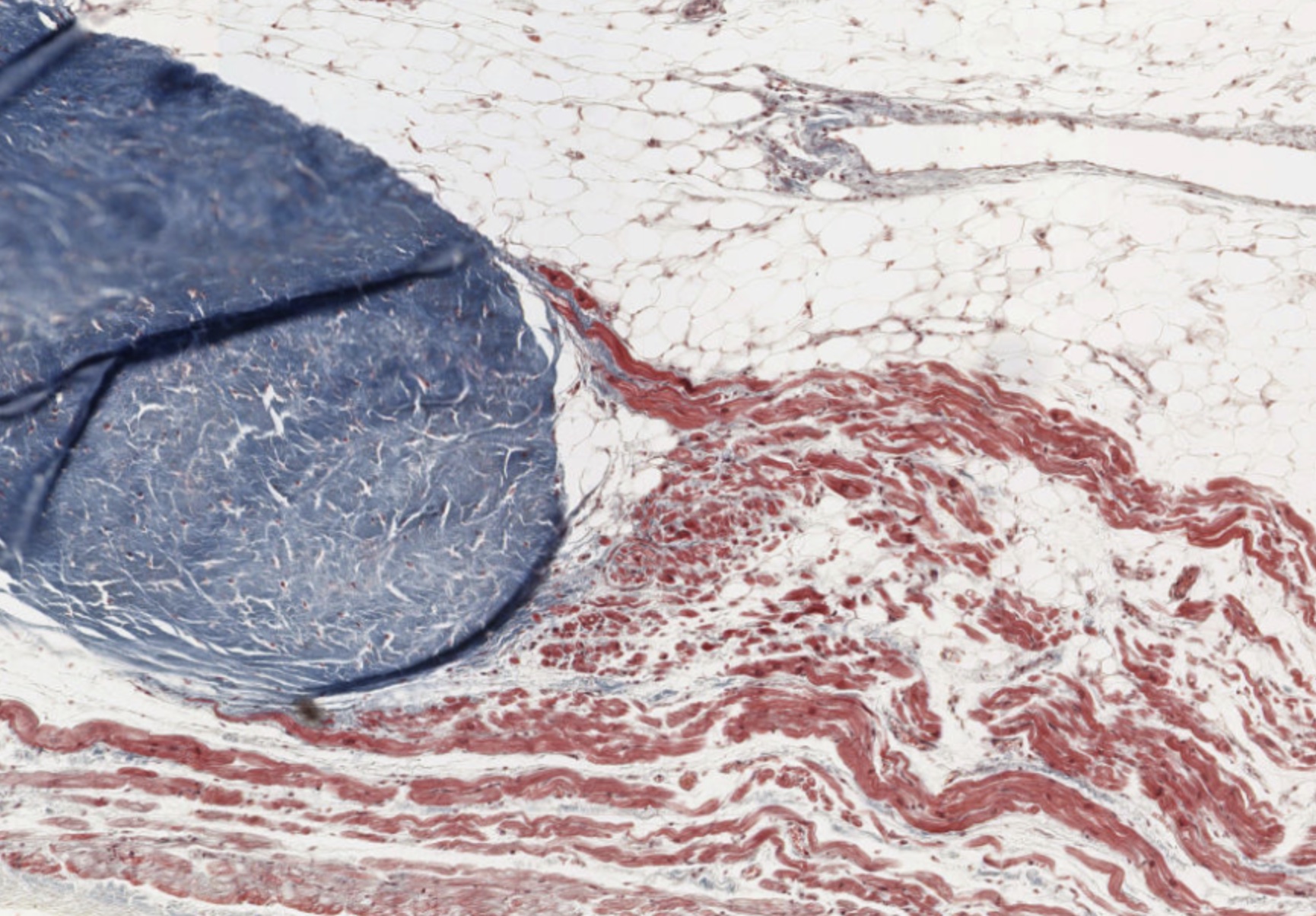}};
    \draw (0, 0.9) node {\tiny Mallory \cite{umich_virtual_slide_list}};
    \end{tikzpicture} &
    \begin{tikzpicture}
    \draw (0, 0) node[inner sep=0] {\picdims[height=\linewidth]{\linewidth}{0.75\linewidth}{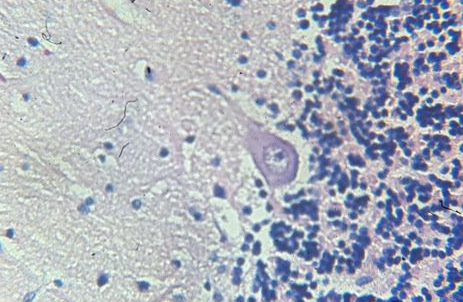}};
    \draw (0, 0.9) node {\tiny Nissl \cite{stain_nissl}};
    \end{tikzpicture} &
    \begin{tikzpicture}
    \draw (0, 0) node[inner sep=0] {\picdims[height=\linewidth]{\linewidth}{0.75\linewidth}{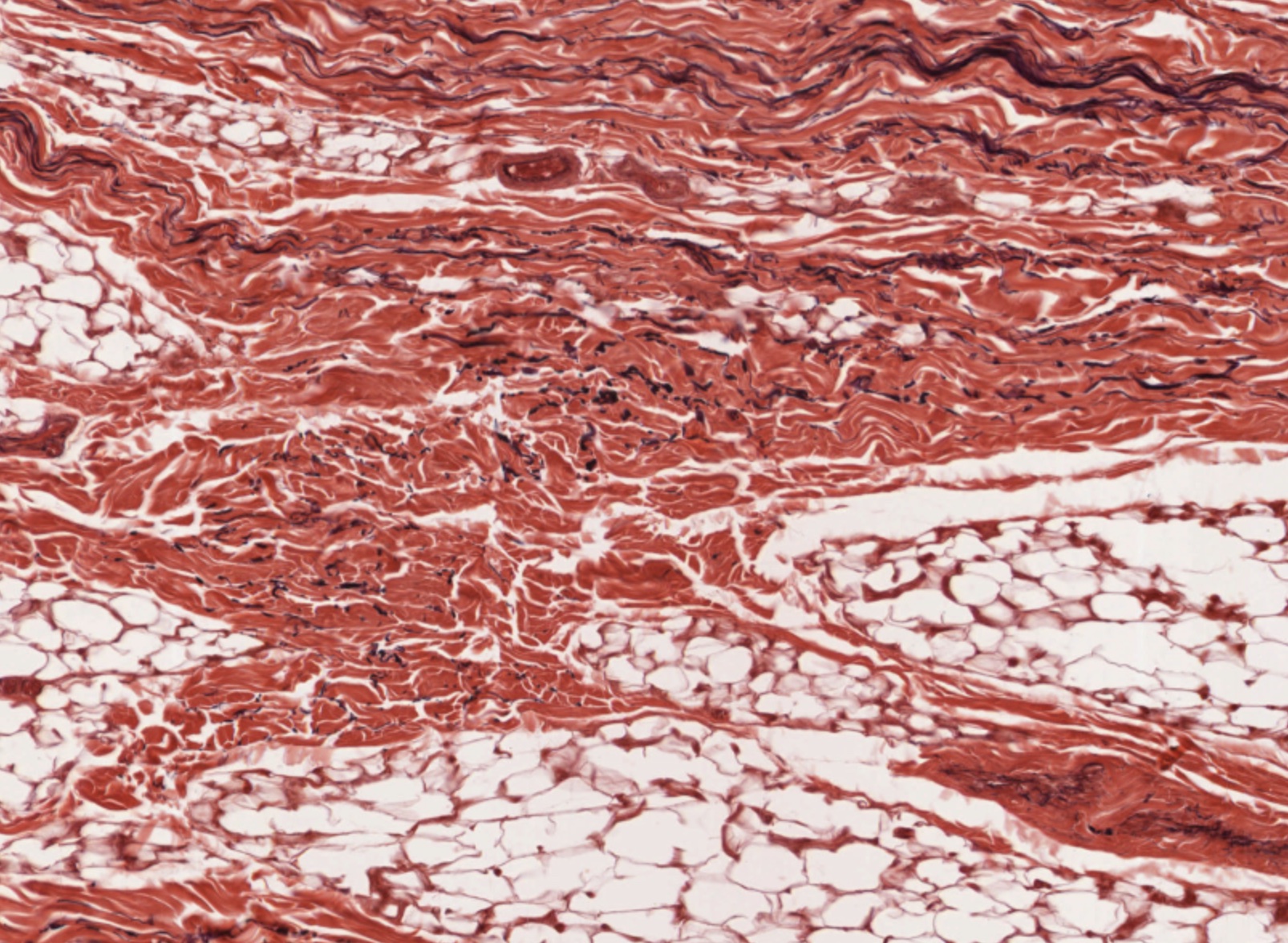}};
    \draw (0, 0.9) node {\tiny Verhoeff \cite{umich_virtual_slide_list}};
    \end{tikzpicture} \\    
    \end{tabular}
\caption{WSI tissue images with different types of histological stains. Each stain highlights different areas and structures of the tissue in order to aid in visualizing underlying characteristics. Amongst this diversity, there is Hematoxylin and Eosin or H$\&$E which is mainly used in studies as most histopathological processes can be understood from this stain. All images provided are under a Creative Commons license, specifics on the license can be found in the references.}
\label{fig:WSIFig2}
\end{figure}
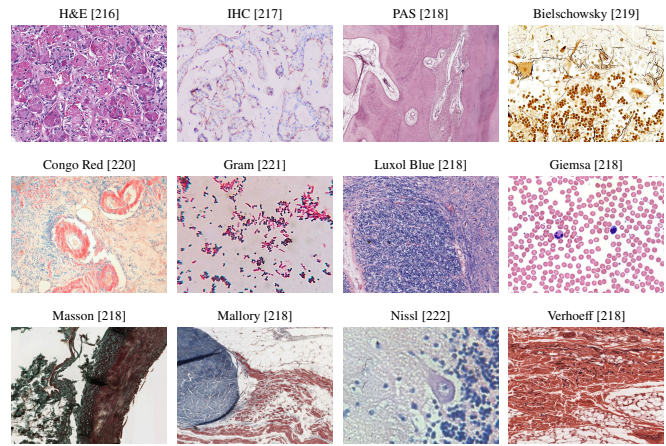
\subsection{Cohort Selection, Scale, and Challenges} \label{selection_criteria}
The data used to create/train CPath CAD tools can greatly impact the performance and success of the tool. Curating the ideal dataset, and thus selecting the ideal set of WSIs for the development of a CAD tool is a nontrivial task. Several works suggest that datasets for deep learning in CPath should include a large quantity of data with a degree of variation and artifacts in the WSIs \cite{35,373}. Some works also recommend the inclusion of difficult or rarely diagnosed cases; other works indicate that inclusion of extremely difficult cases may decrease the performance of advanced models \cite{373, MHISTlearn}.

A study highlighting the results of the 2016 Dataset Session at the first annual Conference on Machine Intelligence in Medical Imaging outlines several key attributes to create an ideal medical imaging dataset \cite{Kohli_2017}, including: having a large amount of data to achieve high performance on the desired task, quality ground truth annotations, and being reusable for further efforts in the field. While the scope of this conference did not include CPath, many of the points made regarding medical imaging datasets are also relevant to the development of CPath datasets. The session also outlined the impact that class imbalances can have on \textcolor{r5col}{ML} models, an issue also prevalent in CPath as healthy or benign regions often outnumber diseased regions by a significant margin \cite{68}. 

Our survey of past works in the literature reveals some trends in CPath datasets. Currently, the majority of datasets presented in the literature for CAD tool development are small-scale datasets \cite{35}, using a small number of images, and/or images from a small number of pathology laboratories. Examples of these smaller datasets include a dataset with 596 WSIs (401 training, 195 testing) from four centres for breast cancer detection \cite{237} and the Breast Cancer Histology (BACH2018) dataset, which has 500 ROI images (400 training, 100 testing) and 40 WSIs (30 training, 10 testing) \cite{2}. Although curating a dataset from fewer pathology laboratories may be simpler, these smaller scale datasets may not be able to effectively generalize to data from other pathology centres \cite{48,280}. An example of this can be seen in which data from different pathology centres are clustered disjointly in a t-distributed stochastic neighbor embedding (t-SNE) representation demonstrated in \cite{373}. Another alternative was proposed in \cite{saldanha2022swarm}: using a swarm learning technique multiple AI models \textcolor{r4col}{were} trained on different small data sets separately and then unified into one central model.

Additionally, stain variations, slide artifacts, and variation of disease prevalence may sufficiently shift the feature space such that a deep learning model may not sustain high performance on unseen data in new settings \cite{280,Willemink_2020}. As artifacts in WSIs are inevitable, with some artifacts, such as ink mark-up on glass slides, being an important part of the pathology workflow \cite{Ali_2019}, the ability of CAD tools to become robust to these artifacts through exposure to a diverse set of images is an important consideration. 

\begin{figure*}[t!]
\centering
\includegraphics[width=1\textwidth]{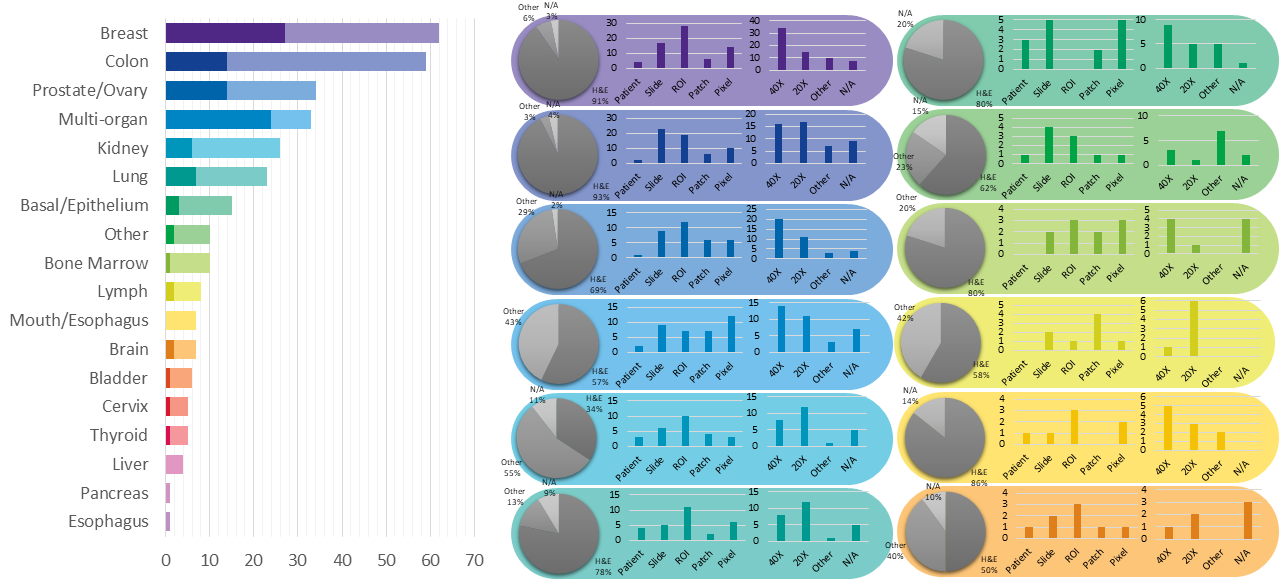}
\caption{(left) shows the distribution of datasets per organ as a capture of the current trend in datasets, although the number of datasets can change over time an understanding of what organs have more available data is important for developing CAD tools. Along the vertical axis, we list different organs, while the horizontal axis shows the number of datasets; wherein the darker color denotes public availability while the light color includes unavailable or by request statuses. (right) Distribution of staining types, annotation levels, and magnification details per organ color coded consistently with the bar graph. Organs have been sorted based on the abundance of datasets. For more details, please refer to Table \ref{DC} in the supplementary section.}
\label{fig:dataset_overview}
\end{figure*}

Compared to the number of studies conducted on small-scale datasets, relatively fewer studies have been performed using large-scale, multi-centre datasets \cite{35,63,373}. One study uses over 44,715 WSIs from three organ types, with very little curation of the WSIs for multi-instance learning detailed in \cite{35}. Stomach and colon epithelial tumors were classified using 8,164 WSIs in \cite{63}. A similar study uses 13,537 WSIs from three laboratories to test a machine learning model trained on 5,070 WSIs and achieves high performance \cite{35}. 

Despite some advancements, there exist major barriers to using such large, multi-centre datasets in CAD development. Notably, for strongly supervised methods of learning, an immense amount of time is needed to acquire granular ground truth annotations on a large amount of data \cite{63}. To combat this, some researchers have implemented weakly-supervised learning by harvesting existing slide level annotations to forego the need for further annotation \cite{35}. Additionally, it may be difficult to aggregate data from multiple pathology centres due to regulatory, privacy, and attribution concerns, despite the improvements that diverse datasets offer. \autoref{sec:models-list} discusses model architectures and training techniques that harness curated datasets of various annotation levels.

\textbf{\textit{Dataset Availability}} \label{data-avail}
\textcolor{r1col}{In general computer vision, progress can be tracked by the increasing size and availability of datasets used to train models, e.g. ImageNet grew from 3.2 million images and ~5000 classes in 2009 to 14 million images and ~21,000  classes in 2021 \cite{imagenet_cvpr09}. We infer a similar trend in dataset growth and availability indicates progress in CPath. In our survey of over 700 CPath papers, we determine the current landscape by noting the dataset(s) used in work, along with dataset details such as the organ(s) of interest, annotation level, and stain type, tabulating the results into Table \ref{DC} of the supplementary materials, with summarized findings from Table \ref{DC} are shown in Figure \ref{fig:dataset_overview}.}

From Figure \ref{fig:dataset_overview} we can clearly see that the majority of datasets used for research developments in computational pathology are privately sourced or require additional registration/request. With organs represented in a small number of datasets, such as the liver, thyroid, brain, etc, having a smaller proportion of freely accessible datasets as compared to the Breast, Colon, or Porstate. This can be problematic when trying to create CAD tools for cancers in these organs due to a lack of accessible data. We additionally note that although data sets requiring registration/request for access can be easily accessible, as in the case of Breast Cancer Histopathological Database (BreakHis) \cite{BreakHis} being used in multiple works \cite{3,39,40}, the need for registration presents a barrier to access as requests may go unanswered or take much time to review. 

In our categorization of CPath datasets, we find that a few prominent datasets have been released publicly for use by the research community. Many such datasets are made available through grand challenges in computational pathology \cite{135}, such as the CAMELYON16 and CAMELYON17 challenges for breast lymph node metastases detection \cite{16,Camelyon17,Camelyon17_1}, and the Gland Segmentation in Colon Histology Images Challenge (GLaS) competition for colon gland segmentation in conjunction with  Medical Image Computing and Computer Assisted Intervention (MICCAI) 2015 \cite{30,85}. Notable amongst publicly available data repositories is the cancer genome atlas (TCGA) \cite{TCGA-GDC}, a very large-scale repository of WSI-data containing many organs and diseases, along with covering a variety of stain types, magnification levels, and scanners. Data collected from TCGA has been used in a large number of works in the literature for the development of CAD tools \cite{190,177, TCGA-Nuclei}. As such, TCGA represents an essential repository for the development of computational pathology. While patient confidentiality is a general concern when compiling and releasing a CPath dataset, large-scale databases such as TCGA prove that it is possible to provide relatively unrestricted data access without compromising patient confidentiality. Further evaluating public source datasets, it seems that the majority of them use data extracted from large data repositories, such as TCGA, without specifying the IDs of the images used, which provides a challenge in comparing datasets or CAD tool performance across works. However, there are a few datasets that are exceptions to that  \textcolor{r4col}{phenomenon} \cite{17,134,146,279}. 

\textcolor{r1col}{Figure \ref{fig:dataset_overview} also provides some insights on the dataset breakdown by organ, stain type, and annotation level. Per organ, it can be seen that the breast, colon, prostate/ovary, and lung tissue datasets are amongst the most common, understandably since cancer occurrence in these regions is the most frequent \cite{Cancer2020}--complying with cancer statistics findings in \ref{sec:cancer_statistics}. Multi-organ datasets are the other most common type, where we have designated a dataset to be multi-organ if it compiles WSIs from several different organs. To note, multi-organ datasets are especially useful for the development of generalized image analysis tools in computational pathology. The annotation level provided in the datasets did not indicate any pattern across most organs.}

\textbf{\textit{Dataset Bias}} \label{data-bias} \textcolor{r5col}{It is also important to note the potential for bias in datasets that may influence the ability of any deep learning algorithm to generalize on unseen data \cite{NAKAGAWA2023100,Chauhan_Gullapalli_2021}. This problem is a prevalent issue in general machine learning applications \cite{Wang_2020_CVPR,xu_tian_bias_2020,FABBRIZZI2022103552,GEORGOPOULOS2020103954}, and CPath is not immune to it. The survey review in \cite{mehrabi_bias_in_ml} reviews a large number of other examples in machine learning that exhibit such bias, both from a dataset-standpoint and an algorithm-standpoint.}

\textcolor{r5col}{Such a lack of generalizability in CPath can impact the ability of machine learning models trained on biased data to meet the needs of patients. As noted in \cite{NAKAGAWA2023100}, minority groups may be disproportionately negatively impacted if care is not taken in curating a diverse dataset that adequately reflects the relevant demographics for the problem to be solved.}

\textcolor{r5col}{Several works have delved into the issue of dataset bias in CPath specifically \cite{bias_tcga_example1,bias_tcga_example2}. A notable example is in \cite{bias_tcga_example2}, where the study was able to demonstrate that deep learning models trained on WSIs from TCGA were able to infer the organization that contributed the slide sample. Notably, some features, such as genetic ancestry, patient prognosis, and several key genomic markers were significantly correlated with the site the WSI was provided from. As the vast majority of data in TCGA is acquired from 24 origin centers \cite{bias_tcga_example1}, such site-specific factors may impact the ability of a DL model to perform well on patient data from different sites.}

\textcolor{r5col}{As discussed previously, having a large set of diverse data may help to mitigate generalization issues \cite{48,280,NAKAGAWA2023100}. Additionally, the study \cite{bias_tcga_example2} makes the suggestion that training data should be from separate sites than validation data, and that per-site performance of a model should be reported when validating a model. In doing this, the robustness of the model to site-specific variation, including both stain and demographic related variation, can be evaluated.}       
\section{Domain Expert Knowledge Annotation}\label{sec:AnnotationLabellingWorkflow}
A primary goal of CPath is to capture and distill domain expert knowledge, in this case the expertise of pathologists, into increasingly efficient and accurate CAD tools to aid pathologists everywhere. Much of the domain knowledge transfer is encompassed within the process of human experts, in this case pathologists, generating diagnostically-relevant annotations and labels for WSIs. It must be emphasized, that without some level of label, a WSI dataset is not directly usable to train a model for most CAD tasks that involve the generation of diagnoses, prognoses, or suggestions for pathologists. Thus, the process of obtaining and/or using annotations at the appropriate granularity and quality is paramount in the field. This section focuses on describing various types of ground-truth annotation to cover the spectrum of weak to strong supervision of labels, discussing the practicality of labeling across this supervision spectrum, and how a labeling workflow can be potentially designed to optimize related annotation tasks.

\subsection{Supervised Annotation} \label{annotation}
In contrast to general computer vision, computer scientists do not have expert-level knowledge of histopathology and thus they are not as efficient at generating annotations or labels of pathology images. Further, labels cannot be easily obtained by outsourcing the task to the general public. As a result, pathologists must be leveraged to obtain labels at some stage of the data collection and curation process, and in many annotation pipelines the first step involves recruiting the help of pathologists for their expertise in labelling. 


\textbf{\textit{Obtaining Expert Domain Knowledge}}
The knowledge of pathologists is essential in the development of accurate ground truth annotations--a process most commonly completed by encircling ROI \cite{237}. However, there are studied instances of inter-observer variance between pathologists when determining a diagnosis \cite{32,57,338}. As obtaining the most correct label is essential when training a model for CAD, this issue must be addressed and a review of the data by several pathologists can result in higher quality ground truth data as compared to that of a single pathologist. As a result, most datasets are curated by involving a group of pathologists in the annotation process. If there exists a disagreement between the expert pathologists on the annotation of a ground truth, one of several methods is usually employed to rectify the discrepancy. A consensus can be reached on the annotation label through discussion amongst pathologists, as is done in the triple negative breast cancer (TNBC-CI) dataset \cite{TNBC-CI}, the Breast Cancer Surveillance Consortium (BCSC) dataset \cite{BCSC} and the minimalist histopathology image analysis dataset (MHIST) dataset \cite{567}. Alternatively, images, where disagreements occur, can be discarded, as is done in some works \cite{4,128}. Further, the disagreement between annotators can be recorded to determine the difficulty level of the images, as is done in the MHIST dataset \cite{MHISTlearn}. This extra metadata aids in the development of CAD tools for analysis.

\begin{figure*}[t!]
\centering
\includegraphics[width=1\textwidth]{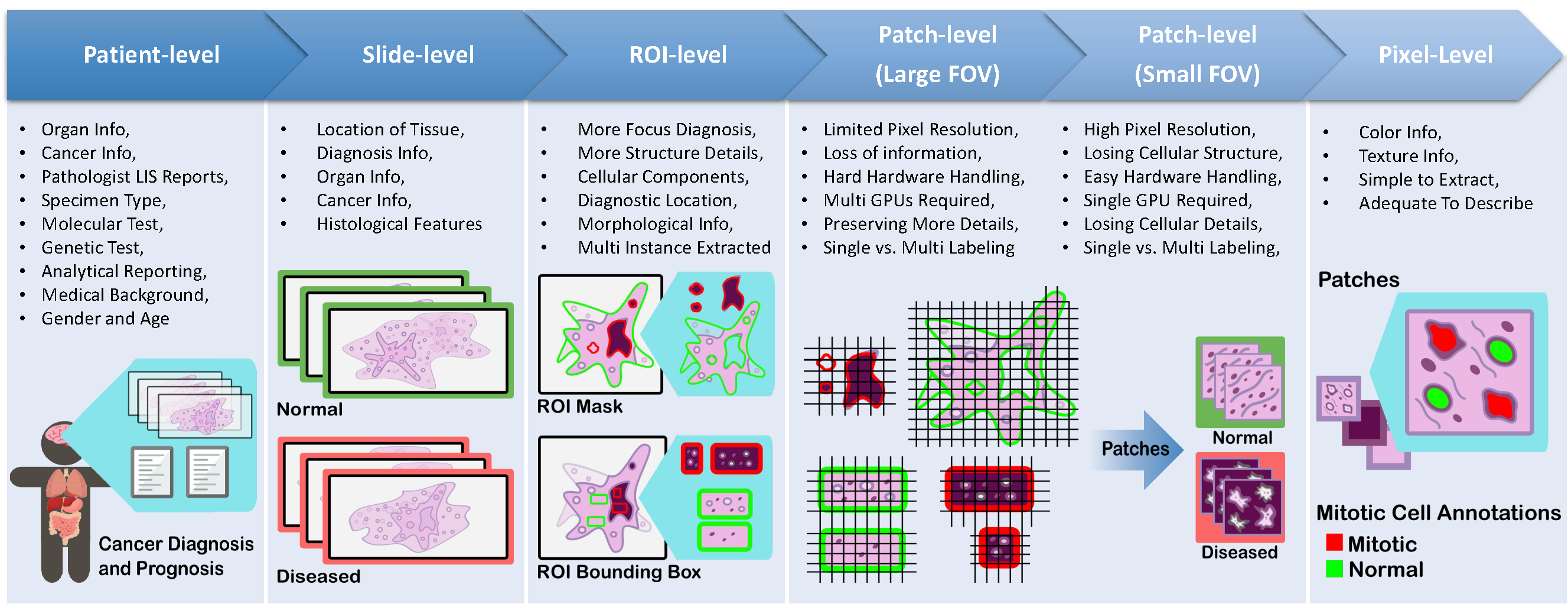}
\caption{Details of the five different types of annotations in computational pathology. From left to right: a) \textbf{Patient-level annotations:} can include high level information about the patient like status of cancer, test results, etc. b) \textbf{Slide-level:} are annotations associated with the whole slide, like a slide of normal tissue or a diseased one c) \textbf{ROI-level annotations:} are more focused on diagnosis and structure details d) \textbf{Patch-level:} are separated into Large FOV (field of view) and small FOV, each having different computational requirements for processing, and finally e)\textbf{ Pixel-level:} includes information about color, texture and brightness }
\label{fig:annotation_overview}
\end{figure*}

Pathologists can also be involved indirectly in dataset annotation. Both the Multi-organ nuclear segmentation dataset (MoNuSeg) \cite{56} and ADP \cite{110} have non-expert labelers annotate their respective datasets. A board-certified pathologist is then tasked with reviewing the annotations for correctness. Alternatively, some researchers have employed a pathologist in performing quality control on WSIs for curating a high-quality dataset with minimal artifacts \cite{111,113}. To enable the large scale collection of accurate annotated data, Lizard \cite{478} was developed using a multi-stage pipeline with several significant ``pathologist-in-the-loop'' refinement steps. 

Existing pathological reports, along with the metadata that comes from public large-scale databases like TCGA, can also be leveraged as additional sources of task-dependent annotations without the use of further annotation. For example, TCGA metadata was used to identify desirable slides in \cite{209}, while pathological diagnostic reports were used for breast ductal carcinoma in-situ grading in \cite{183}.

To note, there are some tasks where manual annotation by pathologists can be bypassed altogether. For instance, IHC was applied to generate mitosis figure labels using a Phospho-Histone H3 (PHH3) slide-restaining approach in \cite{81}, while immunofluorescence staining was used as annotations to identify nuclei associated with pancreatic adenocarcinoma \cite{242}. These works parallel the techniques that pathologists often use in clinical practice, such as the use of IHC staining as a supplement to H$\&$E stained slides for difficult to diagnose cases \cite{31}. They demonstrate high performance on their respective tasks wherein the top-performing models on the Tumor Proliferation Assessment Challenge 2016 (TUPAC16) \cite{TUPAC16} dataset were achieved \cite{81}. Importantly, these techniques still utilize supervision, albeit weakly, by leveraging lab techniques that have been developed and refined to identify the desired regions visually. 

\textbf{\textit{Ground-Truth Diagnostic Information}}
Understanding different annotation levels and their impact on the procedural development of ML pipelines is an important step in solving tasks within CPath. There are five possible levels of annotation, in order of increasing granularity (from weakly-supervised to fully-supervised): patient, slide, ROI, patch, and pixel. Figure \ref{fig:annotation_overview} overviews the benefits and limitations of each level. For additional information regarding each annotation level please refer to Appendix \ref{appendix_annotation}.

\begin{figure}[t]
\centering
\includegraphics[width=0.225\textwidth]{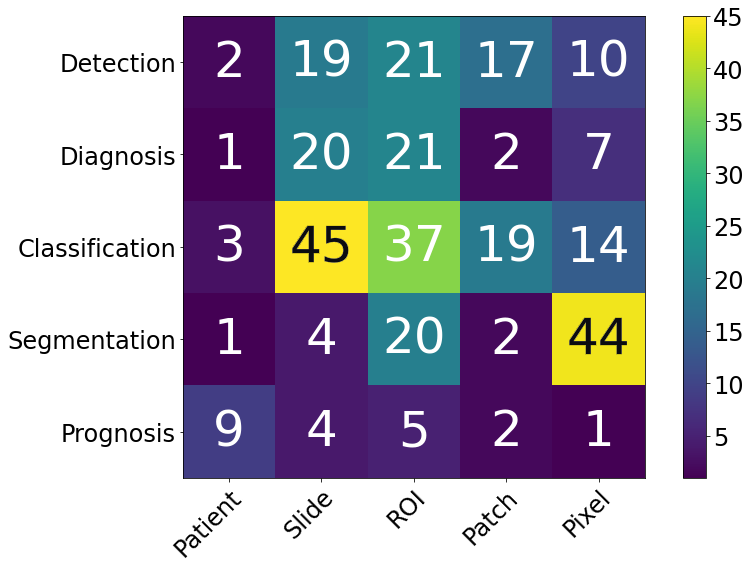}
\caption{A snapshot of the distribution of different annotation levels based on the CPath task being addressed in the surveyed literature for the purposes of highlighting the trend of datasets. The x-axis displays the different annotation levels studied in the papers (from left to right): Patient, Slide, ROI, Patch, and Pixel. The y-axis shows the different tasks (top to bottom): Detection, Diagnosis, Classification, Segmentation, and Prognosis. The height of the bars along the vertical axis measures the number of works that have examined the specific task and annotation level.}
\label{data-annobar}
\end{figure}

\textbf{\textit{Picking the Annotation Level}}
Selecting an annotation level depends largely on the specific CPath task being addressed, as shown in Figure \ref{data-annobar}. For example, segmentation tasks tend to favor pixel-level annotations as they require precise delineation of a nucleus or tissue ROI. Conversely, disease diagnosis tends to favor datasets with ROI-level annotations, as diagnosis tasks are predominantly associated with the classification of diseased tissue, the higher-level annotations may provide a sufficient level of detail and context for this task \cite{603}.

Figure \ref{data-annobar} shows that tasks that use stronger supervision are more likely to be used in CAD tool model development. However, due to the high cost of pixel-level annotation, fully supervised annotations are challenging to develop. Even patch-based annotations often require the division and analysis of a WSI into many small individual sub-images resulting in a similar problem to pixel-based annotations \cite{60,83}. In contrast, WSI data is most often available with an accompanying slide-level pathology report regarding diagnosis thus making such weakly labeled information at the WSI level significantly more abundant than ROI, patches, or pixel-level data \cite{252, 474}. Different levels of annotation can be leveraged together, as demonstrated by a framework to use both pixel and slide level annotations to generate pseudo labels in \cite{499}. Additionally, it is common in CPath to further annotate the slide-level WSIs on an ROI or patch level structure \cite{37,210, 234, 370}.

\textbf{\textit{Active Learning Tools}}
Active learning annotation tools bridge the gap between the need for highly supervised labels and the current abundance of less informative annotations. Such works seek to ease the annotation process by using computational approaches to assist the human annotator. For example, in \cite{281}, a platform was developed for creating nuclei and gland segmentation ground truth labels quickly and efficiently. A convolutional neural network (CNN), trained on similar cohort data, was used to segment nuclei and glands with different mouse actions \cite{281}. Alternatively, \textcolor{r4col}{Awan et al.} \cite{363} presented the HistoMapr\textsuperscript{\texttrademark} platform to assist in diagnosis and ground truth data collection. Through this tool, a pathologist selects one of several proposed classes for each given ROI, thus mitigating the need for hand-drawn annotations or manual textual input \cite{363}. Similarly, an active learning model called the Human-Augmenting Labeling System (HALS) \cite{527} was developed to increase data efficiency by guiding annotators to more informative slide sections. Quick Annotator (QA) \cite{587} is another tool which provides an easy-to-use online interface for annotating ROIs and was designed to create a significant improvement in the annotation efficiency of histological structures by several orders of magnitude.  

There are other active learning annotation tools proposed for different applications in computer vision that can be investigated for use in the pathology datasets. Such examples include methods to produce object segmentation masks for still images \cite{zhang2016instance,zhang2021datasetgan} as well as video \cite{chen2020scribblebox}. One notable example is DatasetGAN \cite{zhang2021datasetgan}; the model is proposed as a training data creator, and it is shown that the model can produce segmentation masks with a small number of labelled images in the training data. While these systems are for general computer vision, they may be adoptable in computational pathology, and would facilitate the necessary relationship between pathologists and computer scientists in the development of CAD tools. As such, they may prove to be a valuable contributor to the CAD system development workflow.

\textbf{\textit{Tissue-Class and Disease Complexity}}
Much of the current CPath research operates under the umbrella of supervised learning tasks, and correspondingly uses labeled data to develop automated CAD tools. We refer to supervised learning to include a diverse spectrum of annotation i.e. weak-supervision (e.g. patient-level) all the way to strong-supervision (e.g. pixel-level). Classes within a dataset can be task-dependent, for example as shown in Table \ref{DC} of the supplementary material, datasets primarily used for segmentation such as MoNuSeg \cite{56} and CPM-17 \cite{70} have classes for each annotated pixel indicating the presence or absence of nuclei. However, classes need not be task-dependent; datasets such as CAMELYON16 \cite{16} outline metastases present in WSIs that can be used for a variety of applications, including disease detection \cite{16} and segmentation tasks \cite{25}.

The current paradigm for dataset compilation in computational pathology, particularly for disease detection and diagnosis, treats different disease tissue types as separate independent classes. For example, BreakHis divides all data into benign/malignant breast tumours \cite{BreakHis}. At the ROI level, GLaS divides colon tissue into five classes: healthy, adenomatous, moderately differentiated, moderately-to-poorly differentiated, and poorly differentiated \cite{30}. So far, this approach to class categorization has resulted in high-performing CAD tools \cite{29,39,40,67,96,105}. However, the treatment of different disease tissue types as an independent class is perceived differently in computer vision domain where the representation learning of normal objects is done differently compared to anomalies. A similar synergy can be found by differentiating healthy tissue classes from diseased ones and one should be mindful about defining meaningful tissue ontology for annotation and labeling. 

\subsection{Optimum Labeling Workflow Design}
\begin{figure}[t]
\centering
\includegraphics[width=0.48\textwidth]{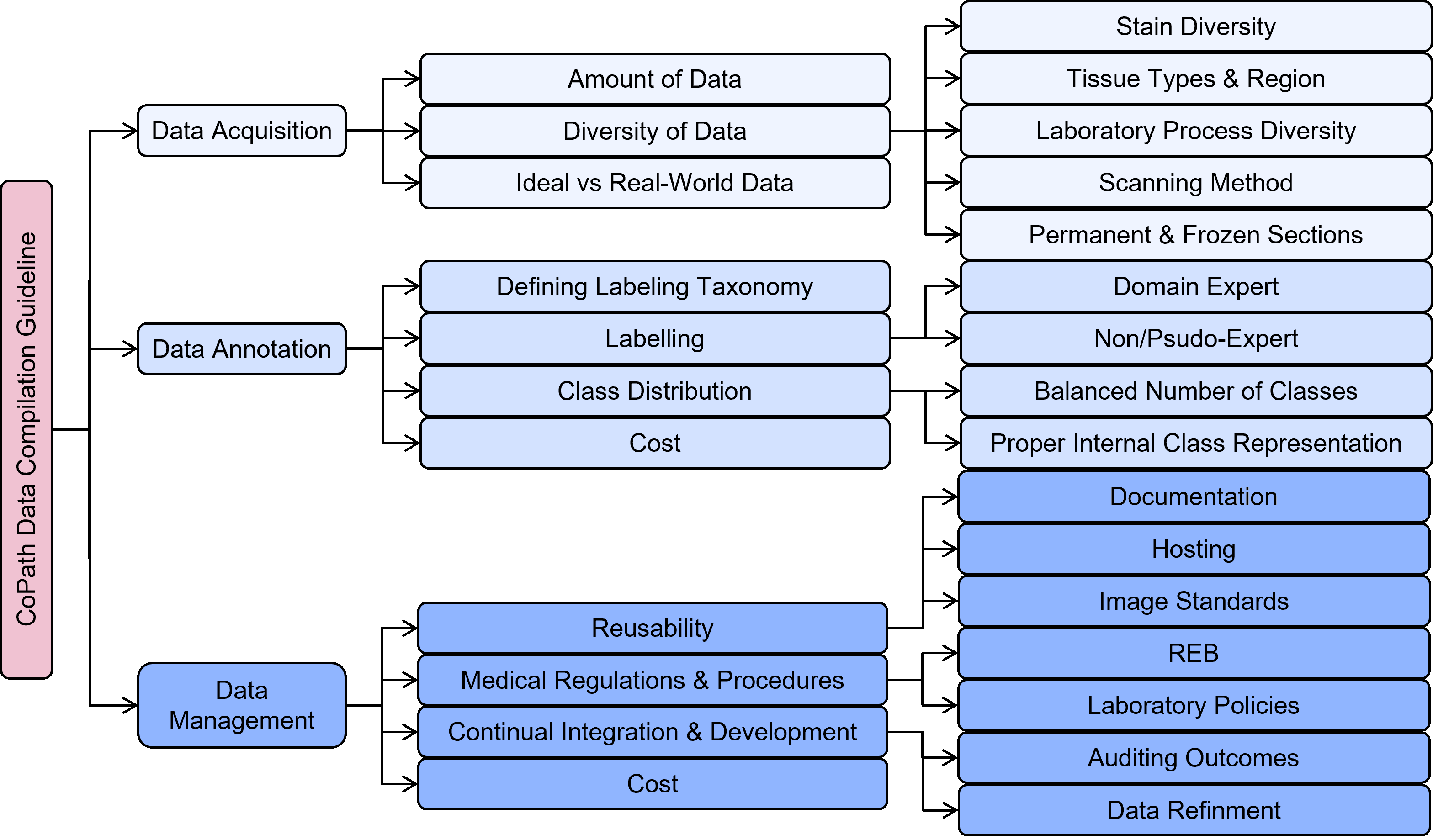}
\caption{Tree diagram for the optimum labeling workflow, where a CPath dataset is divided into tasks and sub-tasks based on its initial characteristics.}
\label{fig:optimum_labelling_workflow}
\end{figure}

This section focuses on the steps required for compilation of a CPath dataset which is broken into three main sub-tasks: Data Acquisition, Data Annotation, and Data Management, as per Figure \ref{fig:optimum_labelling_workflow}. Each sub-task is discussed below with reference to its individual components in the hierarchical structure in Figure \ref{fig:optimum_labelling_workflow}.

\textbf{\textit{Data Acquisition}}
Database compilation starts from data acquisition. When collecting data, it is vital that there are large amounts of data \cite{data_collection_survey}, along with having sufficient diversity \cite{35, 373}. Specifically, diversity in CPath data occurs in multiple ways, such as staining methods, tissue types and regions, laboratory processes, and digital scanners. We advise that CPath researchers consult expert pathologists on the diversity of data required for various tasks. Ideally, all the data acquired in pathology should be perfect without any irregularity and artifacts. However, some level of artifacts and irregularity are unavoidable and introducing realistic artifacts that are representative of real-world scenarios into the data increases the robustness and generalizability of CAD tools.

\textbf{\textit{Data Annotation}}
After collecting sufficient data, the next task is annotation of the data. Data annotation is a costly process in both time and money, thus a budget and schedule should always be established when generating labels. There are often various approaches for annotating different structures \cite{semantic_annotation_cpath}, so a specific labelling taxonomy should be defined \textit{a priori}. As mentioned previously, annotation should involve expert pathologists due to the domain knowledge requirement and importance of label correctness. A table of commonly used commercially available annotation software for annotating different slide formats are show in Table \ref{tab:1cb}, along with their compatible image formats which is important to note when trying to build compatible and accessible datasets.
 
Once the ontology of class-definitions is defined (in collaboration with expert pathologists), there will be two ways to generate labels or annotations in general: domain expert labelling or non-expert labelling. The domain expert labelling refers to having pathologists annotate data that they are specialized at, which is labor-extensive. On the other hand, non-expert labelling can use crowdsourcing techniques to generate weak labels or have non-experts, such as junior pathologists or students, label the data. This process is cheaper and quicker, but it may be harder to maintain the same level of quality as domain expert labeling \cite{semantic_annotation_cpath}. Regardless of the labelling methodology used, labels generated from both should be validated. Finally, to determine the sufficiency of label quantity, one should consider the balance between the number of classes, representation size of each class, and complexity of class representation. Techniques from active-learning could be also leveraged to compensate for lack of resource management as well as maintain the quality of labeling as discussed above. 

\textbf{\textit{Data Management}}
Data management is an important aspect of any dataset creation process, and is the one that is most likely to be overlooked. Proper data management should have considerations for reusability, medical regulations/procedures, and continuous integration and development.

Reusability can be broken down into detailed documentation of the data, accessible and robust hosting of data, and consideration for image standards. Poor cross-organizational documentation can lead to missing metadata, ultimately resulting in discarding entire datasets \cite{DataCascades}. Adherence to an established image standard, such as DICOM, can help resolve some of these issues in reusability. Medical regulations/procedures can be broken down into the construction of a Research Ethics Board (REB) and proper consideration for whom is curating the data. Through incentives for data excellence for medical practitioners, the issue of misaligned priorities between data scientists, domain experts, and field partners can be resolved \cite{DataCascades}. To ensure that models used on actual patients remain relevant and hidden errors do not propagate, continuous integration/development (CI/CD) must be implemented. These systems must include at least two components, a method to audit predictions from the model, as well as a way to refine the training data accounting for discrepancies found through auditing. Several algorithms deployed in high-risk environments, including medical diagnosis, proved to only work well when data was updated after initial deployment \cite{WMD,DataCascades}. Throughout the data management process, consultation with domain experts is a vital step in ensuring the success of data compilations \cite{avoid_pitfalls}.      
\section{Model Learning for CPath}\label{sec:models-list}
\begin{figure}[t]
\centering
\includegraphics[width=0.48\textwidth]{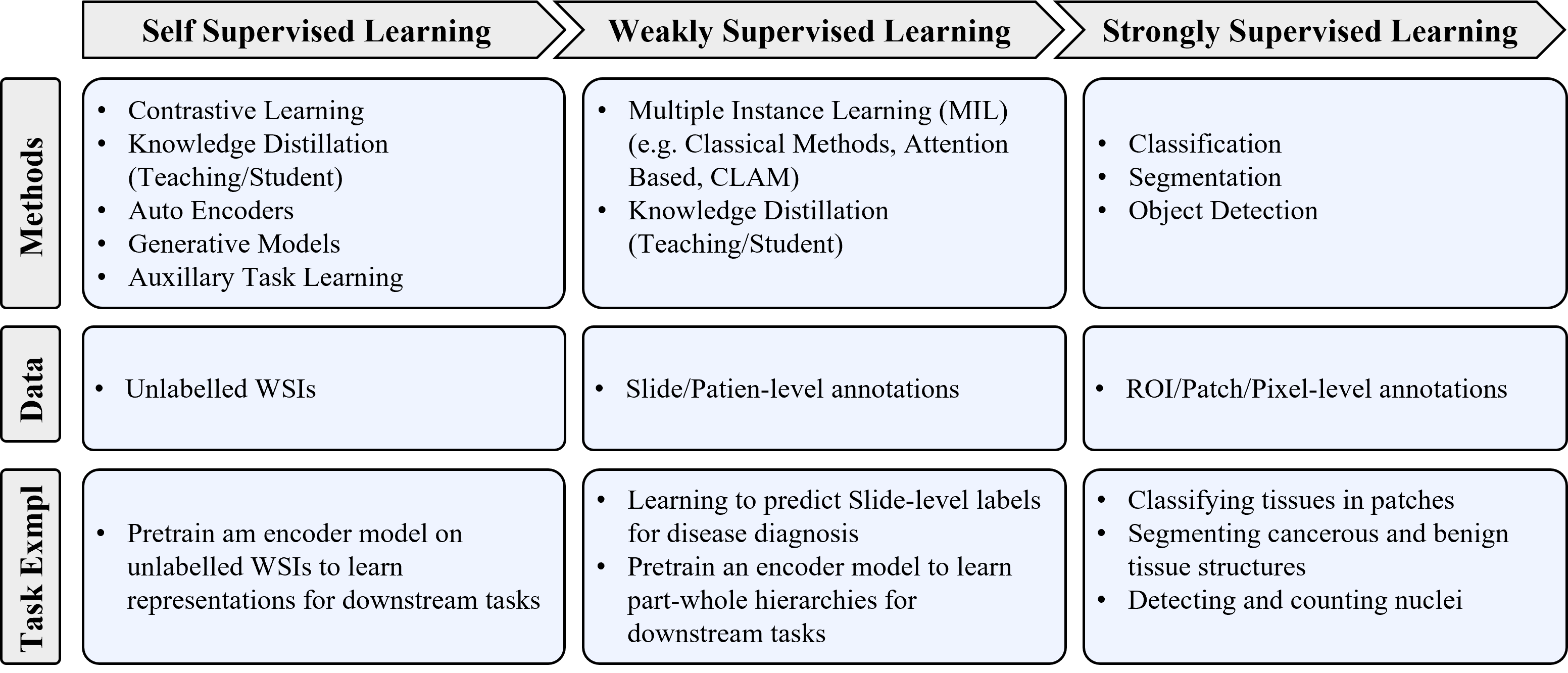}
\caption{Details of types of learning using varying levels of supervision. Note that the types of tasks each type of learning can address vary based on the data that is available, as noted in the \textit{Example Task} portion of the figure. However, from left to right, models trained with less supervision can still learn salient representations of the data that can be used to fine-tune models for tasks requiring more supervision. In that sense, for CPath there is a \textit{spectrum} of supervision from self to strongly supervised learning that aligns well with the annotation levels shown in Figure \ref{fig:annotation_overview}.}
\label{fig:types_of_learning}
\end{figure}

Once an application domain and corresponding dataset have been chosen, the next step of developing a CPath tool involves designing of an appropriate model and representation learning paradigm. Representation learning refers to a set of algorithmic techniques dedicated to learning feature representations of a certain data domain that can be used in downstream tasks \cite{bengio2012representation}. In CPath, the amount of data available for a given annotation level and task are the key determinants to designing a model and learning technique. The last decade has shown neural network architectures to become the dominant method in many machine-learning domains because they are rich enough to avoid handcrafted features and offer superior performance. \cite{lecun2015deep}. The annotation level of the data pertaining to the task corresponds to the level of supervision for the learning technique applied. This relationship between data annotation level and learning supervision level is surveyed in Figure \ref{fig:types_of_learning}.

This section details the various types of models and learning techniques, along with the tasks they have been applied to in CPath. Figure \ref{fig:models} highlights the most common backbone architectures used for feature encoding in \textcolor{r5col}{SOTA} research, based on the corresponding tasks. More details are provided in Table \ref{TT} from the supplementary materials. The selection of architectures is then compared to draw useful insights into accuracy, computational complexity, and limitations. Lastly, existing challenges in model design are investigated. 

\subsection{Classification Architectures}
In CPath, general classification architectures are the most prevalent due to their straightforward applicability to a wide range of tasks including tissue subtype classification, disease diagnosis and detection (more details in Section ~\ref{sec:application} and Figure \ref{fig:organ_overview}). 
\textcolor{r1col}{Architectures commonly used for natural images, in particular CNNs, are widely adopted for CPath. To maximize model performance, it is a common approach to pre-train the model on large datasets like ImageNet before subsequent fine-tuning them to perform well for the specific CPath task, a task known as transfer learning \cite{18,170,227,182,252,62,109,111,112,203,204,226,213,122,26,51,121,32,158,183,493,591,279,609,601,526,590,539,613,479,teh2022learning,su2022deep, yang2022concl}. }
%
%
\textcolor{r1col}{Transfer learning for CPath allows for: 1) improved generalizability, particularly for tasks with limited amount of data; and 2) improved ease in finetuning a model compared to training from scratch \cite{Transfer_Learning_Survey}.}
%

\textcolor{r1col}{Graph Convolutional Neural Networks (GCN) \cite{kipf,guan2022node} is an alternative architecture that can be used to improve the learning of context-aware features across the WSI. GCNs typically consist of nodes representing elements and edges defining relationships between nodes. In \cite{547}, a GCN was defined on a WSI, where the nodes represent patches and edges represent the connections among patches; this work obtained remarkable results on cancer prognosis task outperforming the} \textcolor{r5col}{SOTA} \textcolor{r1col}{in four out of five cancer types  \cite{547}.}
%
%
%

Vision Transformers (ViT) \cite{Vit}, have recently emerged as a direct application of Transformer models \cite{22} to the image domain. In ViT, images are sub-patched and flattened into a 1D embedding along with a positional encoding which is then classified by an MLP head. By using the positional encoding, the model's attention mechanism can focus computation on the most relevant areas of the image. ViT models have been applied with great success to CPath tasks, especially in conjunction with pre-trained CNN models \cite{ViT4}. We refer the reader to a comprehensive survey of transformer methods in medical image analysis for more details \cite{he2023transformers}. 

General classification architectures are also commonly used as a foundation for novel architectural designs. \textcolor{r1col}{For example, Squeeze-and-Excitation (SE) modules were introduced to reduce the number of parameters in ResNet and DenseNet blocks while maintaining high accuracy \cite{71, 40}. A fully-connected conditional random field (CRF) was incorporated on top of a CNN encoder to improve performance while maintaining the same level of computational complexity \cite{101}.} Lastly, patch sampling and pooling were used with AlexNet to perform slide-level disease diagnosis and segmentation \cite{144}.

Finally, in order to achieve superior performance, many researchers often rely on ensemble or multi-stage techniques which combine the predictive power or feature extraction abilities of multiple models to form a final output. These approaches have shown performance improvements compared to traditional single model classifiers \cite{74,202,310,lin2021pdbl,senousy2021mcua,149}. However, this often comes at the expense of higher computational requirements. 
%

\subsection{Segmentation Architectures}
Segmentation is widely used in CPath, as shown in Figure \ref{fig:organ_overview}, and enable localizing the area of interest at the pixel level \cite{UNet}. 
U-Net was initially developed for neuronal structure segmentation in electron microscopy image stacks \cite{UNet}, but has become one of the most common architectures for segmentation in CPath \cite{13,66,84,173,TCGA-Nuclei,348,351,25,74,152,150,371,336,305,225,587,565,578,503,609,545,597,631,618,517,627,580,481}. U-Net has an encoder-decoder structure: an encoder to contract features spatially and a decoder to expand them again to capture semantically related context and generate pixel-level predictions \cite{UNet}. The U-Net model has been used to segment nuclei for creating a novel dataset with unsupervised learning \cite{TCGA-Nuclei}, but it should be noted that this process also relies on the Mask R-CNN framework and pathologists for quality-checking purposes. 

Another common approach for segmentation is to use fully convolutional networks (FCNs) \cite{53,56,342,352,29,175,133,210,105,230,617,580}, customized architectures constructed by combining multiple components of various architectures, or introducing new components to pre-existing architectures \cite{85,219,281,314,349,124,120,132,243,136,113,257,121,105,500,616,486,503,492,545,499,603,552,479}. For example, one work used a custom CNN to predict whether each pixel was benign or malignant, while a second CNN was used to refine the initial prediction through probability fusion \cite{105}.

\subsection{Object Detection Architectures} 
In this section, we specifically focus on architectures that are used for object detection in CPath, where bounding boxes are predicted around regions of interest. \textcolor{r1col}{A major CPath application for object detection is mitosis detection with the primary goal for counting the number of mitosis instances. To this end, a large number of studies has been dedicated to this application \cite{23,48,68,81,104,150,160,190,207,208,250,277,288,350,503,588}.}
%
%
Object detection has been additionally applied for
nuclei \cite{14,44,55,80,231,501,553}, colorectal gland \cite{29,245,565} and glomeruli detection \cite{19,344,362}; however, it can also be applied to the detection of a variety of histopathological objects including tumor-infiltrating lymphocytes \cite{296} or keratin pearls \cite{257}. 

\begin{figure}[t]
\centering
\includegraphics[width=0.3\textwidth]{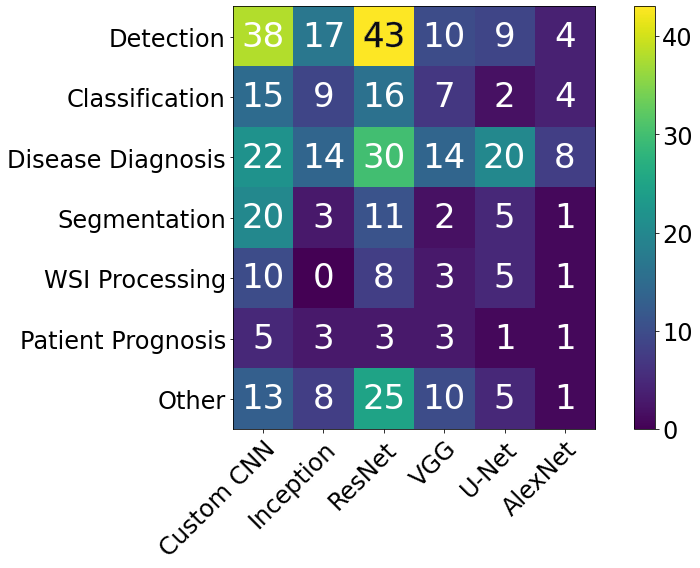}
\caption{Distribution of the most common Neural Network architectures used in the surveyed literature, based on the CPath task. The x-axis displays the Neural Network architectures used in the papers (from left to right): Custom CNN, Inception, ResNet, VGG, U-Net, and AlexNet. The y-axis shows the different tasks (top to bottom): Detection, Classification, Disease Diagnosis, Segmentation, WSI Processing, Patient Prognosis, and Others. For more details, please refer to Table \ref{TT} in the supplementary section.}
\label{fig:models}
\end{figure}

\textcolor{r4col} In CPath, object detection employs a combination of pre-existing off-the-shelf architectures and customized neural networks to perform object detection tasks, as shown in Figure \ref{fig:models}. A model called CircleNet, which uses a deep layer aggregation network as a backbone, was proposed to detect round objects \cite{344}. Their approach involves using an anchor-free ``center point localization'' framework in order to output a heatmap with center points followed by a conversion into a bounding circle for the detection of kidney glomeruli.
%
%
A multi-stage deep learning detection model \textcolor{r1col}{based on Fast R-CNN} was proposed in \cite{160}. 
%
%
First, a modified Fast R-CNN generated region proposals, then a ResNet-50 model eliminated false positives, and a Feature Pyramid Network detected mitosis in sparsely annotated images using a ResNet backbone \cite{350}.

\subsection{Multi-Task Learning}
Multi-task models are individual models predicting for multiple tasks at once (e.g. classification and segmentation), as defined in Section \ref{sec:application}. Multi-task learning (MTL) can be beneficial over independent task learning because sharing representations between related tasks can create more generalizable representations and encourage the task heads to make logically consistent predictions. This type of model, however, is uncommon in CPath, as it requires annotating multiple tasks for each image \cite{280,332,241,288,312,370,feng2021mutual}. We discuss some of these papers in further detail below.  

In one work, a ResNet-50 backbone followed by independent decoders (a pyramid scene parsing network for segmentation and a fully-connected layer for classification) was used to solve 11 different tasks (4 segmentation based and 7 classification based) \cite{280}.
%
%
With significantly less computation, the MTL model achieved comparable or better results to single task learning in classification, but comparatively worse results in segmentation. Similarly, in \cite{241}, a ResNet-50 with two parallel branches to perform segmentation and classification, was able to achieve comparable results on both tasks through an MTL approach.
%
%

%
%
 While the results are impressive, there is still work to be done in this field. One work found that model performance may be sensitive to the number and type of tasks used during training \cite{332}. If the tasks are unrelated, this could deteriorate the performance compared to a single-task setting. How to weigh different task objectives and select optimal tasks to be trained together is yet an active area of research \cite{NEURIPS2018_432aca3a, royer2023scalarization}. MTL represents an interesting field of research \textcolor{r1col} {in CPath as it may reduce the necessity to train multiple deep neural networks to perform different tasks.}

\subsection{Multi-Modal Learning} 

As opposed to multi-task networks where multiple tasks are learned simultaneously, the multi-modal approach involves using network input features from multiple domains/modalities at once \cite{multimodaloriginal}. In the case of CPath, modalities can be represented as pathologists' reports, gene expression data, or even WSI images. Most commonly, immunohistochemistry (IHC) stains alongside the H\&E stain to better visualize specific proteins \cite{mi2021predictive, foersch2023multistain, huang2023artificial}. As a result, models can learn better unified/shared latent representations which capture correlations from multiple indicators, since some information may not be captured by individual indicators \cite{li2022hierarchical}. This approach can be viewed as adding hand-crafted features to boost performance. 
%
 %
 \textcolor{r4col}{While the use of deep learning normally implies using learned features to replace hand-crafted ones, using hand-crafted features can nonetheless improve performance compared to strictly deep learning approaches when data is limited \cite{cpath_2022_dl_survey}. Indeed, many works have obtained best performance by combining manual and learned features \cite{104, 231, Huang2022-vn}. This was  demonstrated in the case of mitotic cell classification when an ensembled classifier model using hand-crafted features set a new record for the MITOS-ATYPIA 2014 challenge with an F-score of $86\%$ \cite{ensemblemitot}.}
However, where data is plentiful, CNNs alone can outperform all other hand-crafted features. In the same MITOS-ATYPIA 2014 challenge, the previous record was broken this way with a new F-score of $96\%$ \cite{hand1}. Although one cannot compare these two works directly as they use different classifier heads and dataset balancing methods, one can argue that the optimal choice of approaches from deep learning, classical ML, and different modalities should depend on the situation. Multi-modal approaches are gaining traction in CPath for specific problems, especially where useful additional data is available \cite{multiN,multiNp1}. For example, gene expression data and WSI images are often combined to improve cancer prognosis prediction \cite{multimodal,multiomodal2} .

\subsection{\protect\textcolor{r4col}{Vision-Language Models}}
\textcolor{r4col}{
Following its successful use in the natural image domain, vision-language data (consisting of histopathology images paired with relevant natural language text) is becoming increasingly prominent in CPath. Whether it be the development of foundational models \cite{huang2023visual} extending to CPath, or fine tuning state-of-art large large models for use in downstream tasks \cite{ikezogwo2023quilt,qu2023exploring,Lu_2023_CVPR}, leveraging the semantic information embedded in the natural language data is becoming more evidently beneficial. It was only recently that foundational language models advanced enough to become useful in CPath, and this has triggered an explosion of interest into building models at the intersection of visual and language information. At the moment, language data is primarily used to address Multi-Instance Learning, although this is still an extremely new field and we anticipate that future works will surely address more advanced tasks (see Section \ref{disc_lang} for further discussion).}

\subsection{Sequential Models} 
Recurrent Neural Networks (RNNs) are typically used in tasks with temporally-correlated sequential data, such as speech or time series \cite{DeepLearningLecun2015}. Since RNNs consider the past through the hidden state, they are suited for handling contextual information. While images are the default data format in CPath (and hence poorly-suited for RNNs), some works opt to combine RNNs with CNNs as a feature extractor \cite{35,42,46,63,95,113,127,128,181,194,216,255,335}, most commonly by aggregating patches or processing feature sequences \cite{35,63,113,194,634}. Another application of RNNs is to consider spatial relations between patches, which can be lost after extracting from the slide \cite{128,216}. 

A particularly exciting use of RNNs is in deciding which region within an image should be examined next \cite{181,254}. In the ``Look, Investigate, and Classify'' 3-stage model, an \textcolor{r5col}{long short-term memory} (LSTM) was used to classify the ROI cropped from the current patch and predict the next region to be analyzed, and achieved good performance while only using $15\%$ of pixels from the original image \cite{181}. Similarly, an LSTM network was used to better predict ROIs by treating state features similar to time-series data, thus identifying only relevant examples to use for training \cite{254}. And an LSTM transformer with ``Feature Aware Normalization'' (FAN) units for stain normalization was used in parallel with a VGG-19 network \cite{22}. \textcolor{r4col}{More recently, transformers using attention mechanisms have been used to allow parallelization and better sequence translation compared to older RNNs or LSTM networks \cite{vaswani2017attention}.}

\subsection{Synthetic Data and Generative Models}
With annotated data difficult to obtain in CPath, especially with granular labels (see Section \ref{data-avail} for more discussion), this is problematic for training generalizable models. Hence, generating \textit{synthetic} data from a controlled environment (either via simulation or a trained model) for augmenting the available training set of annotated data shows much promise. Originally developed for visual style transfer in general computer vision, generative models learn to create novel instances of samples from a given data distribution - they form the dominant approach in CPath.

Initial works primarily utilized Generative Adversarial Networks (GANs) for patch synthesis \cite{13,140,161,477}, stain normalization, \cite{155,21,50,347,530,539}, style transfer \cite{38,164,166,539}, and various other tasks \cite{83}. One unsupervised pipeline relied on a non-GAN model to create an initial patch that was refined by a GAN \cite{13}. In another work, one CycleGAN generated tumor images and another non-tumor, in order to train a classification network \cite{77}. One work used neural image compression to learn the optimal encoder to map image patches to spatially consistent feature vectors \cite{83}. Another work first classified bone marrow cell representations and then used an unsupervised GAN to generate more instances from each cluster \cite{98}. A self-supervised CycleGAN was also used for stain normalization, and shown to improve model performance in subsequent detection and segmentation tasks \cite{347}. Similarly, a CycleGAN pipeline was applied to perform artificial \textcolor{r5col}{IHC} re-staining \cite{IHCGAN}. Recent works in GANs attempt to model spatial awareness of tissues and improve the realism of the generated samples \cite{abousamra2023Topology}.

\textcolor{r4col}{
Lately, diffusion models have become the SOTA in general computer vision and now produce far more semantically plausible and noise-free images than GANs. These improvements promise to make synthetic data finally accepted by pathologists and the broader CPath community as reliable training data and significantly improve the generalizability of models trained on them \cite{aversa2023diffinfinite}.
}

\subsection{Multi-Instance Learning (MIL) Models}
Multi-instance learning (MIL) involves training from data that is labelled as high-level bags consisting of numerous unlabelled instances. In the context of CPath, these labelled bags often represent annotated slides of far more unlabelled patch instances \cite{188}. As labels at the WSI level are much easier to obtain (and hence more prevalent) than patch-level annotations, MIL has been applied to CPath by a significant number of papers \cite{35,194,115,169,60,69,91,118,188,252,330,345,346,377,381,382,538,613,474,547,573,279,538,629,556,564,559,476,636}. Since both utilize coarser annotations for training on massive images, MIL is similar to weakly-supervised learning. However, weak supervision predicts at a finer level (e.g. pixel segmentation from labelled patches) than the provided annotation while MIL prediction is typically at the same level.

One notable work used a two stage approach to first encode patches with a CNN from a slide into feature vectors and then pass the most cancer-likely ones to an slide-level classification RNN. A similar work first detected abnormal regions in the WSI before adaptively fusing the instance-level features with an importance coefficient \cite{169}. Adding additional instance-specific attributes tends to improve MIL performance. One work applied a nuclei grading network to provide a cell-level prediction for each patch, and demonstrate this out-performs hand-crafted cell features for overall slide classification \cite{ViT4}. Recent works explore the morphological and spatial relationships between instances, which conforms with pathologist diagnostic intuitions and have demonstrably improved performance, especially with unbalanced data \cite{636}.

\textcolor{r4col}{
As not all instances are equally relevant to the bag label, many works focus on building attention mechanisms to adaptively focus on more relevant instances. One work used such a mechanism to highlight regions of interest and improve localization relative to other \textcolor{r5col}{SOTA} CNNs \cite{188}. MIL models can be improved by considering multi-scale information: one work notably used embeddings from different magnification levels and self-supervised contrastive learning to learn WSI classifiers \cite{474}. Some works explicitly encode the patient-slide-patch hierarchy in the attention mechanism \cite{Li2023-my,Wong2023-hq}, with one work using a cellular graph for top-down attention \cite{Nakhli_undated-tb}. Graph Neural Networks (GNNs) have been explored to leverage intra- and inter-cell relationships, enabling cancer grading \cite{Abbas2023-ny}, classification \cite{Wong2023-cr}, and survival prediction \cite{Azadi2023-ms,Nakhli2023-yq}. These hierarchy- and morphology-aware models are the current SOTA and pave the way for future improvements.
}

One persistent challenge with using MIL in CPath, compared to natural image computer vision, is the lack of large-scale WSI datasets \cite{campanella2018}. One recent work addressed issues related to small sample cohorts by splitting up large bags (and their labels) into smaller ones through pseudo-bags \cite{zhang2022dtfd}.

\subsection{Contrastive Self-Supervised Learning for Few-Shot Generalization}
The idea of using contrastive learning (CL) for self-supervised learning (SSL) dates back to 2005, yet only recently gained momentum in CPath \cite{closs,closs1,474,536,608,518,526,543}. By using a contrastive loss, a feature embedding is learned to ensure similar (positive) examples are close in vector space, while dissimilar (negative) examples are distant \cite{closs,closs1}. Contrastive learning is an attractive approach for CPath because when used as self-supervision for few-shot learning \cite{474,536,526}, it does not require labelling the massive self-supervision image set but only labelling the small subset used for training on the downstream task, an approach that has recently achieved \textcolor{r5col}{SOTA} performance in a wide array of tasks in CPath \cite{526}. SimCLR was originally proposed to learn representations invariant to different augmentation transforms (such as crop, noise) for natural images \cite{simclr}, and when applied to CPath, was found to match or outperform \textcolor{r5col}{SOTA} supervised techniques \cite{526}. \textcolor{r4col}{Self-supervised pre-training has been shown to perform best against fully-supervised pre-training when applied to small but visually-diverse datasets \cite{526}. Recent works have focused on transferring the self-supervised representations to the downstream task more intelligently: through latent space transfers \cite{yang2022towards}, with an awareness of the patient-slide-patch hierarchy \cite{Jiang_2023_CVPR, Gildenblat_2023_Miccai}, or with semi-supervised pseudo-label guidance \cite{Basak_2023_CVPR}.}

\subsection{Novel CPath Architectures} 
In this section, we discuss papers that made significant changes to the model design or completely designed an architecture from scratch for CPath tasks \cite{70,72,80,126,279,314,251,132,532}. Typically, model architectures are adapted from the natural image domain and minor changes applied for CPath tasks, rather than being designed from scratch for CPath directly. Unfortunately, general computer vision architectures typically require large computational resources not necessarily available in clinical settings and are prone to overfitting on smaller CPath training sets \cite{72}.

More importantly, CPath tasks often comprise of multiple specialized sub-tasks not addressed by common architectures - in such cases, CPath-specific architectures perform better. ``PlexusNet'' achieved \textcolor{r5col}{SOTA} performance with significantly fewer parameters \cite{72} and ``Hover-Net'' used a three-branched architecture for nuclei classification and instance segmentation \cite{70}. Path R-CNN similarly used one branch to generate epithelial region proposals and another to segment tumours \cite{80}.

In other cases, custom architectures are designed to obtain better performance with respect to certain metrics \cite{37,204,288,344,371} or improve computational efficiency and speed \cite{334}, where since model inference can be a bottleneck for WSI processing. To automate architecture design, neural architecture search (NAS) is often used. This is an umbrella term covering evolutionary algorithms (EA), deep learning (specifically Reinforcement Learning), and gradient-based NAS searches. There are two approaches to EA: (1) Neuroevolution, which more generally optimizes at the neuron-level to find optimal weights, and (2) Evolutionary-Algorithms based NAS (EANAS), which searches for optimal combinations of mid-sized neural network blocks and conducts training after this architectural search \cite{neuro, neuros}. In CPath, reinforcement learning-based NAS has designed models for cancer prediction which were found to train faster, have fewer parameters, and perform comparably with manually designed models \cite{NAS}. Another work demonstrated that a significantly smaller model can outperform existing SOTA models on a variety of CPath tasks using an adaptive optimization strategy \cite{532}.

We hypothesize that NAS has yet to be explored significantly in CPath due to the lack of annotated data (see Section \ref{data-avail}) and its relative recency as a research area. According to the ``no free lunch theorems'' \cite{nofree1, nofree2}, no single model can perform best on all tasks. However, computationally efficient but performant models are crucial for CPath applications, and NAS is the most promising approach to computationally design such architectures without manual engineering.

\subsection{Model Comparison}
The various model architectures and types discussed above can and should be compared on common benchmarks to determine the best models for a given task \cite{laleh2022benchmarking}. Numerous papers have conducted such benchmarking work on CNNs. One work comparing GoogLeNet, AlexNet, VGG16, and FaceNet on breast cancer metastasis classification found that deeper networks (i.e. GoogLeNet) predictably performed better \cite{48}. Another work found that using ResNet-34 with a custom gradient descent performed best \cite{19}. Finally, VGG-19 performed best in colorectal tissue subtype clasification, showing that deeper SOTA networks do not necessarily perform better universally. Which CNN performs best depends on the task, the nature of the data, the metrics used, training time, hyperparameters, and/or hardware constraints.

Likewise, third parties have organized ``grand-challenges'' to facilitate the fair comparison of different techniques on a common CPath task and dataset. In some cases, SOTA CNNs achieve the best results, such as the adapted GoogLeNet that obtained the highest AUC \cite{16} and the AlexNet that achieved highest accuracy \cite{211} for breast cancer detection in the CAMELYON16 challenge. Likewise, SqueezeNet, which is an existing \textcolor{r5col}{SOTA} network performs best in colorectal tissue subtype classification \cite{iandola2016squeezenet}. On the contrary, the best performing models for mitosis detection in the TUPAC16 \cite{48} and MITOS12 \cite{371} challenges both relied on custom CNN architectures. For breast cancer diagnosis, a novel Hybrid CNN achieved the best results in the BACH18 (ICIAR18) dataset \cite{365} while the two teams achieving the best classification accuracy in the BreakHis dataset used differing approaches: one directly used ResNet-50 \cite{62} and the other used an ensemble of VGG networks \cite{125}. For nuclei segmentation on the Kumar-TCGA dataset, a novel framework using ResNet and another existing model achieved the highest F1-score \cite{314}. Lastly, a custom CNN achieved the best results for gladn segmentation on the GLaS dataset \cite{281}.

However, as mentioned in Section \ref{selection_criteria}, many grand challenges use private datasets or even extract data from larger public repositories without referencing the original WSIs used. Furthermore, benchmark datasets address different tasks and lack standardization. As models that are hyper-optimized to for specific sets of data continue to be released, the lack of more standardized benchmark datasets and model comparison studies make it impossible to systematically compare new models against existing ones or assess their robustness in clinical settings, thus impeding model development in CPath.       
\section{Evaluation and Regulations}\label{sec:EvalReg}
\subsection{Clinical Validation}
Within the domain of CPath, clinical validation is essential for substantiating the decisions produced by deep learning models so that they are more readily accepted by the medical community. Generally, acceptable clinical criteria are determined by authoritative professional guidelines, consensus, or evidence-based sources. However, in CPath, prediction results are generated by the computer scientists and engineers who build the model, and may not be completely aware of where their work fits into the clinical pathology workflow--the clinical implications of this arrangement are often unknown \cite{363}. By incorporating pathologist expertise, clinical validation can better align the technical work with clinical objectives.

Despite the importance of this step for real-world deployment, very few works have performed clinical validation with expert pathologists. We identify three prominent types of clinical validation in the CPath literature: (1) direct performance comparison of CAD tools with pathologists on a similar task, (2) impact of CAD tool assistance on pathologist performance, and (3) pathologist validation of CAD tool outputs. Each topic is further discussed in the sections alongside notable results.

\textbf{Direct Performance Comparison with Pathologists}
To validate the benefits of deep learning methods, it is desirable that they equal or surpass the performance humans to gain the trust of pathologists in their decisions and their willingness to use them as a second opinion \cite{65}. With this in mind, many papers directly compared their models with pathologists in tasks such as prognosis and diagnosis. 

One study on cancer detection found that the top computational models from the CAMELYON16 challenge out-performed the 11 pathologists with a two-hour time constraint and performed similarly to the expert pathologist without a time constraint \cite{16}. This suggests that deep learning models could be particularly useful in clinical scenarios with excessive numbers of time-critical cases to diagnose. Similarly, for tissue subtype classification, another study performed similarly to, or slightly better than individual pathologists. The proposed model agreed with all pathologists $66.6\%$ of the time and agreed with two-thirds of pathologists $76.7\%$ of the time \cite{10}. An additional study claimed their deep learning model outperformed pathologists without gynecology-specific training in ovarian carcinoma classification \cite{253}. This pushes the idea that CAD predictions can be used as a second opinion due to the potential for human error by individual pathologists.

One paper on diagnosis \textcolor{r4col}{\cite{4} demonstrated that }deep learning models \textcolor{r4col}{can} correctly classify images that even individual pathologists failed to correctly identify. However, another paper found that $50\%$ of the examples misclassified by their model were also misclassified by at least one pathologist \cite{5}. This suggests that deep learning models can aid pathologists in decision-making, but as they tend to achieve a specificity and sensitivity similar to pathologists, they must be applied cautiously to avoid reinforcing the biases or errors of individual pathologists.

Deep learning models for prognosis have been shown to achieve performance similar to or better than experts as well \cite{17,44,227}. In one study, the best model for renal clear cell carcinoma classification achieved $83\%$ accuracy, outperforming the inter-pathologist accuracy of $80\%$ \cite{227}. This shows that deep learning models and pathologists may perform similarly on patient prognosis. 

Overall, AI approaches are not perfect but have approached expert-level ability in a variety of tasks. Deep learning could play an important role as a second opinion and in democratizing the knowledge distilled from many pathologists to other pathology centres. Specifically, deep learning models appear to be best used as a tool to enhance the pathologist workflow, and could provide aid in making quick decisions with high accuracy \cite{echle2021deep}. 

\textbf{Impact of CAD Tool Assistance}
Much of CPath research is conducted under the assumption that the resulting AI tools will be intuitive, usable, and beneficial to pathologists and patients. However, CAD tools that are  \textcolor{r4col}{ developed without feedback from pathologists} could fail to integrate into a realistic pathologist workflow or impact the most significant diagnostic tasks. Thus, a valuable validation experiment is to compare and comprehend the performance of expert pathologists in clinical tasks before and after being given the assistance of a CAD tool.

In one study, a CAD system called Paige Prostate Alpha leveraged a weakly-supervised algorithm to highlight patches in a WSI with the highest probability of cancer \cite{35}. When used by pathologists, the model significantly improved sensitivity, average review time, and accuracy over unaided diagnosis. Likewise, another study using the LYNA algorithm examined the performance of six pathologists on breast cancer tumor classification before and after being able to see the LYNA-predicted patch heatmaps. The results indicate using LYNA substantially improved sensitivity, average review time, and the subjective ``obviousness'' score for all breast cancer types \cite{141, 138}.

These studies suggest that integrating CAD tools into the clinical workflow will greatly improve pathologist efficiency. However, there is a general lack of research on the impact of CAD tools on pathology efficiency. Such studies would shed more light on the impact of CAD tools and identify approaches for implementation in clinical settings.

\subsection{FDA Regulations}
Despite the ongoing development of CAD tools in CPath and its potential for triaging cases and providing second opinions, the regulations regarding this technology pose an obstacle to the testing and deployment of these devices. The FDA currently provides three levels of clearance on AI/ML-based medical technologies: 510(k) clearance, premarket approval, and the De Novo pathway. While one source lists 64 AI/ML-based medical solutions that are currently FDA-approved or cleared, none of these are in the field of CPath \cite{benjamens2020state}. A few companies, such as Paige AI, hold the 510(k) clearance for their digital pathology image viewer; however, an automated diagnostic system has yet to be approved. This may indicate a reluctance to change, and the lack of clarity in the process of FDA approval has prevented numerous impactful technologies from being deployed. There is a need for collaboration between researchers, doctors, and governmental bodies to establish a clear pathway for these novel technologies to be validated and implemented in clinical settings. 
\section{\protect\textcolor{r4col}{Emerging Trends in CPath Research}}\label{sec:emerging-trends}

\textcolor{r4col}{
Computational pathology research has seen a sudden shift of focus in the past year of 2023. Driven by recent technological advances in the field of computer vision for natural images and the release of capable foundational models in natural language processing, formerly difficult research problems in CPath have been solved, opening up exciting new avenues of research, especially the difficulty of training models on adequate annotated data. We will discuss the main research trends below in further detail and make simple predictions of where the field is headed.
}

\subsection{\protect\textcolor{r4col}{Contrastive Self-Supervised Learning becomes Mainstream}}
\textcolor{r4col}{
Data annotation for CPath is a persistent problem - it is easy to collect large amounts of visual data but much harder to annotate them. Transfer learning can help but it is difficult to transfer a model trained on one dataset to generalize to another. Whereas past efforts focused on carefully engineered methods, the recent development of contrastive self-supervised learning \cite{simclr} means that it has become the mainstream approach in CPath \cite{526, 474,536}. Not only does it utilize the massive amounts of unlabelled images typically available in CPath, but it also as a result only requires finetuning on a small set for the downstream task. We anticipate that this will lead to the development of general-use foundational models in the future to perform the most common CPath tasks, as more pathology images are collected and models become more advanced.
}

\subsection{\protect\textcolor{r4col}{Prediction becoming increasingly High-Level}}
\textcolor{r4col}{
We noticed that recent research works are increasingly addressing higher level prediction tasks than before. Whereas patch classification \cite{18,170,227,182,252} or pixel segmentation \cite{UNet,13,84,173,TCGA-Nuclei} was formerly mainstream, these problems appear to have been largely solved, and now there is far more research into higher-level problems dominate, such as multiple-instance learning \cite{188, 35, Li2023-my, Wong2023-hq, Nakhli_undated-tb}. As computational methods continue to improve, it is natural that they are applied not merely as attention aids for pathologists (i.e. at the pixel or patch level), but furthermore are used to make intelligent slide- and patient-level decisions on their own. Indeed, they promise to vastly improve pathologist efficiency when used with human pathologists in the loop to validate the automated decisions, especially when paired with modern natural language capabilities.
}

\subsection{\protect\textcolor{r4col}{Spatial and Hierarchical Relationships receiving Attention}}
\textcolor{r4col}{
Inspired by the approach taken for natural image computer vision, the mainstream approach in CPath currently requires breaking up large WSIs into smaller patches and perceiving them independently (see Figure \ref{fig:annotation_overview}). However, this ignores the spatial relationships between cells and tissues or between the patches and their parent slides in histopathology images, which are often relevant or even crucial when making decisions. Many works have recently found success in explicitly encoding an awareness of these inter-cell relationships \cite{Abbas2023-ny, Azadi2023-ms, Nakhli2023-yq} and the patch-slide-patient hierarchy \cite{Nakhli_undated-tb, Li2023-my,Wong2023-hq}, especially using Graph neural networks, but these suffer from higher latency than conventional CNNs. We anticipate future works will seek to speed up GNNs for tasks where spatial and hierarchical relationships are important and continue developing hierarchy-aware attention for MIL techniques.
}

\subsection{\protect\textcolor{r4col}{Vision-Language Models for Explainable Predictions}}\label{disc_lang}
\textcolor{r4col}{
One persistent problem in CPath has been developing models that can explain their decisions for human validation. One obvious route is to develop models that produce natural language output (and even converse with the human user to explain their decisions), but until recently, this would have required collecting massive amounts of pathology text paired with images. With foundational vision-language models widely available and able to generalize to great effect in the natural image domain \cite{clip, zhai2022lit, singh2022flava}, recent works have shown that they perform excellently when applied with minimal re-training to CPath images \cite{ikezogwo2023quilt,qu2023exploring,Lu_2023_CVPR}. Further advances require collecting more pathology-specific data, but we anticipate that crowd sourcing of public pathology annotations will become mainstream and this will lead to the development of foundational vision-language models. As natural language capabilities continue improving, we also anticipate that synoptic report automation will become feasible and reinforcement learning from human feedback (RLHF) \cite{rlhf} will become common for improving CPath language models.
}

\subsection{\protect\textcolor{r4col}{Synthetic Data now Realistic Enough}}
\textcolor{r4col}{
Whereas one way to combat the difficulty of annotating CPath data is to develop models that require fewer annotations, another trend is to generate more annotated data for training. Whereas concerns were previously raised about their realism, new advances in generative image models have now been leveraged to produce realistic histopathology images and pixel-accurate annotations simultaneously. However, current works are limited by specific tissues, organs, diseases \cite{13, 77, 83, 98}, or stains \cite{347, IHCGAN} and are limited by their inability to easily expand to other histopathology content. We note that generating synthetic data via game engines and 3D model assets is a recent trend in the natural image domain \cite{carla, unityperception, isaacgym}, but visual modelling of histopathology entities is little explored. We anticipate that future works will attempt to improve synthetic histopathology image generation by: (1) attempting to create generative models that can generalize to a broad variety of histopathology images and (2) create simulation software to generate realistic histopathology images without learned models.
}       
\section{Existing Challenges and Future Opportunities}\label{sec:Outlook}
\subsection{CPath as Anomaly Detection}
Typically in computer vision, the various classes represent distinct normative entities, such as airplanes or bears \cite{Caltech-256, ImageNet}. There exist abundant ``normal'' samples and potentially few ``anomalous'' samples, which are considered data points significantly dissimilar to the majority within a given class \cite{anomaly}. These anomalies are not only out of distribution from the samples in a dataset, there is also a lack of consensus on understanding anomalous representations as effectively identifying anomalies requires ML models to learn a feature space encompassing all ``normal'' samples within each class \cite{anomaly}. 

In other words, in general computer vision, each class cannot simply be considered as an anomalous version of any other class. However, in CPath, since each class is often considered a different disease state on a single tissue type, diseased classes are essentially extensions of the ``normal'' healthy class into the ``anomalous'' zone. From a pathologist's perspective, similar to the general computer vision approach, the curriculum learning process of a resident pathologist first involves training on histology and gaining a mastery of normal tissue identification, and then train on diseased tissues, so they are able to flag the sample as anomalous and follow up with possible diagnosis.

In light of this, it may be illuminating to approach the problem from an anomaly detection viewpoint: provided a model has sufficient variety of healthy tissue, any anomalies must then be diseased. The output of such an anomaly detection algorithm is dependent on the task at hand. One source describes several meaningful output types that may be produced \cite{anomaly}: an anomaly score which describes how anomalous a sample is and a binary label indicating whether a sample is normal or anomalous. If only identifying anomalous samples is enough, a binary classification procedure may be sufficient. However, if it is necessary to identify the particular stage of progression of a disease type, then a more granular approach in assigning some \textit{anomaly score} may be more appropriate as explored in a previous work \cite{ECCV}. This work found that the confidence score in tissue classification was inversely correlated with disease progression, thus the confidence score may act as a proxy for an anomaly score. Theoretically, such approaches may better replicate the behaviour of pathologists. While several works have used an anomaly detection approach on medical image data outside of CPath \cite{Anomaly-Skin1, Anomaly-Skin2, Anomaly-GI}, few works tackle the problem for WSI data in CPath.

\subsection{Leveraging Existing Datasets}
As mentioned in Section \ref{selection_criteria} of this paper, a minority of datasets in CPath are available to be freely used by the public. Additionally, the level of annotations varies for each dataset. However as can be noted in Table \ref{DC} of the supplementary material, for prominent public datasets such as CAMELYON16, CAMELYON17, GlaS, BreakHis, and TCGA, there is far more available data annotated at the slide level as opposed to more granular predictions. For example, considering breast datasets, there are 399 WSIs annotated at the Slide and ROI levels in CAMELYON16 \cite{Camelyon16} and 1399 WSIs annotated at the Patient, Slide, and ROI levels in CAMELYON17 \cite{Camelyon17}, in contrast, the TCGA-BRCA dataset contains $1163$ diagnostic slides and $1978$ tissue slides that are accompanied with labels at the Patient and Slide levels and diagnostic reports with labels for tissue features and tumor grades \cite{TCGA-GDC}. 

The lack of publicly available datasets with granular annotations is a major challenge in CPath. To address this lack some training data, techniques have been proposed to efficiently obtain labels, such as an active deep learning frameworks that use a small amount of labelled data to suggest the most pertinent unlabelled samples to be annotated \cite{103}. Alternatively, other works propose models to synthetically create WSI patches, usually with the use of GANs. For example, Hou et al. \cite{13} introduced an unsupervised pipeline that was capable of synthesizing annotated data at a large scale, noting that even pathologists had difficulty distinguishing between real and synthesized patches. However, despite these promising results, the issue of acquiring accurate and large datasets remains a prevalent issue within CPath. 

Generally, tasks such as tissue classification or gland segmentation require labels at the ROI, Patch, or Pixel levels. However, existing data annotated at the patient and slide levels can be used for these tasks by leveraging weakly supervised techniques such as MIL \cite{279, 188}, or by learning rich representations using self-supervised techniques such as DINO \cite{caron2021emerging, chen2022scaling} and contrastive learning \cite{526} that can be used in downstream tasks. Specifically, work is being done to develop training methodologies and architectures that are more data efficient for patient- and slide-level annotations, such as CLAM, which is a MIL technique that is used to train a performant CPath model with as little as $25\%$ of the training data to get over $0.9$ AUC \cite{279}. Another recent work used self-supervised learning on WSIs without labels to train a hierarchical vision transformer and used the learned representations to fine-tune for cancer subtyping tasks. This finetuned model outperformed other \textcolor{r5col}{SOTA} methods that used supervised learning methods on both the full training set and when all models used only $25\%$ of the training set. These examples demonstrate a recent trend in the application of weakly and self-supervised learning techniques to leverage pre-existing and available data with weak labels, showcasing that a large amount of granular labels are not necessarily required for achieving \textcolor{r5col}{SOTA} performance. We urge researchers in the CPath field to follow this trend and focus on how to leverage existing weakly labelled datasets, especially to learn rich representations as a pre-training step for learning on smaller strongly labelled datasets. 

\subsection{Creating New Datasets}
Although we mention the availability of many datasets and comment on how to leverage this existing data, there is still a need for new CPath datasets that address overlooked clinical and diagnostic areas. Therefore, creation of new CPath datasets should focus on addressing two main goals: (1) tasks that are not addressed adequately by existing datasets and (2) accumulating as large a dataset as possible with maximal variety.

Regarding the first goal, there are still organs, diseases, and pathology tasks without freely available data or sufficient annotations to develop CAD tools. For example, in Figure \ref{fig:dataset_overview}, we see that whereas breast tissue datasets are abundant, there are few public datasets for the brain and none for the liver. Collecting and releasing datasets for these organs would have significant impact in enabling further works focusing on these applications. Further, analysis of specific organ synoptic reports can guide CPath researchers to build CAD tools to identify or discriminate the most impactful diagnostic parameters. In the case of the prostate, which is discussed in Section \ref{sec:organs_diseases}, the synoptic report requires distinguishing IDC from PIN and PIA as it correlates to high Gleason scores. \textcolor{r4col}{This is important, as} high-grade PIN is a cancer precursor requiring follow-up sessions for screenings. These parameters are identified and noted in the report by the pathologist and factor into the final diagnosis and grading. Thus, collecting annotated datasets for such parameters can be crucial to developing CAD tools that are relevant to clinical workflows and can enrich learned representations.

The second goal is concerned with the scaling laws of deep learning models with respect to the amount of data available and their application to diverse clinical settings. As seen in the general computer vision domain, larger datasets tend to improve model performance, especially when used to learn rich model representations through pre-training that can be used for downstream tasks such as classification and semantic segmentation \cite{sun2017revisiting}. Additionally, ensuring that datasets capture the underlying data distribution and thus sufficiently encompass the test distribution has been shown to be especially important in the medical domain \cite{althnian2021impact}. For CPath, this means ensuring a dataset captures the expected variations in tissue structure, disease progression, staining, preparation artifacts, scanner types, and image processing. \textcolor{r4col}{Collecting a sufficiently large dataset continues to be problematic, however, so recent works have focused on using crowd sourcing to annotate histopathology data posted publicly on Twitter and YouTube \cite{huang2023visual,qu2023exploring}, a practice that is similar to that commonly used for natural images.}

\subsection{Pre- and Post-Analytical CAD Tools}
In recent years, advances in image analysis, object detection, and segmentation have motivated new approaches to support the analytical phase of the clinical workflow, especially in the two steps where CAD tools could significantly increase efficiency and accuracy: (1) specimen registration and (2) pathology reports. This need is highlighted by a study determining that the pre-analytical and post-analytical phases (as shown in Figure \ref{fig:clinical_workflow} account for up to $77\%$ of medical errors in pathology \cite{abdollahi2014types}. Likewise, Meier et al. classify $14\%$ of medical errors as diagnostic errors, with an even smaller proportion being misinterpreted diagnoses in their study \cite{meier2011study}. Other authors attribute approximately $15-25\%$ of diagnostic errors to slide interpretation \cite{darcy2016test, nakhleh2005prelude, nakhleh2006error, nakhleh2008patient, nakhleh2016interpretive}. These results reinforce the need for CPath applications that address more than just the analytical phase \cite{nakhleh2015role}. Considering post-analytical step of compiling a pathology report, a few natural language processing efforts have been used to analyze completed pathology reports \cite{odisho2020natural, lopez2022natural, kim2020validation}, extract primary site codes from reports\cite{qiu2017deep}, and generate of captions or descriptive texts for WSI patches \cite{95}. However, to the best of our knowledge, there are no works that reliably extract clinical data from service requests and electronic medical records to automatically generate synoptic or text reports. Developing such a tool that could explicitly identify the most significant parameters for its decisions would directly improve clinical workflow and increase the interpretability of the results at the same time. We encourage the field of CPath to expand its efforts in creating tools for the pre- and post-analytical steps in order to reduce the large proportion of clinical errors attributed to those phases, and suggest some potential applications in Figure \ref{fig:clinical_workflow}.

\subsection{Multi Domain Learning}
Despite being particularly well-suited for CPath, multi-domain learning (MDL) is still a relatively unexplored topic. MDL aims to train a unified architecture that can solve many tasks (e.g. lesion classification, tumour grading) for data coming from different domains (e.g. breast, prostate, liver). During inference, the model receives an input image and the corresponding domain indicator and is able to solve the corresponding task for the given domain. There are two reasons that make MDL attractive for CPath. The first is that the additional information from a source domain (coming from a related organ such as the stomach) can be informative for improving performance in the target domain (e.g. colon). By sharing representations between related domains, the model is enabled to generalize to other domains. The second motivation is to alleviate the data sparsity problem where one domain has a limited number of labeled data. Through MDL, the domain with limited data can benefit from the features that are jointly learned with other related tasks/domains \cite{zamir2018taskonomy, zhang2021survey}.


\subsection{Federated Learning for Multi-Central CPath}
Data-driven models require a large amount of data to yield strong performance. In CPath, this requires incorporating diverse datasets with varying tissue slide preparations, staining quality, and scanners. An obvious solution to train such models is to accumulate the data from multiple medical centers into a centralized repository. However, in practice, data privacy regulations may not permit such data sharing between medical institutions, especially between countries. A possible solution lies in privacy-preserving training algorithms, such as federated learning \cite{bonawitz2019towards,kairouz2021advances}, which can make use of decentralized data from multiple institutions while maintaining data privacy. In federated learning, training starts with a generic machine learning model in a centrally located server. But instead of transferring data to a centralized server for training, copies of the model are sent to individual institutions for training on their local data. The learning updates are encrypted and sent to the central server and then aggregated across the institutions. Ming Lu et al. \cite{381} demonstrated the feasibility and effectiveness of applying federated, attention-based weakly supervised learning for general-purpose classification and survival prediction on WSIs using data from different sites. Using such algorithms for CPath can facilitate cross-institutional collaborations and can be a viable solution for future commercial solutions that need to continuously augment and improve their ML models using decentralized data. 


\subsection{CPath-specific Architecture Designs}
Many deep learning architectures are not designed for CPath specifically, which raises a serious question about the optimality of using ``borrowed'' architectures from general computer vision. For instance, \cite{72} notes that traditional CV architectures may not be well suited for CPath due to a large number of parameters that risk overfitting. Additionally, the field of pathology has much domain-specific knowledge that should be taken into account before choosing an ML model. For example, under varying magnifications different morphological patterns are captured, from cellular-level details to tissue architecture features \cite{149}. Naively applying an architecture without considering such details could discard key visual information and lead to deteriorated performance. 

Unlike natural images, WSIs exhibit translational, rotational, and reflective symmetry \cite{PCam_paper} and CNNs for general vision applications do not exploit this symmetry. The conventional approach to overcome this issue is to train the model with augmented rotations and reflections, but this increases training time and does not explicitly restrict CNN kernels to exploit those symmetries. Rotation-equivariant CNNs, which are inherently equivariant to rotations and reflections were introduced for digital pathology \cite{PCam_paper}, significantly improving over a comparable CNN on slide level classification and tumor localization. Similarly, Lafarge et al. \cite{277} designed a group convolution layer leveraging the rotational symmetry of pathology data to yield superior performance in mitosis detection, nuclei segmentation, and tumor classification tasks. These results motivate the application and further research of rotation-equivariant models for CPath.

In general, we note that the \textcolor{r5col}{SOTA} computer vision architectures used in computational pathology have tended to lag behind those used for natural images by a couple of years. This delay in knowledge propagation from the mainline computer vision research in natural images may be due to the data-centric nature of the CPath field. As data labelling is specialized and expensive to conduct, annotating more data or clever training tweaks to finetune established architectures is more attractive than developing advanced, specialized architectures. However, we recommend that CPath researchers should still use the most powerful relevant models available for the simple reason that they tend to perform best given the computational resources available. While computational efficiency is generally not as important during training, it is imperative at inference time if models are to be run in real-time on medical devices with limited computational resources.



\subsection{Digital and Computational Pathology Adoption}
Despite the numerous advantages to the clinical workflow and applications offered by using digital pathology and CPath, the adoption of digital pathology remains the first barrier to clinical use. A major reason for adoption hesitancy is the common opinion that digital slide analysis is an unnecessary step in a pathologist’s workflow which has been refined over decades to produce reproducible and robust diagnoses without digitization \cite{lujan2022challenges, liu2019digital, griffin2017digital, jara2010digital}. In terms of clinical efficiency, studies have shown mixed results, with two finding that digitization actually decreased efficiency (by increasing turnaround time by $19\%$) \cite{schuffler2021integrated, hanna2019whole}. However, another study demonstrated a clear increase in productivity and reduction of turnaround-time \cite{baidoshvili2018evaluating}. One of the co-authors (B.N.) has implemented digital pathology at a public tertiary institution, which began as a pilot study over three years including three experienced academic pathologists and showed that digitization reduced turnaround time by $18\%$ for biopsies and $25\%$ for resections, and increased case output by $17\%$. These trial results led to all pathologists not retiring within two years to transition to a digital pathology workflow in 2019. Due to the varied nature of results and outcomes in studies analyzing the effectiveness of digital pathology there is more work to be done to have a multi-institution and lab analysis for more general and concrete results.

\textcolor{r2col}{A major factor in the adoption of digital or computational pathology practices is the source of funding and the pay structure of pathologists. A few cost-analysis studies show that the transition to digital pathology becomes financially advantageous in 2 years, with savings projected to be up to \$5M after 5 years in a sizeable tertiary center \cite{hanna2019implementation, griffin2017digital, ho2014can, evgenievna2022analysis}. The financial impact will also be viewed differently in public vs private healthcare settings. Public healthcare is primarily limited by funding and universal access to healthcare whereas for private lab networks improvements in processes and services are directly linked to the prospects of obtaining additional contracts and increased profitability. However, studies considering multiple institutions and funding settings are still required to fully characterize the financial impact compared to clinical benefit. Additionally, on an individual pathologist level, compensation structures can affect buy in for implementation. For example, at our co-authors' B.N. and V.Q.T institution, a fee-for-service structure is used to compensate pathologists thus an increase in throughput and productivity has a direct correlation to increased pay. We propose that this fee-for-service model contributes to the widespread embracement of DP at this institution. In contrast, pathologists in a salary-based environment are paid based based on a combined package of services which includes diagnostics, research, teaching, administration, quality control, etc. An increase in clinical productivity would technically not benefit them directly, as it would translate to a high number of rendered diagnostics over the same amount of time.}

Integration CPath into the clinical workflow is relatively understudied as few papers have actually deployed, or performed clinical validation of their results. Works in this area have either proposed methods to deploy their models in the clinic or developed tools to enable the use of their research in the clinic \cite{18,65,223}. However, as a primary goal of CPath is the use of CAD tools in clinical settings, more works should consider how to integrate models and tools into the clinical workflow, especially in conjunction with expert clinicians.

\subsection{Institutional Challenges}
Several institutional challenges may affect the implementation of CPath tools, and similar challenges in implementing digital pathology workflows at medical institutions have been well-described by many studies \cite{isaacs2011implementation, pare2016impacts, thorstenson2014implementation, cheng2016enabling, stathonikos2013going}. As noted by multiple studies considering the digital transition of pathology laboratories \cite{isaacs2011implementation, pare2016impacts, thorstenson2014implementation, cheng2016enabling, stathonikos2013going}, the importance of a common shared goal and frequent communication between the involved parties is necessary to successfully deploy a digital system. These lessons are likely extendable in the context of CPath and CAD development as well. Specifically, Cheng et al. \cite{cheng2016enabling} reported on their experiences and lessons learned as a 7-point-based system to efficiently deploy a digital pathology system in a large academic center. We believe similar systematic approaches will need to be developed to implement CPath applications in a clinical setting. 

Another institutional challenge concerns the regulatory oversight at the departmental, institutional, accrediting agencies, pathology association, state/provincial, and federal levels.  Regulatory measures underlying WSI scanners are well established, as well as the technical and clinical validation of their use \cite{pantanowitz2013validating, abels2017current, parwani2014regulatory}. On the other hand, patient confidentiality, ethics, medical data storage regulations, and data encryption laws are equally, if not more, time-consuming and intensive to comply with. These issues can be mitigated by deploying a standardized digital pathology system throughout multiple institutions at the state/provincial level. For example, our co-author (B.N.) has obtained governmental approval and funding to distribute a set of digital pathology systems throughout the province's public anatomical pathology laboratories. Similarly, a unified set of standards for processing and digitizing slides, along with unifying storage and access to WSIs for research use in collaborative efforts is paramount in moving forward in both the development and implementation of CAD systems. 

\subsection{Clinical Alignment of CPath Tasks}
Researchers in the CPath field must ensure that the CAD tools they create are clinically relevant and applicable to pathology so that effort and resources are not allocated towards extraneous or clinically irrelevant tasks. For example, certain CADs have been proposed to facilitate case triaging and reduce turnaround time for critical diagnoses \cite{AJ-IDC, 141, 147, 171, sankarapandian2021pathology}]. However, several regulatory agencies in pathology aim for $90\%$ of cases to be completed within a timeframe of 72 hours for signing-out resection specimens and up to 48 hours for biopsies \cite{alshieban2015reducing, nakhleh2005quality}. In this context, triaging becomes extraneous, as signing out cases faster than 48-72 hours has no clinical impact. However, in the context of an institution operating at longer turnaround times or struggling to keep up with the caseload, this method could be lifesaving. Alternatively, identifying mitotic figures and counting positive Ki-67 nuclei are appreciated tools already in use in multiple digital pathology settings, despite these tools being seldom applied to the large caseload proportion of most practicing pathologists.

As noted previously, the overall number of pathologists in the USA has decreased $17\%$ from 2007 to 2017 and caseloads have increased by $41.7\%$ \cite{metter2019trends}. This trend places further emphasis of developing CAD tools towards specific challenges encountered by pathologists and where sub-specialists may not be readily available. For example, a large consortium generated a prostate cancer CAD that achieved a $86.8\%$ concordance with expert genitourinary pathologists \cite{bulten2022artificial}, a significant breakthrough for healthcare settings where prostate biopsies are not signed out by sub-specialists. Additionally, targeting specific diagnoses with high rates of medical errors and inter-observer variance, notably in dermatological, gynecological, and gastrointestinal pathology, should be prioritized and integrated into practice quickly to support patient care \cite{peck2018review}. Finally, advanced CAD tools capable of diagnosing features out of reach by conventional pathology could have a great impact. For example, identifying the origin of metastases from morphological cues on the WSIs without added \textcolor{r5col}{IHC} \cite{382} or CADs capable of calculating the exact involvement of cancer on a biopsy core for prognostic purposes \cite{bulten2022artificial}.

\subsection{Concluding Remarks}
Bringing pathologists and computer scientists together and initiating meaningful collaborations with shared gains between all parties is likely the most efficient path forward for CPath and CAD integration. Opportunities to facilitate collaborations should be promoted by parties such as the Pathology Innovation Collaborative Community and the Digital Pathology Association. Furthermore, we encourage involved pathologists and computer scientists to communicate and collaborate on studies towards the common goal of providing patients with fast, reproducible, and high-quality care.            
\section*{Acknowledgment}
Authors would like to thank Huron Digital Pathology for providing support and insightful discussions related to digital pathology hardware infrastructures. Authors would like to also thank Xin Zhao, Koosha E. Khorasani, Mona Sharifi, Kiana Abtahi, Alejandra {Zambrano Luna}, Cassandre Notton, Tomas Pereira, and Amirhossein Mohammadi for their help in initial preparation of our model-cards. This work was partly supported  from NSERC-CRD (CRDPJ515553-17) funding grant and MITACS-Elevate award (Ref. IT14369).

\bibliographystyle{unsrt}
\bibliography{citations}

\pagebreak
\onecolumn
\textbf{\section{\label{sec:supp}Appendix}}
\subsection{Clinical Pathology Workflow}
\label{sec:appendix_clinical_workflow}
\textbf{\textit{Sample Collection}}
Tissue samples are categorized into two general types a) \textit{small specimens}, normally obtained to diagnose disease and guide subsequent treatment and b) \textit{large specimens}, surgically removed to treat the disease after diagnosis. \textit{Small specimen} biopsies are performed by different specialties in different settings, which can vary from family doctors sampling skin lesions to head and neck specialists performing panendoscopic biopsies under anesthesia. Based on the type of sample required and its originating site, small specimen samples are obtained by different methods: 1) core biopsies, 2) cytological specimens, 3) small excisions, and 4) pincer biopsies. In contrast, \textit{large specimens} are mostly resections performed by surgeons for treatment purposes once the diagnosis has already been obtained. Large specimens are significantly more complex than small specimens and require a high-level of expertise to process before reaching the microscopic interpretation step.

\textbf{\textit{Accessioning}}
The patient-care team fills out request forms which are tagged to the pathology specimen and sent along with the sample to the pathology department to enter the specimen and patient information into the laboratory information system (LIS). Depending on the institution, the LIS can then be linked to the electronic medical records (EMR) system to populate basic demographic data, and to a slide tracking system (STS) to locate and time each event in pathology. Ensuring that accessioning is done correctly is essential, as specimen mix-ups or incorrect data entry from request forms are a large source of errors in the worfklow \cite{abdollahi2014types}. 

\textbf{\textit{Specimen Preparation}}
Most samples arrive at the pathology lab having already been preserved in a 4\% formalin solution. Other preservation media are used for specific pathology objectives; notably using \textcolor{r5col}{Roswell Park Memorial Institute medium} (RPMI) to preserve cells for flow cytometry and glutaraldehyde preservation for electronic microscopy \cite{smith2011cytology}. Sometimes, the samples are sent to the pathology lab without a preservation medium, and in this ``fresh'' state the tissue will undergo cold ischemia changes until the pathologist can assess the sample, perform specimen preparation, and fix the tissue in the preservation media. Indeed, each of these events has an impact on the quality of the tissue, which in turn significantly affects the quality of the final WSI. Specifically, large specimens are almost always processed by a pathologist who will open the container, take basic measurements, and open the organ to ensure uniform formalin penetration. 

At some medical centers, intra-operative consultations for resection samples are processed in a frozen section procedure, which allows for more rapid diagnosis of the tissue specimens while trading off diagnostic accuracy when compared to fixation in formalin \cite{Brender_2005, ayyagari2021analysis}. Frozen section samples will be rapidly frozen after the proceeding \textit{grossing} step, therefore they arrive in the lab unprocessed and remain unprocessed until after grossing, undergoing the aforementioned cold ischemia changes in the meantime.

\textbf{\textit{Grossing}}
Once the basic specimen preparation has occurred, the tissue is analyzed by the pathology team without the use of a microscope; a step called grossing. Smaller specimens are often grossed by technicians who fill out a template-based description notably highlighting the number, size, and appearance of the fragments. After grossing, the tissue fragments are placed into tissue cassettes for final fixation. Grossing larger specimens is much more complex, and is usually performed by pathologists, pathologist-assistants, medical residents, or fellows who have undergone extensive training. The grossing starts by orienting the specimen according to the surgical procedure. By cross-referencing the clinical findings and the EMR reports, the operator will localize the disease, locate the pathological landmarks, describe these landmarks, and measure the extent of the disease. Specific sampling of these landmarks is performed, and these samples are then put into cassettes for the final fixation.

\textbf{\textit{Final Tissue and Slide Preparation}}
For non-frozen section samples, the final fixation step will accrue all the samples put into cassettes and will add a last phase of fixation in formalin. Afterwards, the tissue cassettes are removed from formalin and inserted into a tissue processor which dehydrates the tissue through alcohol gradients, subsequently replacing the liquid with melted paraffin wax. These samples are removed from the tissue processor and placed into a mold filled with paraffin which solidifies into a block.  A technician then cuts the blocks with a microtome into $4\mu{m}$ slices and places them onto positively charged slides. Slides are then deparaffinized, rehydrated by an inverse alcohol gradient, and stained with pigments such as Hematoxylin and Eosin (H\&E), or further processed for ancillary testing by immunohistochemistry. Once these steps are complete, the tissue is dehydrated, cleared, and mounted with a coverslip. Note that significant variations will affect the final quality of the slides based on the performance of the prior steps, the experience of the technician who cut the slides, and batch effects from the reagents which are often reused for multiple slides prior to being replaced--known as cross-contamination. 

Frozen section samples are rapidly frozen using a freezing medium such as liquid nitrogen, dry ice, or isopentane \cite{Brender_2005, Suvarna_2019, Peters_2016, ayyagari2021analysis}. After freezing, the tissue is cut using a microtome and fixed immediately, most often with formalin \cite{Suvarna_2019, Peters_2016}. Slides are then stained and covered with a glass coverslip and stored at $-80\degree$C \cite{Peters_2016}.

\textbf{\textit{Slide Scanning and Image Processing}}
The stained slides are cleaned of excess mounting media and scanned with a whole-slide scanner. They are then loaded onto the image management software, which has previously been linked to the LIS and the STS \cite{jahn2020digital}.

\textbf{\textit{Interpretation}}
After a slide is processed and prepared, a pathologist views the slide to analyze and interpret the sample. The approach to interpretation varies depending on the specimen type. Interpretation of smaller specimens is focused on diagnosis of any disease. Analysis is performed in a decision-tree style approach to add diagnosis-specific parameters, e.g. esophagus biopsy $\rightarrow$ type of sampled mucosa $\rightarrow$ presence of folveolar-type mucosa $\rightarrow$ identify Barrett’s metaplasia $\rightarrow$ identify degree of dysplasia. Once the main diagnosis has been identified and characterized, the pathologist sweeps the remaining tissue for secondary diagnoses which can also be characterized depending on their nature. Larger specimens are more complex and \textcolor{r4col}{\sout{focus on}} \textcolor{r4col}{the focus is on} characterizing the tissue and identifying unexpected diagnoses beyond the prior diagnosis from a small specimen biopsy. Microscopic interpretation of large specimens is highly dependent on the quality of the grossing and the appropriate detection and sampling of landmarks. Each landmark (e.g., tumor surface, tumor at deepest point, surgical margins, lymph node in mesenteric fat) is characterized either according to guidelines, if available, or according to the \textcolor{r4col}{\sout{pathologist’s judgment}} \textcolor{r4col}{pathologists' judgment}. After the initial microscopic interpretation additional deeper cuts (``levels''), special stains, immunohistochemistry, and/or molecular testing may be performed to hone the diagnosis by generating new material or slides from the original tissue block.

\textbf{\textit{Pathology Report}}
The pathologist synthesizes a diagnosis by aggregating their findings from grossing and microscopic examination in combination with the patient’s clinical information, all of which are included in a final pathology report. The classic sections of a pathology report are patient information, a list of specimens included, clinical findings, grossing report, microscopic description, final diagnosis, and comment. The length and degree of complexity of the report again depends on the specimen type. Small specimen reports are often succinct, clearly and unambiguously listing relevant findings which guide treatment and follow-up. Large specimen reports depend on the disease, for example, in cancer resection specimens the grossing landmarks are specifically targeted at elements that will guide subsequent treatment.

In the past, pathology reports had no standardized format, usually taking a narrative-free text form. Free text reports can omit necessary data, include irrelevant information, and contain inconsistent descriptions \cite{renshaw_synoptic}. To combat this, synoptic reporting was introduced to provide a structured and standardized reporting format specific to each organ and cancer of interest \cite{renshaw_synoptic, hewer2020oncologist}. Over the last 15 years, synoptic reporting has enabled pathologists to communicate information to surgeons, oncologists, patients, and researchers in a consistent manner across institutions and even countries. The College of American Pathologists (CAP) and the International Collaboration on Cancer Reporting (ICCR) are the two major institutions publishing synoptic reporting protocols. The parameters included in these protocols are determined and updated by CAP and ICCR respectively to remain up-to-date and relevant for diagnosis of each cancer type. For the field of computational pathology, synoptic reporting provides a significant advantage in dataset and model creation, as a pre-normalized set of labels exist across a variety of cases and slides in the form of the synoptic parameters filled out in each report. Additionally, suggestion or prediction of synoptic report values are a possible CPath application area.

\subsection{Diagnostic Tasks}
\label{sec:appendix_diagnostic}
\textcolor{r1col}{Here we provide some examples of diagnostic tasks where CPath has been applied, for the reader to understand the variety of diagnostic problems that CPath can be used to address.}

\textbf{\textit{Detection}}
\textcolor{r1col}{A machine learning framework for detecting cancerous tissue regions and predicting scan-level diagnosis is proposed in \cite{603}, wherein thresholding and statistical analysis used to abstain from making a decision in uncertain cases.In contrast to directly predicting the presence of cancers, feature-focused detection tasks can be highly useful in patient diagnosis and treatment planning. For example, identifying microsatellite instability is a crucial factor in determining if immunotherapy will be effective on a patient, and deep learning methods were shown to be effective at detecting microsatellite instability in \cite{18}. Similarly, the detection of fibrous regions in liver WSIs is a precursor step to liver tumor classification and a computational approach to detection was demonstrated in \cite{178}. Furthermore, the first automatic detection algorithm for keratin pearls, which are valuable biomarkers for oral squamous cell carcinoma grading is presented in \cite{257}. Future research into automated detection methods for similar cancer biomarkers could be a valuable step towards developing AI-based pathologist support tools. As an example, lymphocytes, a type of white blood cell, can be detected and quantified to assess the overall health of the immune system. However, manually detecting these cells is a time-consuming task and pathologists rarely identify and count lymphocytes. \textcolor{r4col}{Thus several computational approaches, including open source tools such as QuPath \cite{bankhead2017qupath}} and deep learning based approaches, are used to provide lymphocyte counts to pathologists \cite{627}. Likewise, counting nuclei can contribute towards diagnoses, however, nuclei detection is a difficult task because of the large variations in the shape of different types of nuclei, such as nuclear clutter, heterogeneous chromatin distribution, and irregular and fuzzy boundaries. Addressing these issues, for example, spatially constrained context-aware correlation filters with hierarchical deep features extracted from multiple layers of a pre-trained network were proposed to accurately detect nuclei in \cite{553}.}

\textit{\textbf{Tissue Subtype Classification}}
\textcolor{r1col}{Deep neural networks have been shown to be effective at extracting molecular tumor features from histopathology images, opening new avenues for deep learning applications in computational pathology  \cite{191}. As an extension of the tissue subtype classification task, ML models are often able to identify important correlations between tissue structures and disease. Work on nuclei classification suggests that features regarding the nuclear inner texture are most relevant for high classification accuracy \cite{178}. Additionally, a classifier discovered unique chromatin patterns associated with specific types of thyroid follicular lesions in \cite{33}. The potential discovery of similar associations makes tissue subtype classification a relevant task to pursue. Another work presented a computational pathology framework that can localize well-known morphological features on WSIs without the need for spatial labels for each feature using attention-based multiple-instance learning on WSI classification \cite{279}. This method outperforms standard weakly-supervised classification algorithms and is adaptable to independent test cohorts, biopsy/resection samples, and varying tissue content. Additionally, the co-representation learning for classification (CoReL) framework is proposed in \cite{588} to improve state-of-the-art classification performance for nuclei classification, mitosis detection, and tissue type classification with less data \cite{588}. }

\textit{\textbf{Disease Diagnosis}}
\textcolor{r1col}{As stated in \ref{sec:application}, disease diagnosis can be considered a fine-grained classification
problem which subdivides the general positive disease class
into finer disease-specific labels based on the organ and patient
context. Under this paradigm, research tends to be focused on maximizing performance for reliable clinical applications~ \cite{65}\cite{93}\cite{95}. Recently, works have begun implementing different emulations of pathologist behaviour in their proposed models. For instance, multi-scale receptive fields were proposed for use in networks to simulate the pathologist viewing process of slides at varying zoom levels \cite{149}. Alternatively,  weighted slide-level features were used to classify Barett's esophagus and esophageal adenocarcinoma, similar to a pathologist assessing the overall impact of various cancer biomarkers \cite{170}. To emulate how pathologists isolate and focus on salient regions of the slide, the concept of visual attention can be applied to identify the most important regions of tissue slides, thus ignoring diagnostically-irrelevant image regions \cite{181}\cite{188}\cite{254}. Such methods indicate a positive step towards the clinical implementation of AI-based CAD tools by reinforcing and emulating tested methodologies in pathology. Further, differential diagnoses in complicated cases of metastatic tumors and cancer of unknown primary (CUPs) can require many clinical tests to narrow a differential diagnosis, and a method called Tumor Origin Assessment via Deep Learning (TOAD) is introduced as an assistive tool to assign a differential diagnosis \cite{493}. This work uses digitized H\&E slides of tumors with known primary origins to train a model with transfer learning and weakly supervised multitask learning to simultaneously identify the tumorous or metastatic regions and predicts the site of origin. }

\textit{\textbf{Segmentation}}
\textcolor{r1col}{Segmentation CAD tools can capture characteristics of individual glands, nuclei, and tumor regions. The wide generalizability of this task to various disease types makes it a particularly suitable tool for computational pathology, on which many studies have been conducted \cite{13}\cite{53}\cite{66}\cite{70}\cite{84}. For example, models that use segmentation to determine nuclear characteristics including size and shape can help pathologists distinguish between various cell types and consequently, disease severity \cite{85}\cite{481}. \textcolor{r4col}{\sout{To address segmentation challenges in histopathology tissue, a}In \cite{580}, a } generalized deep learning-based framework was proposed \textcolor{r4col}{\sout{in \cite{580}, using} which uses} a sequence of novel techniques \textcolor{r4col}{for\sout{in the}} preprocessing, training, and inference steps which in conjunction improve the efficiency and the generalizability of model. Similarly, a new framework for WSI analysis in colonoscopy pathology, including lesion segmentation and tissue diagnosis was developed and includes an improved U-Net with a VGG net as the backbone, as well as two training and inference schemes to address the challenge of high resolution images analysis \cite{609}.}

\textcolor{r1col}{There are also some instances of segmentation in different organs. For example, an interactive segmentation model was proposed in which the user-provided squiggles guide the model toward semantic segmentation of tissue regions \cite{486}. Also, they proposed four novel techniques to automatically extract minimalistic and human-drawn-like guiding signals from Ground Truth (GT) masks so that they can be used during the model's training. Similarly for the eye, macular edema (ME) is a common disease where analyzing the fluid lesions is a critical stage of the diagnostic process. The optical coherence tomography (OCT) technique can potentially investigate three fluid types and a novel pipeline for segmentation of the three types of fluid lesions in OCT was proposed in \cite{617}. They presented a multi-layer segmentation to detect the ROI and presented an FCN architecture with attention gate (AG) and spatial pyramid pooling (SPP) module to improve the feature extraction. To predict cellular composition from images, ALBRT is proposed in \cite{518}, using contrastive learning to learn a compressed and rotation-invariant feature representation which first detects the presence of different cell types in an image patch and then provides cell counts for each type. Another novel deep learning model was developed for simultaneous nuclei instance segmentation in \cite{552}. The model is based on an encoder-decoder architecture design that performs nuclei segmentation by predicting the distance of pixels from their nuclei centers along with the nuclei probability masks and predicts nuclei classes when nuclei type annotations are available. Another work in nuclei segmentation is the hard-boundary attention network (HBANet), which identifies hard-boundaries between nuclei, a difficult problem due to overlapped nuclei \cite{615}. It presents a background weaken module (BWM) to improve the model’s attention to the foreground, and integrates low-level features containing more detailed information into deeper feature layers. Furthermore, a gradient-based boundary adaptive strategy (GS) is designed to generate boundary-weakened data as extra inputs and train the model in an adversarial manner. Finally, segmentation has also been applied to delineate tumorous tissue regions for a variety of cancer types, such as breast \cite{25}\cite{143}, colorectal \cite{136}\cite{144}, and prostate cancer \cite{72}\cite{189}. Such works assist in the efficient isolation of tumor tissue, which is a crucial task for making accurate disease predictions. }

\subsection{Prognosis}
\textcolor{r1col}{Prognostic models must predict the likely development of a disease based on patient features. For instance, a prognostic model was developed by adjusting a tumor microenvironment-based spatial map with clinical variables such as patient age, gender, health history, and cancer stage \cite{223}. This multi-domain data analysis approach is advanced by another work, which uses both histopathological image data and cancer genomic data in their novel deep learning framework \cite{126}\cite{chen2020pathomic}. In \cite{635}, the authors discussed the correlation between platelets and other haematological measures to cancer by assessing patient status and considering the patient features in the primary care dataset, such as age and sex. They demonstrate the model performance with the plot of survival analysis per age group for platelets. Experiments on the publicly available TCGA data demonstrates that prognostic accuracy was maximized when both forms of data were simultaneously considered. Merging information from multiple WSIs of a patient allowed a hybrid aggregation network (HANet), consisting of a self-aggregation module and a WSI-aggregation module, to predict survival risks \cite{546}.}

\subsection{Prediction of Treatment Response}
\textcolor{r1col}{\sout{ A deep learning-based biomarker using H$\&$E-stained images was developed to predict pathological complete response (pCR) of breast cancer patients receiving neoadjuvant chemotherapy, which demonstrates a strong prediction ability for guiding treatment decisions \cite{li2021deep}. The developed model outperforms conventional biomarkers including stromal tumor-infiltrating lymphocytes and subtype. In another work, the authors propose a method using convolutional neural networks to discover image-based signatures for ipilimumab response prediction in malignant melanoma patients \cite{harder2019automatic}. An immunotherapy response prediction in patients with non-small cell lung cancer from H$\&$E- stained images was proposed in \cite{madabhushi2021predicting}. Similarly, a deep learning method was developed to predict the treatment response to neoadjuvant chemoradiotherapy in local advanced rectal cancer patients, which can provide assistance in making personalized treatment plans \cite{zhang2020predicting}. }}

Oral epithelial dysplasia (OED) segmentation is critical for early identification and effective treatment and HoVer-Net+ is a model to simultaneously perform nuclear instance segmentation (and classification) and semantic segmentation of epithelial layers based on H$\&$E stained histopathology slides of the oral mucosa \cite{500}. This model achieves the state-of-the-art performance in both tasks ($0.839$ dice score) and is the first method for simultaneous nuclear instance segmentation and semantic tissue segmentation.

\subsection{Cancer Statistics}\label{sec:cancer_statistics}

\begin{figure}[htp]
\includegraphics[width=0.46\textwidth]{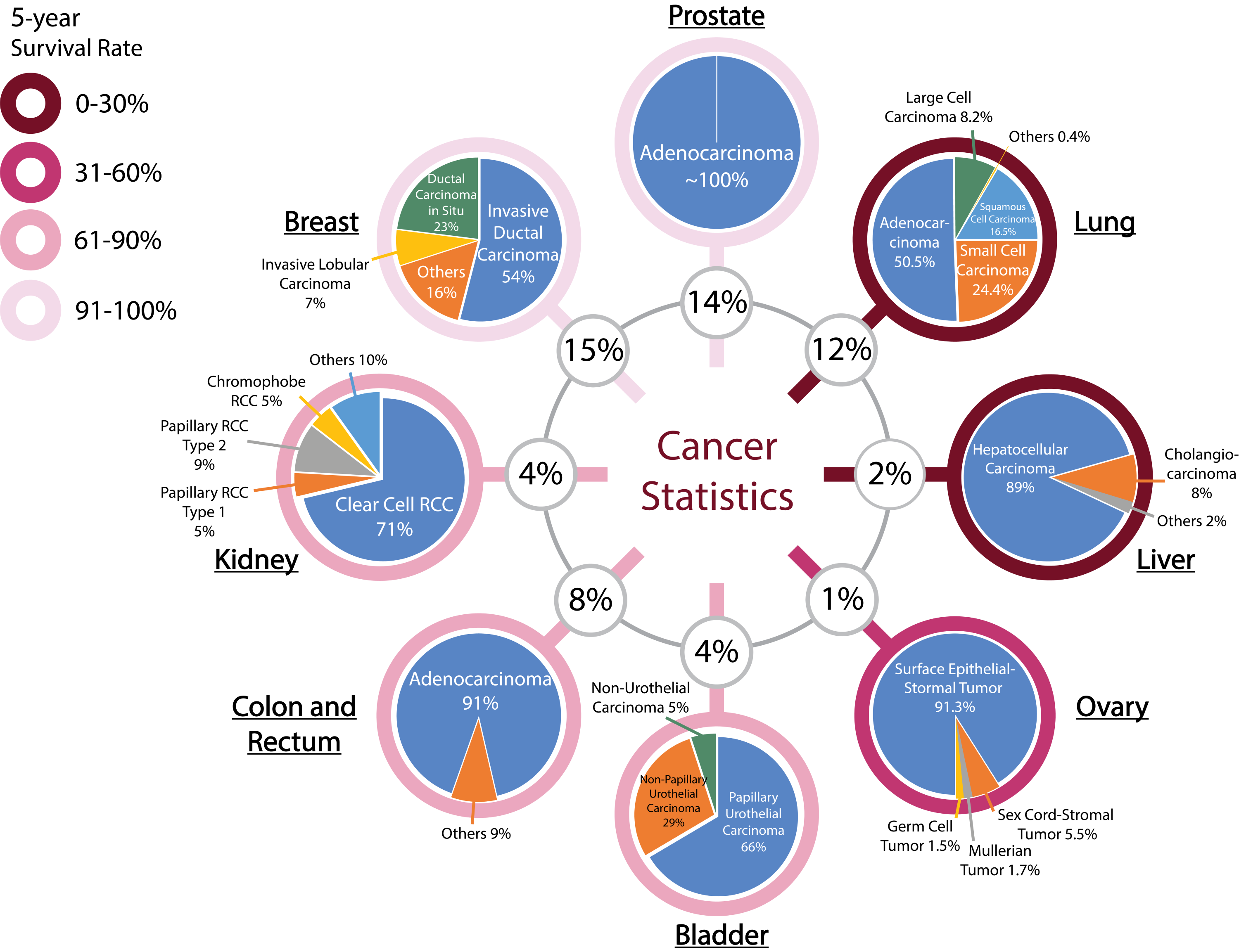}
\caption{Demonstration of the cancer statistics, featuring both the 5-Year Survival Rate and Incidence of each cancer in addition to incidence percentage of each subtype. The grey inner circle shows the incidence percentage of the respective cancer. The colored circle around each cancer corresponds to the respective 5-Year Survival Rate bin, showcasing the severity of the cancer. Darker shades (lower survival rate) means fewer people will survive the cancer after 5 years period and the cancer has poor prognosis. On the other hand, lighter shades (higher survival rates) mean more people will survive after 5 years and the cancer has good prognosis. \cite{per-organ-incidence,kenfield2008comparison,breast_cancer_prognosis,bladder-survival2,bladder-survival,andreassen2016incidence,liver-survival,goodman2007neoplasms,ries2007seer,ovari-survival,kidney-survival,bottaro2005multifocal,srougi2009incidence,colon-survival,remo2019morphology,what-prostate,prostate-prognosis,robinsandcotran,lung_cancer_prognosis}}
\label{fig:cancer-statistics}
\end{figure}

\textcolor{r1col}{Cancer remains the leading cause of global mortality in 2020, claiming nearly 10 million lives or approximately 1 in 6 deaths \cite{cancer-overview}. The grey circle in Figure \ref{fig:cancer-statistics} illustrates the prominence of Breast, Prostate, Colon, and Rectum, as well as Lung cancers, which collectively account for half of all diagnosed cases. The mounting volume of pathology cases poses a significant challenge in clinical workflows. This underscores the pivotal role of computational pathology in streamlining processes, aiding pathologists in coping with overwhelming workloads. Notably, certain cancers not only exhibit high prevalence but also contribute substantially to the overall mortality rates. Lung cancer, for instance, represents approximately $12\%$ of cancer cases in the United States, with its prognosis falling within the lowest range, as depicted in Figure \ref{fig:cancer-statistics} ($0-30\%$). Computational pathology proves instrumental, particularly in lung cancer, by facilitating classification and prognosis tasks due to distinct variations among its types in terms of presentation, prognosis, and treatment strategies. Conversely, there are less prevalent cancers like liver cancer characterized by poor prognoses. CPath's ability to compile specialized datasets for such cancers not only aids pathologists but also supports clinicians in devising personalized treatment plans. Understanding disease statistics and severity is paramount when designing a Computer-Aided Diagnosis (CAD) tool in Computational Pathology (CPath) or curating datasets. By factoring in disease prevalence, mortality rates, and severity across various cancer types, CAD tools can be optimized to prioritize detection, prognosis, and treatment planning for the most prevalent and severe cases, aligning computational pathology advancements with the urgent needs of patients and healthcare practitioners.}
\subsection{Whole Slide Imaging}
\label{appendix_wsi}
Generally, a WSI scanning device is composed of four major components \cite{Zarella_2018}: (1) a light source; (2) a slide stage; (3) an objective lens; and (4) a digital camera. In order to produce a WSI that is in focus, which is especially important for CPath works, appropriate focal points must be chosen across the slide either using a depth map or by selecting arbitrarily spaced tiles in a subset \cite{Indu_2016}. Once focal points are chosen, the image is scanned by capturing tiles or linear scans of the image, these individual components are then stitched together to form the full image known as the big flat TIFF image  \cite{Zarella_2018,Indu_2016}. To reduce the area needed to be scanned, a segmentation algorithm can be used within the scanner to separate tissue regions from extraneous background regions \cite{133}. Additionally, slide can also be scanned at various magnification levels depending on the downstream task and analysis required. The vast majority of WSIs are scanned at $20\times$ ($\sim0.5\mu{m}$/pixel) or $40\times$ ($\sim0.25\mu{m}$/pixel) magnification as these are the most useful in practice for general pathologist \cite{Zarella_2018}.

\textbf{\textit{WSI Storage and Standards}}

\begin{table}[t]
\caption{ The following table lists commercially available WSI Scanners grouped by manufacturing company and their respective available compression slide formats. }
\scriptsize
\label{tab:1ca}       
\begin{tabular*}{\columnwidth}{l} 
\specialrule{.2em}{.1em}{.1em} 
Company: Scanner Model (Slide Format)\\
\noalign{\smallskip}
\specialrule{.2em}{.1em}{.1em} 
    Leica Biosystems:  \href{https://www.leicabiosystems.com/digital-pathology/scan/}{\color{CiteColor}{Aperio AT2 / CS2 / GT450}} (TIFF (SVS)) \\ \hline     
    Hamamatsu:  \href{https://nanozoomer.hamamatsu.com/jp/en/product/search.html}{\color{CiteColor}{Nanozoomer SQ / S60 / S360 / S210 }} (JPEG)  \\ \hline 
    F. Hoffmann-La Roche AG:  \href{https://diagnostics.roche.com/nz/en_us/products_and_solutions/pathology/digital-pathology.html}{\color{CiteColor}{Ventana DP200 / iScan HT / iScan Coreo}} (BIF, TIFF, \\JPG2000, DICOM)  \\ \hline 
    Huron Digital Pathology: \href{http://www.hurondigitalpathology.com/}{\color{CiteColor}{TissueScope IQ / LE / LE120}} (BigTIFF, DICOM compliant) \\ \hline 
    Philips: \href{https://www.usa.philips.com/healthcare/product/HCNOCTN442/ultra-fast-scanner-digital-pathology-slide-scanner}{\color{CiteColor}{Ultra-Fast Scanner}}(iSyntax Philips proprietary file)  \\ \hline 
    3DHistech: \href{https://www.3dhistech.com/products-and-software/hardware/}{\color{CiteColor}{Pannoramic Series}} (MRXS, JPG, JPG2000) \\ \hline 
    Mikroscan Technologies: \href{https://www.mikroscan.com/mikroscan-sl5/}{\color{CiteColor}{SL5}} (TIFF) \\ \hline 
    Olympus: \href{https://www.olympus-lifescience.com/en/microscopes/virtual/vs120/}{\color{CiteColor}{SL5}} (JPEG, vsi, TIFF) \\ \hline 
    Somagen Diagnostics: \href{http://www.somagen.com/products/visiontek/}{\color{CiteColor}{Sakura VisionTek}} (BigTIFF, TIFF, JPG2000) \\ \hline 
    Akoya Biosciences: \href{https://www.akoyabio.com/phenoptics/mantra-vectra-instruments/vectra-polaris/}{\color{CiteColor}{Vectra Polaris}} (JPEG, single-layer TIFF, BMP, or PNG) \\ \hline 
    Meyer Instruments: \href{https://www.meyerinst.com/motic-easyscan-pro-one-and-6-digital-slide-scanners/}{\color{CiteColor}{EASYSCAN PRO 6}} (SVS, MDS, JPEG, JPEG2000) \\ \hline 
    Kfbio: \href{https://kfbio2018.en.made-in-china.com/product/avJxPZjrsHRp/China-Digital-Pathology-Slide-Scanner-with-2-Slides-Module.html}{\color{CiteColor}{KF-PRO}} (JPEG, JPEG2000, BMP, TIFF) \\ \hline
    Motic: \href{https://www.motic.com/AM_DigitalScanner/}{\color{CiteColor}{EasyScan Pro}} (JPEG, JPEG2000, Aperio Compatible) \\ \hline 
    Precipoint: \href{https://www.precipoint.com/o8-oil-microscope-scanner/}{\color{CiteColor}{PreciPoint O8}} (GTIF) \\ \hline 
    Zeiss: \href{https://www.zeiss.com/microscopy/int/products/imaging-systems/axio-scan-z1.html}{\color{CiteColor}{Zeiss Axio}} (Not specified) \\ \hline 
    Objective Imaging: \href{http://www.objectiveimaging.com/Glissando/Glissando-Scanner.php}{\color{CiteColor}{Glissando}} (SVS, BigTIFF) \\ \hline 
    Microvisioneer: \href{https://www.microvisioneer.com/}{\color{CiteColor}{manualWSI}} (Not specified) \\ 
\noalign{\smallskip}
\specialrule{.2em}{.1em}{.1em}
\end{tabular*}
\end{table}

WSIs are in giga-pixel dimension format \cite{CPsurvey1,Herrmann_2018}. For instance a tissue in $1cm\times1{cm}$ size scanned $@0.25\mu{m}$/pixel resolution can produce a $4.8$GB image (uncompressed) with a $50,000\times50,000$ pixels. Due to this large size, hardware constraints may not support viewing entire WSIs at full resolution \cite{MarquesGodinho_2017}. Therefore, WSIs are most often stored in a tiled format, so that only the viewed portion of the image (tile) is loaded into memory and rendered \cite{MarquesGodinho_2017}. Additionally, to support efficient zooming, WSIs are stored in a pyramidal structure, where higher levels of the pyramid represent lower magnification levels. The highest level of the pyramid is generally a very low resolution thumbnail snapshot of the WSI, while the lowest is the full resolution image (i.e. big-flat TIFF). In this way,  different zoom levels can be captured and displayed in a WSI format efficiently--artificially replicating different zooming levels from optical microscopy \cite{Singh_2011,MarquesGodinho_2017}. Additionally, WSIs can be compressed before storage to reduce their filesize, often using JPEG, JPEG 2000, or LZW algorithms, which can reduce an image size by more than seven times \cite{Zarella_2018}. Alongside the WSI, metadata regarding patient, tissue specimen, scanner, and WSI information is stored for reference \cite{CPsurvey1, Herrmann_2018,Clunie_2020}. Due to their clinical use and importance, it is important to develop effective storage solutions for these WSI data files and metadata, allowing for robust data management, querying of WSIs, and efficient data retrieval \cite{Wang_2012, Rao_2018}. 

To develop CPath CAD tools in a widespread and general manner, a standardized format for WSIs and their corresponding metadata is essential \cite{Herrmann_2018}. However, there is a general lack of standardization for WSI formats outputted by various scanners, as shown in Table \ref{tab:1ca}, especially regarding metadata storage. The Digital Imaging and Communications in Medicine (DICOM) standard provides a framework for biomedical image format and data management and has been extended to the CPath field through Supplement 145 \cite{Clunie_2020,Singh_2011}. Some research has shown that the use of the DICOM standard allows for efficient data access and greater interoperability between different centres and different CPath-related devices \cite{Herrmann_2018}. However there is currently a lack of widespread adoption \cite{Herrmann_2018,CPsurvey1,Clunie_2018,Clunie_2020}, reflected in Table \ref{tab:1ca} where only two recorded scanners are DICOM-compliant. Notably, with regards to metadata, DICOM provides a systematic format detailing a variety of relevant medical information, consistent with DICOM standards in other medical imaging fields \cite{Clunie_2020}. The expansion through Supplement 145 also adds pyramid-tiling for WSIs, a format that is directly beneficial to creation of CPath and CAD tools. While many scanners have adopted the pyramid-type scheme for image data, they have not fully adopted the DICOM image format, outputting in either TIFF, BigTIFF format, or in TIFF-derivatives such as SVS or GTIFF. While the TIFF format allows for semi-structured metadata \cite{MarquesGodinho_2017}, the consistency in metadata structure offered by DICOM is an advantage over the former \cite{Clunie_2020}. 

Apart from storage format, a general system for storing and distributing WSIs is also an important pillar for CPath. Whereas in other medical imaging fields such as radiology, images are often stored in a picture archiving and communications systems (PACS) in a standardized DICOM format, with DICOM storage and retrieval protocols \cite{MarquesGodinho_2017}, the need for standardization persists in pathology for WSI storage solutions. Few works have proposed solutions to incorporate DICOM-based WSIs in a PACS, although some research has successfully implemented a WSI PACS consistent using the DICOM standard using a web-based service for viewing and image querying \cite{MarquesGodinho_2017}.

\subsection{Organs and Diseases}\label{sec:organs_and_diseases}

The following Appendix section is a supplement to section \ref{sec:organs_diseases}. Details are provided for several notable works in CPath for the organ types listed in the subsection.

\textbf{\textit{Breast}}

\begin{itemize}
    \item A subset of breast-focused research studies the correlation between tissue morphology and molecular differences. Molecular testing can often be preferred over tissue morphology assessments when selecting breast cancer treatments as it provides objective and reproducible disease classifications. For instance, the connection between epithelial patterns with various molecular predictions and heatmaps was used to clearly visualize this correlation in \cite{38}.

    \item In \cite{578} a  proposed U-NET based architecture called piNET is used for cell detection and classification in order to calculate the proliferation index (PI) of the Ki67 biomarker. The network classifies cells as Ki67+/Ki67- and uses this classification to calculate the PI. The architecture is able to achieve a PI accuracy of $85.2\%$, higher than the accuracies of other models \cite{578}.

    \item A two-stage CNN, one for patch-level feature extraction and the other for classification, was proposed and achieved $95\%$ accuracy for classifying normal, benign, in situ carcinoma, and invasive carcinomas \cite{47}.

    \item In \cite{516}, several well-known models (DenseNet121, ResNet50, VGG16, and Xception), are compared with their own CNN model named lightweight convolutional neural network (LCNN) for the detection of breast cancer metastasis to axillary lymph nodes (ALN).

    \item Further research has proposed an automated patient-level tumor segmentation and classification system that takes full use of diagnosis information hints from pathologists \cite{616}. A multi-level view DeepLabV3+ (MLV-DeepLabV3+) was created to investigate the differentiating aspects of cell characteristics between tumor and normal tissue for tumor segmentation. Furthermore, expert segmentation models were chosen and merged using Pareto-front optimization to mimic expert consultation and provide a flawless diagnosis.
\end{itemize}

\textbf{\textit{Prostate}}

\begin{itemize}

    \item Most works focus on prostate aim to classify cancer based on Gleason scoring. To aid in that effort, a U-Net model for object semantic segmentation is created in \cite{618}, with the goal of precisely labeling each pixel in an image as belonging to either foregrounds, which may contain glands, or background.

    \item A method is proposed in \cite{80} that trains a model on both epithelial cell detection and Gleason grade prediction tasks to achieve better performance in both tasks than models trained on either of the tasks alone. Further, some works have investigated epithelial cell detection to explore data augmentation and stain/color normalization techniques \cite{48,187} which demonstrates the importance of epithelial cell features as an indicator for prostate cancer detection.

    \item Work in \cite{614} focused on classifying glands, gland boundary regions, and stroma. The authors opted for two classic classifiers: support vector machines (SVM) and Convolutional Neural Networks (CNNs), finding the SVM to perform best by offering high accuracy and good indicators of regions which are present with high probability. The output of the SVM classifier could help pathologists locate existing glands, saving them a significant amount of time from actively searching for them.
\end{itemize}

\textbf{\textit{Lung}}

\begin{itemize}
    \item To further aid diagnosis, prognosis, and treatment decisions, a novel method for nuclei detection and characterization is introduced in \cite{85} using an unsupervised autoencoder network to learn without the use of any annotations. The unsupervised autoencoder is used to construct a CNN that only requires $5\%$ of training data to generate comparable results to the SOTA on supervised lymphocyte and nuclei tasks, thereby reducing the need for extremely large annotated datasets.

    \item In \cite{5}, the work not only classifies adenocarcinoma and squamous cell carcinoma, but also predicts the 10 most commonly mutated genes in adenocarcinoma. The findings indicate the presence of genotype-phenotype correlations for lung cancer tissues, and paves the way for cancer classification and mutation predictions of other types of less common lung cancers.
\end{itemize}

\textbf{\textit{Colon and Rectum}}

\begin{itemize}
    \item An interesting detection application for MSI is implemented in \cite{18}. In this work, ResNet-18 is used to predict MSI from H\&E slides with an AUC of $0.84$, although performance was reduced with tissue samples from different ethnicities. As MSI requires extensive additional testing which is not always performed \cite{18}, the study highlights the applicability of deep learning in detecting this important prognosis indicator.

    \item The work in \cite{530} presents a weakly-supervised model named the Slide Level Annotation Model (SLAM) based on ShuffleNet \cite{shufflenet} that can be trained to detect genetic/molecular changes, including MSI or BRAF mutation, in colorectal WSIs. The results show improvement over SOTA models, and a visualization heatmap is generated which allow for improved result interpretability and analysis.
\end{itemize}

\textbf{\textit{Bladder}}

\begin{itemize}
    \item In \cite{95}, bladder cancer grade classification is explored using a large dataset of 915 WSIs which ends up outperforming 17 pathologists by an average of $10\%$. The model in this study has been integrated into an end-to-end diagnostic tool that provides interpretable cellular-level ROI visualization and natural language descriptions of histology slides. However, there was also a diagnostic disagreement of $23\%$ between the system and pathologists, which could hinder the diagnostic process and consequently limit the overall productivity.
\end{itemize}

\textbf{\textit{Kidney}}

\begin{itemize}
    \item For effective patient prognosis, tissue microarray analysis is typically used to identify biomarkers. Currently, this process is time-consuming and prone to error, especially due to the heterogeneity of nuclei. A random forest classifier was proposed to more efficiently detect cancerous nuclei in MIB-1 stained tissue micro-array spots and predict the survival rate for renal cell carcinoma (RCC) patients in \cite{44}. The results show that there is a significant difference in survival times for patients with high and low proliferating tumors, and further state MIB-1 staining as a key prognostic factor for the survival chance of RCC.
    
    \item To aid in determing donor organ acceptance in kidney transplants, \cite{105} uses frozen kidney sections as input data to identify the percentage of glomerulosclerosis.
\end{itemize}

\textbf{\textit{Brain}}
\begin{itemize}
    \item The achieved state-of-the-art \textcolor{r5col}{SOTA} classification accuracy on the 2014 MICCAI Grand Challenge dataset was achieved in \cite{34}. However, on completely unseen datasets, performance varied from $84\%$ to $93\%$. This shows the complexity of the diseases; low grade glioma (LGG) versus glioblastoma multiforme (GBM) classification is not a trivial task. Not only do their appearances vary in pathological samples, the diagnosis is often made from a few distinct features in a small slide region \cite{34}.
\end{itemize}

\textbf{\textit{Liver}}
\begin{itemize}
    \item A notable work for liver cancer classification evaluates a pathologist's performance using a liver cancer diagnostic tool for the diagnosis of HCC and cholangiocarcinoma \cite{65}. Despite lower performance than pathologists, the tool's decisions directly affected pathologists' decisions. Correct model predictions increased pathologists' average accuracy, while incorrect predictions lowered average accuracy. Furthermore, pathologists frequently consulted the model's predictions for difficult cases. This confirms the potential use of deep learning models as an AI diagnostic tool to provide knowledgeable second opinions.
\end{itemize}

\textbf{\textit{Lymph Nodes}}

\begin{itemize}

    \item One of the earliest histopathology challenges, ICPR2010, targets lymphocyte and centroblast counting \cite{gurcan_madabhushi_rajpoot_2010}.

    \item The importance of the lymph nodes in cancer diagnosis is notably addressed in the CAMELYON16 and CAMELYON17 Challenges, in which participants classify lymph node metastases \cite{31}. Metastases in breast regional lymph nodes are classified based on size: micrometastasis, macrometastasis, and isolated tumour cells (ITCs). The ITC classification accuracy was less than $40\%$ for all top teams. This indicates that there was extreme difficulty in detecting ITCs, most likely due to the small size and variability. These results suggest that further improvements can be made by introducing more true positives of ITC data, or incorporating IHC stain information as an additional layer of information to improve detection robustness.
\end{itemize}

\textbf{\textit{Organ Agnostic}}

\begin{itemize}
    \item Nuclei segmentation of epithelial, inflammatory, fibroblast and miscellaneous tissues is performed across seven different organs in \cite{70}. The method attempts to generalize across a large variety of datasets for increased usability and scalability in a clinical setting.
\end{itemize}
\subsection{Ground Truth Labelling and Annotation}
\label{appendix_annotation}
\textbf{\textit{\textit{Patient-level Annotation}}}
Patient-level annotations assign a single label to a single patient and come from case reports that can address multiple WSIs from a primary organ site. In addition to the WSIs, the Laboratory Information System (LIS) may also contain additional metadata, diagnostic information, and analytical or synoptic report information \cite{pantanowitz2007medical}. Notably, the LIS can store specimen type, molecular and genetic tests, patient medical history, and clinical variables such as the patient's age and gender \cite{550, 611}. 


\textbf{\textit{Slide-level Annotation}}
Slide-level annotations designate labels for a single WSI, which encompasses diagnosis and cancer information \cite{493, 474}. In comparison to the patient-level, this level of focus provides a more precise tissue location for the provided diagnosis \cite{547, 573, 639}.

\textbf{\textit{ROI-level Annotation}}
ROI-level annotation identifies regions within a slide that can be of either diagnostic or analytical relevance to a pathologist. Regions themselves can be designated using two methods: (1) bounding boxes \cite{363, 216, 237} or (2) pixel-wise masks that are augmented on the WSIs \cite{92, 2, Camelyon17}. Importantly, each ROI is considered to be a single class \cite{wang2019pathology}, but the labels represent more detailed tissue structures providing more specific and detailed diagnostic information than patient and slide levels \cite{279}, ultimately being more applicable in disease diagnosis tasks \cite{500}. 

\textbf{\textit{Patch-level Annotation}}
Patch-level annotation is done on mosaic tiles (usually in a square shape) extracted from the WSI/ROI with a given field of view (FOV). Most deep-learning models are trained at the patch level, which contains anatomical structures of tissues and cells. Patches are either single-labeled or multi-labeled according to the taxonomical labeling workflow \cite{14}. One key aspect for patch-level annotation is determining the optimum FOV to encompass enough tissue classes \cite{tokunaga2019adaptive}, as considering smaller or bigger FOV can provide different advantages, as demonstrated in Figure \ref{fig:annotation_overview} for the patch-level.

\textbf{\textit{Patch-Size Selection}}
The choice of the patch size is limited by the computational complexity of the hardware that is used for training CAD tools. For instance, the majority of deep learning pipelines accept image sizes of less than 300$\times$300 pixels \cite{chang2019artificial, xiang2021dsnet, 536, 607, 279}. The size of the FOV needs to be determined such that acceptable levels of morphological tissues will be covered within that patch. Accordingly, the pixel resolution is determined given a certain FOV and patch size. Given the above factors, if a larger FOV is required, then pixel resolution is limited which translates to information loss. In comparison, if higher pixel resolution is required, then the FOV will be limited accordingly which may exclude cellular/architectural relevance pertaining to the underlying class representation \cite{588}. To mitigate this tradeoff, larger image dimensions are required which consequently increases the computational power required for patch processing (e.g. high RAM GPU memory or parallelized multi-GPU processing) \cite{tokunaga2019adaptive}.

\textbf{\textit{Pixel-level Sizing}}
Pixel-level annotation requires labelling each pixel as a specified class. In this level, features are simple to extract and sufficient for describing the images as they encompass color and texture information \cite{shirazi2018automated}. However, there is a lack of biological interpretability as the other levels of annotation more appropriately describe characteristics of the cellular and tissue structures \cite{499}. A solution based on human-interpretable image features can include histological knowledge and expert annotations that can describe different cell anatomies such as the stroma, the nuclei of the cells, and the size and shape of tumor regions, as well as the texture of the tissues and the location of tumor-infiltrating lymphocytes \cite{572}.

Pixel-wise masks are differentiated from pixel-level annotations in that when an ROI mask of this type is tiled into multiple instances (i.e. patches), each sample is considered a single class. The ROI-level is in contrast to the pixel-level annotations, wherein the latter is defined to include all annotations where each patch can contain several class types.

\begin{table}[t]
\caption{Commercially available annotation software along with their manufacturing company and available input slide formats.}
\label{tab:1cb}
\scriptsize
\begin{tabular}{l}
\specialrule{.2em}{.1em}{.1em} 
Company: Annotation Tool (Input Format)\\
\noalign{\smallskip}
\specialrule{.2em}{.1em}{.1em} 
    Leica Biosystems: \href{https://www.leicabiosystems.com/digital-pathology/manage/}{\color{CiteColor}{Aperio eSlide Manage}} (JFIF, JPEG2000, PMM) \\
    \hline 
    Pathcore: \href{https://pathcore.com/sedeen/}{\color{CiteColor}{Sedeen Viewer}} (Aperio SVS, Leica SVN, TIFF, JPEG2000) \\ \hline
    Indica: \href{https://indicalab.com/}{\color{CiteColor}{Halo}} (TIFF/SVS) \\
    \hline 
    Objective Pathology: \href{https://www.objectivepathology.com/}{\color{CiteColor}{MyObjective}} (Scanner-wide compatibility) \\
    \hline 
    ASAP: \href{https://computationalpathologygroup.github.io/ASAP/}{\color{CiteColor}{ASAP}} (Multiple formats through OpenSlide) \\
    \hline 
    SiliconLotus: \href{https://www.siliconlotus.com/}{\color{CiteColor}{SiliconLotus}} (Not specified) \\
    \hline 
    Augmentiqs: \href{https://www.augmentiqs.com/microscope-annotation-software/}{\color{CiteColor}{Annotation Software Suite}} (Not specified)\\
    \hline 
    QuPath: \href{https://qupath.github.io/}{\color{CiteColor}{QuPath}} (Multiple formats, Bio-formats and OpenSlide) \\
    \hline 
    Proscia: \href{https://proscia.com/concentriq-research/}{\color{CiteColor}{Concentriq}} (Not specified) \\
    \hline 
    Visiopharm A/S: \href{https://visiopharm.com/visiopharm-digital-image-analysis-software-features/viewer/}{\color{CiteColor}{VisioPharm}} (Not specified) \\
    \hline 
    Hamamatsu: \href{https://nanozoomer.hamamatsu.com/jp/en/Software/index.html}{\color{CiteColor}{NDP}} (JPEG) \\
    \hline 
    Roche: \href{https://diagnostics.roche.com/global/en/products/product-category.html#080-00}{\color{CiteColor}{Ventana Companion Image Analysis}} (BIF, TIFF, JPG2000, DICOM compliant) \\
    \hline 
    Huron: \href{http://www.hurondigitalpathology.com/resources/}{\color{CiteColor}{HuronViewer}} (BigTIFF, FlatTIFF, DICOM compliant) \\
    \hline
    Philips: \href{https://www.usa.philips.com/healthcare/resources/landing/philips-intellisite-pathology-solution}{\color{CiteColor}{Intellisite}} (iSyntax Philips proprietary file) \\
    \hline 
    3DHistech: \href{https://www.3dhistech.com/products-and-software/software/digital-microscopes-viewers/}{\color{CiteColor}{CaseViewer}} (JPG, PNG, BMP, TIFF) \\
    \hline 
    {AnnotatorJ} \cite{hollandi2020annotatorj}: \href{https://github.com/spreka/annotatorj}{\color{CiteColor}{AnnotatorJ}} (JPG, PNG, TIFF) \\
    \hline 
    {NuClick} \cite{281}: \href{https://github.com/navidstuv/NuClick}{\color{CiteColor}{NuClick}} (Not specified) \\ 
\noalign{\smallskip}
\specialrule{.2em}{.1em}{.1em}
\end{tabular}
\end{table}



\subsection{\label{subsec:supp_dat}Surveyed Datasets}
\subsubsection{\label{subsec:supp_dat_criteria}Table Creation Details}
For each dataset recorded in the literature, a collection of information was collected. This information was organized into 10 categories, listed below. The full table is given in Table \ref{DC}: 
\begin{enumerate}
    \item \emph{Dataset Name}: The name of the dataset, if given. If no name is given, then a name was given for book-keeping purposes. 
    \item \emph{References}: The works that use this dataset are listed.
    \item \emph{Availability}: A hyperlink to the dataset, when publicly available or available for request directly is provided. 
    \item \emph{Stain type}: The type of stain used.
    \item \emph{Size}: Describes the number of WSIs, where this information is available, or the number of patches present in the dataset.
    \item \emph{Resolution ({\textmu}m)/ Magnification}: Presents the resolution, in micrometers along with the magnification in the format {\textmu}m/Magnification. If a piece of information is unavailable (either resolution or magnification) this information is omitted from the table.
    \item \emph{Annotation Type}: Describes the annotation granularity present in the dataset (patient, slide, ROI/ROI mask, patch, pixel) where available.
    \item \emph{Label Structure}: Whether each image in the dataset has a single label associated with it, or multiple. Datasets, where each image has only a single label associated with it, are labeled with \emph{S}, whereas those with multiple labels are labeled with \emph{M}. 
    \item \emph{Classes}: The number of classes available, where this count is meaningful. Where it is more helpful to describe the format of ground truth (ex. nucleus pixel locations), this is written instead.
    \item \emph{Class Balance (CB)}: Datasets which are balanced are marked with a \emph{B}, whereas those which are imbalanced are marked as \emph{I}. Those where this information is unavailable are marked with an \emph{U}. 
\end{enumerate}

\subsection{\label{subsec:supp_organ}Organ Overview}
For each paper recorded in the literature, a collection of information about their specific goal was collected. This information was categorized by organ and arranged into a table, the organs being: Basal/Epithelium, Bladder, Brain, Mouth/Esophagus, Breast, Liver, Lymph Node, Prostate/Ovary, Kidney, Lung, Pancreas, Thyroid, Stomach/Colon. Below will be the explanation of each column in Table \ref{OV}:

\begin{enumerate}
    \item \emph{References:} Reference number of the paper that involved the specified task.
    \item \emph{Tasks:} Specific target goal that the work wanted to achieve, this range from different types of detection, classification, and segmentation to prognosis and diagnosis.  
    \item \emph{Disease Specification:} Describes the pathology of the target goal of the paper.
    \item \emph{Methods:} Define the different machine learning methods used to achieve the proposed target task of the paper.
  
\end{enumerate}

\subsection{\label{subsec:supp_tech}Technicalities by Task}
For each paper recorded in the literature, a collection of information on the Neural Network architectures used was organized and categorized by its specific task. It was found that across the majority of papers, the following five tasks were the most prevalent: Detection, Disease Diagnosis, Segmentation, WSI Processing, and Patient Prognosis. At the end of the table, an \textbf{Other Task } section was added to attach other works that don't follow the selected tasks. Below will be the explanation of each column in Table \ref{TT}:

\begin{enumerate}
    \item \emph{References:} Reference number of the paper that involved the specified task.
    \item \emph{Tasks Specification:} Describes the pathology of the target goal of the paper.
    \item \emph{Architecture:} Defines the different Neural Network architectures used to achieve the proposed target task of the paper.
    \item \emph{Datasets:} Name of the datasets used for the specified task (see Table \ref{DC} for information on datasets).
  
\end{enumerate}

\newpage

\footnotesize	
\scriptsize{
\begin{landscape}

}
\end{landscape}
\newpage
\onecolumn
\subsection{\label{subsec:supp_model_card}Model Card Categorization}
For comprehensive review on model cards and updated information, please refer to our \href{https://github.com/AtlasAnalyticsLab/CPath_Survey}{GitHub} repository for more information. The following lists examples of our model cards used in preparation of our survey paper.
\subsubsection{\label{subsec:supp_model_card_temp}Template}
\textbf{Model-Card for Categorizing Computational Pathology Papers} \\
\textbf{Step-1) Paper Summarization: } \\
Summarize the paper in terms of (A) Goal/Problem of the paper to be solved; (B) Why the problem introduced by the authors is important to the community in terms of Technical Novelty, Comprehensive Experiment, New Insights, Explainability; and (C) Overall conclusion of the paper. \\
\textbf{Step-2) Model Card Table Categorization: } \\
The following is a model-card for each paper to populate the table accordingly. Find relevant information within each category that is reported in the paper. Try to compile it efficiently and populate each sub-type within each category. \\
\begin{center}
\begin{tabularx}{\textwidth}{ |l|X| } 
 \hline
    \textbf{Keywords} & comma separated list \\
    \hline
    \textbf{Organ Application} & \textbf{Organ:} \\
    & \textbf{Task:} \\
    \hline
    \textbf{Dataset Compilation} & \textbf{Name:} \\
    & \textbf{Availability:} \\
    & \textbf{Dataset Size:} (\#patches/\#slides/\#images) \\
    & \textbf{Image Resolution:} \\
    & \textbf{Staining Type:} \\
    & \textbf{Annotation Type:} (region/patch/slide-level) \\
    & \textbf{Histological Type:} (cellular/tissue ROI/etc, I.e., on what basis is it labeled) \\
    & \textbf{Label Structure:} (single label/multi label) \\
    & \textbf{Class Balance:} (is size of dataset balanced across each classes) \\
    \hline
    \textbf{Technicality} & \textbf{Model:} (architecture/transfer learning/output format)  \\                      
    & \textbf{Training Algorithm:} (end-to-end/separately staged) \\
    & \textbf{Code Availability:} (give source) \\
    \hline
    \textbf{Data Processing} & \textbf{Image Pre-processing:} (patching, data augmentation, color normalization) \\
    & \textbf{Output Processing:} \\
    \hline
    \textbf{Performance Summary} & \textbf{Evaluation Metrics: } \\
    & \textbf{Notable Results:} Numerical result for strongest performing model. \\
    & \textbf{Comparison to Other Works:} Comparison to state-of-the-art models (one sentence) \\
    \hline
    \textbf{Novelty} & \textbf{Medical Applications/Perspectives: } \\
    & \textbf{Technical Innovation:} (algorithms for processing or deep learning, new metrics) \\
    \hline
    \textbf{Explainability} & \textbf{Visual Representations: }(feature distribution, heatmaps, tsne, gradCAM, pseudocode, etc.) \\
    \hline
    \textbf{Clinical Validation} & \textbf{Usage in Clinical Settings:} Has the work been used by pathologists in clinical setting? \\
    & \textbf{Suggested Usage:} How can the work be used by pathologists? \\
    & \textbf{Performance Comparison:} Has the model performance been compared to that of pathologists? \\
    \hline
    \textbf{Caveats and Recommendations} & \textbullet{ Personal comments on the paper} \\
    & \textbullet{ Relevant info from other papers} \\
    & \textbullet{ Criticism and limitations of the work} \\
    \hline
\end{tabularx}
\end{center}

\textbf{Step-3) Citation:} BibTeX Citation.

\newpage

\subsubsection{\label{subsec:supp_model_card_samp}Samples} 

\textbf{Paper: Pathologist-level classification of histologic patterns on resected lung adenocarcinoma slides with deep neural networks} \\
\textbf{Summary: } \\
Classification of histological structures in lung adenocarcinoma tissue is important for patient prognosis and treatment plans. Some histological patterns (such as lepidic patterns) are associated with better survival rates, whereas others (micropapillary and solid patterns) are associated with poor prognoses. The identification of these histological patterns is a challenge, as 80\% of adenocarcinoma tissue samples contain a mixture of different patterns, and the qualitative classification criteria can result in variance in diagnosis between different pathologists. Automated analysis and classification of tissue structures through convolutional neural networks has been a compelling area of research. This paper presents a variant of ResNet, ResNet18 to perform a patch-based classification amongst the different lung adenocarcinoma histological patterns. A heatmap for the entire WSI is made using the probability score per patch, with low-confidence patches being discarded. The performance of the model is compared with 3 expert pathologists to determine relative performance. The study concludes that the model has high performance, with results on par with expert pathologists. \\\\
\textbf{Categorization:} 
\begin{center}
\begin{tabularx}{\textwidth}{ |l|X| } 
 \hline
    \textbf{Keywords} & Deep learning, convolutional neural networks, lung adenocarcinoma, multi class, ResNet \\
    \hline
    \textbf{Organ Application} & \textbf{Organ:} Lung\\
    & \textbf{Task:} Histologic pattern classification in lung adenocarcinoma\\
    \hline
    \textbf{Dataset Compilation} & \textbf{Name:} Dartmouth-Hitchcock Medical Centre in Lebanon, New Hampshire\\
    & \textbf{Availability:} Unavailable due to patient privacy constraints. Anonymized version available upon request. \\
    & \textbf{Dataset Size:} 422 WSIs, 4 161 training ROIs, 1 068 validation patches \\
    & \textbf{Image Resolution:} 20$\times$ magnification\\
    & \textbf{Staining Type:} H\&E \\
    & \textbf{Annotation Type:} Region-level training set, patch-level validation set, slide-level test set \\
    & \textbf{Histological Type:} Lepidic, acinar, papillary, micropapillary, solid, and benign. \\
    & \textbf{Label Structure:} Single label \\
    & \textbf{Class Balance:} Imbalanced, with significantly fewer papillary patterns in all data. \\
    \hline
    \textbf{Technicality} & \textbf{Model:} ResNet model with 18 layers \\      
    & \textbf{Training Algorithm:} \\
    & \textbullet{ Multi-class cross-entropy loss} \\
    & \textbullet{ Initial learning rate of 0.001} \\
    & \textbullet{ Learning rate decay by factor of 0.9 per epoch} \\
    & \textbf{Code Availability:} \href{https://github.com/BMIRDS/deepslide}{https://github.com/BMIRDS/deepslide} \\
    \hline
    \textbf{Data Processing} & \textbf{Image Pre-processing:} \\
    & \textbullet{ Created training ROIs by selectively cropping regions of 245 WSIs.} \\
    & \textbullet{ Spliced 34 validation WSIs into 1 068 224x224 patches. }\\
    & \textbullet{ Colour channel normalization to mean and standard deviation of entire training set.} \\
    & \textbullet{ Data augmentation by rotation; flipping; and random colour jittering on brightness, contrast, hue, and saturation.} \\
    & \textbf{Output Processing:} Low-confidence predictions filtered out for predictions below a threshold. Thresholds are determined by a grid search over classes, optimizing for similarity between the trained model and the validation data.\\
    \hline
    \textbf{Performance Summary} & \textbf{Evaluation Metrics: }F1-Score, AUC\\
    & \textbf{Notable Results:} F1-Score of 0.904 on validation set, AUC greater than 0.97 for all classes. \\
    & \textbf{Comparison to Other Works:} ResNet18, 34, 50, 101, 152 compared for performance to choose optimal depth. All had similar accuracies on validation set, so chose ResNet18 for lower model complexity.\\
    \hline
    \textbf{Novelty} & \textbf{Medical Applications/Perspectives: }Potential platform for quality assurance of diagnosis and slide analysis.\\
    & \textbf{Technical Innovation:} First paper to attempt to classify based on histological lung adenocarcinoma subtypes. \\
    \hline
    \textbf{Explainability} & \textbf{Visual Representations: }Heatmaps for patterns detected, AUC curve for each class \\
    \hline
    \textbf{Clinical Validation} & \textbf{Usage in Clinical Settings:} N/A \\
    & \textbf{Suggested Usage:} \\
    & \textbullet{ Could be integrated into existing lab information management systems to provide second opinions to diagnoses. } \\
    & \textbullet{ Visualization of a slide could highlight important tissue structures.} \\
    & \textbullet{ Could help facilitate tumour diagnosis process by automatically requesting genetic testing based on histological data for patient.} \\
    & \textbf{Performance Comparison:} \\ 
    & \textbullet{ On par with pathologists for all evaluated metrics} \\
    & \textbullet{ Model in agreement 66.6\% of the time with pathologists on average, with robust agreement (agreement with 2/3 of the pathologists) 76.7\% of the time. } \\
    & \textbullet{ WSI region annotation differences between pathologist and model are compared for a sample slide.} \\
    \hline
    \textbf{Caveats and Recommendations} & \textbullet{ Data taken from one medical centre, so may not be representative of lung adenocarcinoma morphology} \\
    & \textbullet{ Dataset relatively small compared to other deep learning datasets, with some classes having very few instances} \\
    \hline
\end{tabularx}
\end{center}
\textbf{Citation:} \\
@article\{wei2019pathologist, \\
  title=\{Pathologist-level classification of histologic patterns on resected lung adenocarcinoma slides with deep neural networks\}, \\
  author=\{Wei, Jason W and Tafe, Laura J and Linnik, Yevgeniy A and Vaickus, Louis J and Tomita, Naofumi and Hassanpour, Saeed\}, \\
  journal=\{Scientific reports\}, \\
  volume=\{9\}, \\
  number=\{1\}, \\
  pages=\{3358\}, \\
  year=\{2019\}, \\
  publisher=\{Nature Publishing Group UK London\} \\
\}

\newpage

\textbf{Paper: Classification of lung cancer histology images using patch-level summary statistics } \\
\textbf{Summary:}\\
The classification of non-small cell lung cancer WSIs as either lung adenocarcinoma (LUAD) or lung squamous cell carcinoma (LUSC) is an important task in diagnosis and treatment planning. Manually classifying these WSIs is a laborious and subjective task that is often complicated by poorly differentiated tissue structures within the slide. Automated classification of WSIs may facilitate the analysis of non-small cell lung cancers. This paper proposes a new 3-class network for effective classification of tissue regions within a WSI. It uses a modification of the ResNet50 architecture, ResNet32 to create probability maps of LUAD/LUSC/non-diagnostic pixels in the WSI. Features from these probability maps are next extracted and fed into a random forest classifier for the final classification. The model achieves the greatest accuracy of 0.81 in the Computational Precision Medicine Challenge and provides a new method of classification for non-small lung cancer histological images. \\\\
\textbf{Categorization: }
\begin{center}
\begin{tabularx}{\textwidth}{ |l|X| } 
 \hline
    \textbf{Keywords} & Non-small cell lung cancer, histology image classification, computational pathology, deep learning \\
    \hline
    \textbf{Organ Application} & \textbf{Organ:} Lung\\
    & \textbf{Task:} Classification between non-small cell lung cancer types\\
    \hline
    \textbf{Dataset Compilation} & \textbf{Name:} Computational Precision Medicine at MICCAI 2017\\
    & \textbf{Availability:} Unavailable, link on MICCAI 2017 website unreachable\\
    & \textbf{Dataset Size:} 64 WSIs \\
    & \textbf{Image Resolution:} 20$\times$ magnification \\
    & \textbf{Staining Type:} H\&E \\
    & \textbf{Annotation Type:} Pixel-level and Slide-level \\
    & \textbf{Histological Type:} \\
    & \textbullet{ At pixel-level, classifies as lung adenocarcinoma (LUAD), lung squamous cell carcinoma (LUSC), and non-diagnostic (ND)} \\
    & \textbullet{ At slide-level, LUAD or LUSC} \\
    & \textbf{Label Structure:} Single label\\
    & \textbf{Class Balance:} Balanced dataset at the slide level, 32 LUAD and 32 LUSC \\
    \hline
    \textbf{Technicality} & \textbf{Model:} \\ 
    & \textbullet{ Ensemble ML model}\\
    & \textbullet{ Variant of ResNet50, called ResNet32 with 32 layers and 3x3 kernel, as compared to 7x7 kernel with ResNet50. }\\
    & \textbullet{ 50 statistical and morphological features extracted from probability maps generated by ResNet32. The top 25 are selected for best class separability and used as input to a random forest.} \\
    & \textbf{Training Algorithm:} Separately staged, ResNet32 creates probability maps, then random forest generates final prediction for each WSI \\
    & \textbf{Code Availability:} Unavailable \\
    \hline
    \textbf{Data Processing} & \textbf{Image Pre-processing:} \\
    & \textbullet{Splicing of slides into 256$\times$256 patches, then random cropping into 224$\times$224 patches } \\
    & \textbullet{Reinhard stain normalization } \\
    & \textbullet{Random crop, flip, rotation data augmentation }\\
    & \textbf{Output Processing:} N/A\\
    \hline
    \textbf{Performance Summary} & \textbf{Evaluation Metrics: }Accuracy\\
    & \textbf{Notable Results:} \\
    & \textbullet{ ResNet32 with Random Forest achieves 0.81 accuracy over WSI} \\
    & \textbullet{ Results superior to ResNet32 with Maximum Vote, which had 0.78 accuracy. Features for the random forest are tailored for WSI classification, and so can achieve higher performance. } \\
    & \textbf{Comparison to Other Works:} Compared ResNet32 to VGG, GoogLeNet, and ResNet50, with higher average classification accuracy.  \\
    \hline
    \textbf{Novelty} & \textbf{Medical Applications/Perspectives: }Automated distinguishing of LUAD tissue from LUSC could be done at scale to assist pathologists in diagnosis and treatment planning for patients.\\
    & \textbf{Technical Innovation:} \\
    & \textbullet{ First 3-class network for classification of WSI into diagnostic/nondiagnostic areas} \\
    & \textbullet{ Ensemble method resulted in greatest accuracy at the MICCAI 2017 competition.} \\
    \hline
    \textbf{Explainability} & \textbf{Visual Representations: }Probability maps for each pixel-level class\\
    \hline
    \textbf{Clinical Validation} & \textbf{Usage in Clinical Settings:} N/A \\
    & \textbf{Suggested Usage:} Automated distinguishing of LUAD and LUSC slides could aid pathologists in treatment planning.\\
    & \textbf{Performance Comparison:} N/A \\
    \hline
    \textbf{Caveats and Recommendations} & \textbullet{Because features for random forest training are chosen based on categorization of lung tissue samples, may not be able to generalize well to other tissue types.} \\
    \hline
\end{tabularx}
\end{center}
\textbf{Citation:} \\
@inproceedings\{graham2018classification, \\
  title=\{Classification of lung cancer histology images using patch-level summary statistics\}, \\
  author=\{Graham, Simon and Shaban, Muhammad and Qaiser, Talha and Koohbanani, Navid Alemi and Khurram, Syed Ali and Rajpoot, Nasir\}, \\
  booktitle=\{Medical Imaging 2018: Digital Pathology\}, \\
  volume=\{10581\}, \\
  pages=\{327--334\}, \\
  year=\{2018\}, \\
  organization=\{SPIE\} \\
\}

\newpage

\textbf{Paper: Digital pathology and artificial intelligence (Review)}\\
\textbf{Summary:} \\
The advent of cheaper storage solutions, faster network speed, and digitized WSIs has greatly facilitated the presence of digital pathology in modern pathology. Particularly, WSIs allow for the development and integration of automated AI tools for histopathological analysis into the pathologist’s workflow. AI tools have the potential to increase the efficiency of diagnostics and improve patient safety and care. However, histological analysis comes with several challenges, including the large size, different potential image magnifications, presence, and variation of stain color information, and z-axis information (in the thickness of the slide). These challenges make it difficult for a human viewer to extract all available information and provide important issues that an AI tool must overcome. This paper outlines several different areas in which AI may be applied in digital pathology, namely education, quality assurance (QA), clinical diagnosis, and image analysis. The potential uses are outlined as follows:

\begin{itemize}
    \item Education: Through the digitization of slides, education can be enhanced. As slides no longer need to be viewed through a microscope, and images can be zoomed into and panned, convenience can be increased without sacrificing the quality of education. Synthetic tissue sample images using GANs can be used to easily create test material in trainees, as well as evaluate cognitive biases in practicing pathologists.
    \item Quality Assurance: Can help pathologists remain updated in their field and check for lab proficiency of diagnoses, as well as monitor for inter-observer variance.
    \item Clinical Diagnosis: AI can aid in the preparation of digital slide imagery, such as in reducing the frequency of out-of-focus areas in slides. Color and stain normalization methods using AI-based models are another possible area of application.
    \item Image Analysis: AI can be used to process the data, including in nuclear segmentation and ROI detection.
\end{itemize}

AI systems have several different limitations. AI models have been criticized as being black box models. Explainability of decisions will need to be increased. While visualization techniques are being developed, these tend to reduce performance. Additionally, regulatory and economic effects of AI-based systems are unknown at this time. Some areas of future research for AI applications in computational pathology include one-shot learning and reinforcement learning. \\

\textbf{Citation:} \\
@article\{niazi2019digital, \\
  title=\{Digital pathology and artificial intelligence\}, \\
  author=\{Niazi, Muhammad Khalid Khan and Parwani, Anil V and Gurcan, Metin N\}, \\
  journal=\{The lancet oncology\}, \\
  volume=\{20\}, \\
  number=\{5\}, \\
  pages=\{e253--e261\}, \\
  year=\{2019\}, \\
  publisher=\{Elsevier\} \\
\} \\

\end{document}